\documentclass[12pt]{article}
\usepackage[a4paper, total={7in, 10in}]{geometry}
\usepackage{graphicx}	
\usepackage{amsmath}	
\usepackage{amssymb}	
\usepackage{subfigure}
\usepackage{sidecap}
\usepackage{tikz}
\usepackage{multirow}
\usepackage{authblk}
\usetikzlibrary{fadings}
\usetikzlibrary{decorations.pathreplacing}
\usepackage{hyperref}
\usepackage{txfonts}
\usepackage{newtxtext,newtxmath}
\usepackage{movie15}
\usepackage[T1]{fontenc}
\date{}

\makeatletter
\newcommand\ackname{Acknowledgements}
\if@titlepage
   \newenvironment{acknowledgements}{%
       \titlepage
       \null\vfil
       \@beginparpenalty\@lowpenalty
       \begin{center}%
         \bfseries \ackname
         \@endparpenalty\@M
       \end{center}}%
      {\par\vfil\null\endtitlepage}
\else
   \newenvironment{acknowledgements}{%
       \if@twocolumn
         \section*{\abstractname}%
       \else
         \small
         \begin{center}%
           {\bfseries \ackname\vspace{-.5em}\vspace{\z@}}%
         \end{center}%
         \quotation
       \fi}
       {\if@twocolumn\else\endquotation\fi}
\fi
\makeatother

\begin{document} 
\title{Predicting future astronomical events using deep learning}
\author[1,a]{Shashwat Singh\thanks{shashwat98singh@gmail.com}}
\author[2,a]{Ankul Prajapati}
\author[1,a]{Kamlesh N. Pathak}

\affil[1]{Sardar Vallabhbhai National Institute of Technology, Surat - 395007, India}
\affil[2]{Sorbonne Université, Paris - 75005, France}
\affil[a]{Bose.X TRIAC\thanks{https://www.bosex.org/}}

\maketitle

\begin{abstract}
In a quest towards an intelligent decision-making machine, the ability to make plausible predictions is the central pillar of its intelligence. A predicting algorithm's central idea is to understand the governing physical rules and make plausible and apt predictions based on the same governing laws. Extending the study towards the astrophysical phenomenon puts the model's ability to test since the model has to understand various parameters that govern the dynamics of the event and understand the spatial and temporal evolution by applying the plausible laws. This work presents a deep learning model to predict plausible future events that maintain spatial and temporal coherence. We have trained over two broad classes, the evolution of \texttt{Sa, Sb, S0}, and \texttt{Sd} galaxy mergers and evolution of gravitational lenses with a higher redshift of the foreground galaxy having $15M_{\odot}$. We extended our work towards developing a direct measure of the performance metric for any prediction algorithm. We thereby introduce a novel metric, Correctness Factor (CF), which directly outputs how accurate a prediction is.
\end{abstract}

\section{Introduction}

A theoretical framework, if correct, is sufficient enough to predict events in the near future and at the cosmic scale. It even enables to trace back and reason the circumstances that lead to the existing conditions. Thus it tends to develop a deterministic approach towards studying the cosmos. However, fluctuations at the minute level can lead to a domino wherein a completely different condition unfolds. This deviation from the prediction is extremely important and governs the direction and pace of our understanding. Thus astrophysical predictions have always been crucial but are extremely challenging essentially due to the two primary factors: rarity of the event and cosmic timescales at which these events occur. Secondly, even if any event occurs, the probability of a successful observation largely depends on the instrument's accuracy and precision. A well established theoretical framework can assist in the same in a sense by creating an alert prior to the event.

Astronomical data is usually a time-series sequence of particular spatial parameters that evolves during the given time sampled. A 1-dimensional signal is a time series evolution of certain fixed parameters evolving over discrete time samples. For a 2-dimensional sequence, which is the current focus of our work, is a spatiotemporal sequence wherein the data has space and time components. The quality of data and information corresponding to the data increases significantly on further increasing the dimensions. In terms of Deep Learning (DL), such sequence prediction problems are dealt with by learning from past events and making predictions based on such events. However, such problems are nontrivial due to the high dimensionality of the spatiotemporal sequences, especially when multi-step predictions have to be made. Moreover, building an effective prediction model for astronomical data is even more challenging due to the chaotic nature that governs the astrophysical systems' dynamics.

In this work of ours, we present a DL model that has been successfully employed in making astronomical prediction over two broad classes: \texttt{Sa, Sb, S0}, and \texttt{Sd} galaxy mergers and evolution of gravitational lenses with a higher redshift of the foreground galaxy. Working along the same lines, we also propose a novel metric, the Correctness Factor (CF), which directly measures how accurate a prediction is, bridging the gap between the mathematical formalism for metric for a spatiotemporal sequence-prediction requiring the least human interference. The paper has been structured such that Sec-\ref{overview} develops a general overview of the problem and how we attempt to deal with the same. Sec-\ref{Related works} discussing the previous works in the field, Sec-\ref{DP} is devoted to giving insights to our datasets and preparation methodology. In Sec-\ref{Proposed Model} we present the DL model that has been employed for predicting spatiotemporal sequence, and Sec-\ref{metric} we give insights to how we concluded our metric and validate it with test and experimentations in Sec-\ref{tt} and Sec-\ref{exp}. Finally, we conclude our results in Sec-\ref{conclude}.

\begin{figure*}[t]
\centering
    \subfigure{\includegraphics[width=2cm,height=2cm]{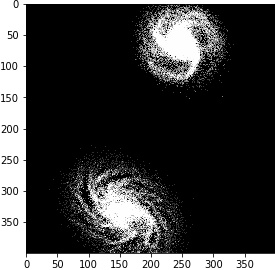}}
    \subfigure{\includegraphics[width=2cm,height=2cm]{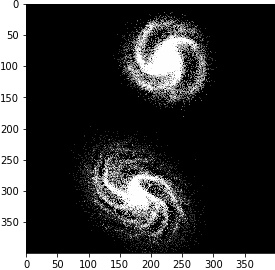}}
    \subfigure{\includegraphics[width=2cm,height=2cm]{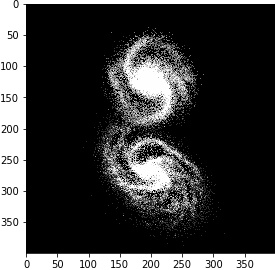}}
    \subfigure{\includegraphics[width=2cm,height=2cm]{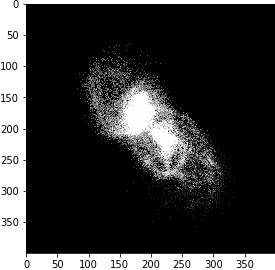}}
    \subfigure{\includegraphics[width=2cm,height=2cm]{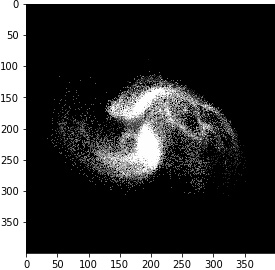}}
    \subfigure{\includegraphics[width=2cm,height=2cm]{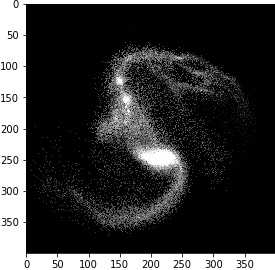}}
    \subfigure{\includegraphics[width=2cm,height=2cm]{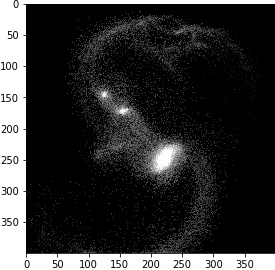}}
    \subfigure{\includegraphics[width=2cm,height=2cm]{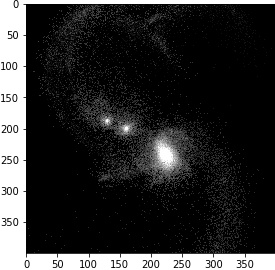}}

    \subfigure{\includegraphics[width=2cm,height=2cm]{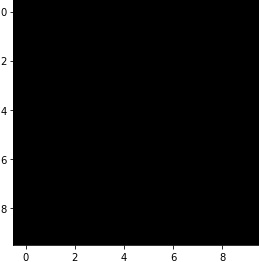}}
    \subfigure{\includegraphics[width=2cm,height=2cm]{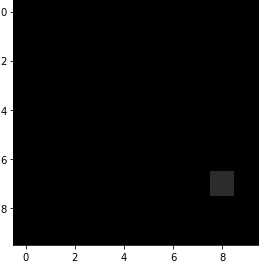}}
    \subfigure{\includegraphics[width=2cm,height=2cm]{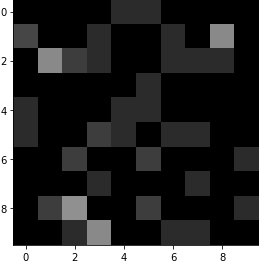}}
    \subfigure{\includegraphics[width=2cm,height=2cm]{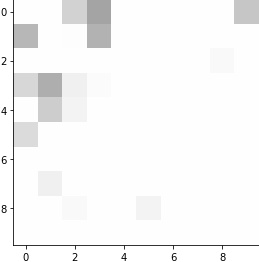}}
    \subfigure{\includegraphics[width=2cm,height=2cm]{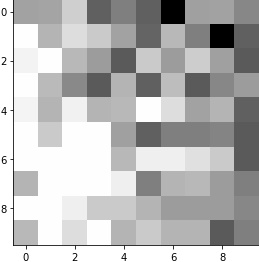}}
    \subfigure{\includegraphics[width=2cm,height=2cm]{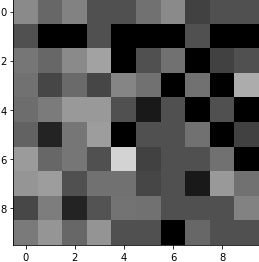}}
    \subfigure{\includegraphics[width=2cm,height=2cm]{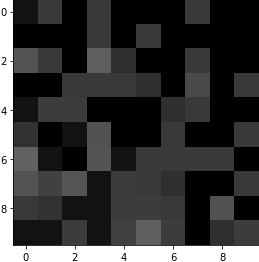}}
    \subfigure{\includegraphics[width=2cm,height=2cm]{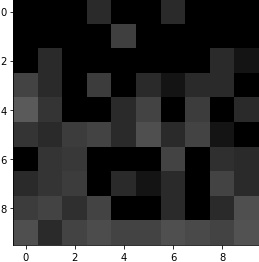}}
    \label{fig2}
    \caption{For each 8 frames (top-row) of a spatiotemporal sequence (gSa-gSb merger with (i,$\phi$,$\theta$) = (0, 30, 60)) a 10$\times$10 zoomed-section from the center has been highlighted (bottom row) to show how similar the variations are locally but the consecutive frames as a whole would give a visually different but semantically coherent sequence.}
    
\end{figure*}

\begin{figure*}[t]
\centering
    \subfigure{\includegraphics[width=2cm,height=2cm]{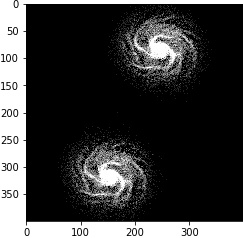}}
    \subfigure{\includegraphics[width=2cm,height=2cm]{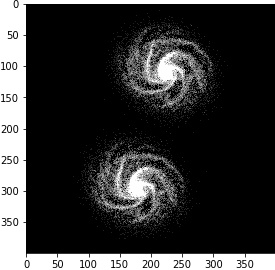}}
    \subfigure{\includegraphics[width=2cm,height=2cm]{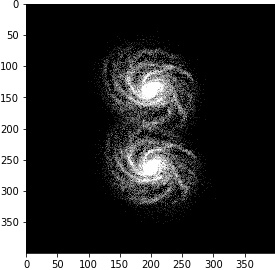}}
    \subfigure{\includegraphics[width=2cm,height=2cm]{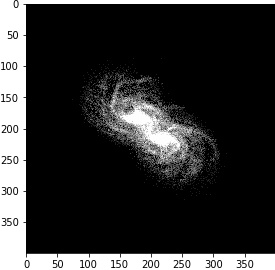}}
    \subfigure{\includegraphics[width=2cm,height=2cm]{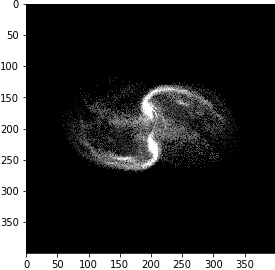}}
    \subfigure{\includegraphics[width=2cm,height=2cm]{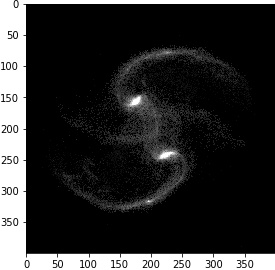}}
    \subfigure{\includegraphics[width=2cm,height=2cm]{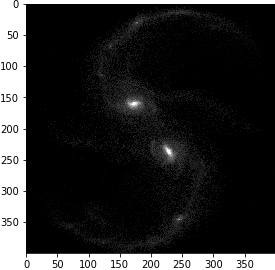}}
    \subfigure{\includegraphics[width=2cm,height=2cm]{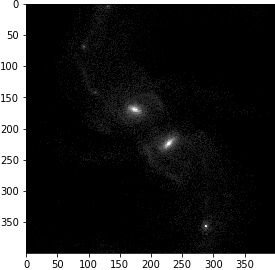}}

    \subfigure{\includegraphics[width=2cm,height=2cm]{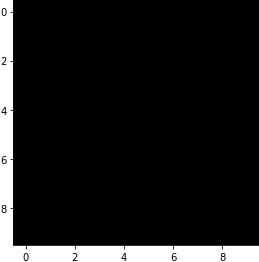}}
    \subfigure{\includegraphics[width=2cm,height=2cm]{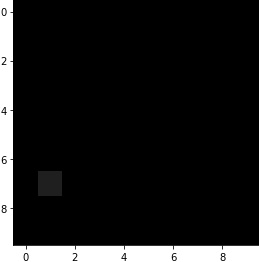}}
    \subfigure{\includegraphics[width=2cm,height=2cm]{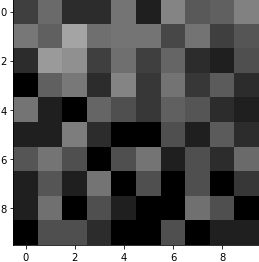}}
    \subfigure{\includegraphics[width=2cm,height=2cm]{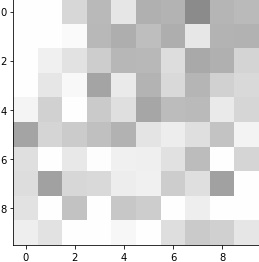}}
    \subfigure{\includegraphics[width=2cm,height=2cm]{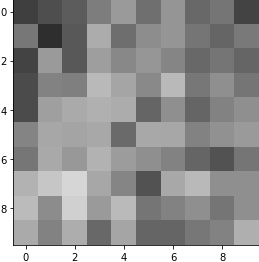}}
    \subfigure{\includegraphics[width=2cm,height=2cm]{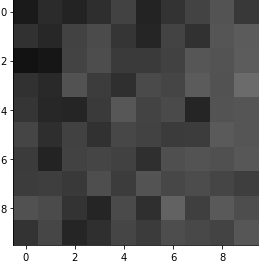}}
    \subfigure{\includegraphics[width=2cm,height=2cm]{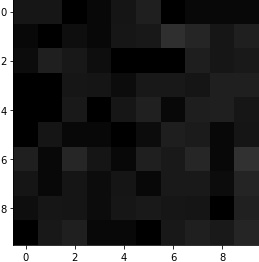}}
    \subfigure{\includegraphics[width=2cm,height=2cm]{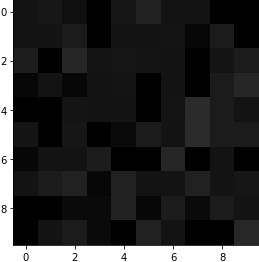}}

    \caption{For each 8 frames (top-row) of a spatiotemporal sequence (gSa-gSb merger with (i,$\phi$,$\theta$) = (0, 30, 90)) a 10$\times$10 zoomed-section from the center has been highlighted (bottom row) to show how similar the variations are locally but the consecutive frames as a whole would give a visually different but semantically coherent sequence.}
    \label{fig1}
\end{figure*}

\section{Overview}
\label{overview}
\subsection{Problem Statement}
In this section, we shall discuss the problem by focusing on a set of spatiotemporal sequences extracted from our dataset. Each data point in the sequence is an arrangement of $M \times N$ grid consisting of $M$ rows and $N$ columns with $C$ color channels. Thus, the observation at any time can be represented by a tensor $X \in R^{M \times N \times C}$, where $R$ denotes the domain of the observed features. If we record the observations periodically, we will get a sequence of tensors $X_{1}, X_{2}, . . . , X_{t}$ (representation in Fig-\ref{fig1} and Fig-\ref{fig2})\footnote{The resulting sequence can be viewed as 60 frames per second at: \url{https://github.com/SSingh087/seq-pred}}. The spatiotemporal sequence prediction problem is to predict the most probable length-$K$ sequence in the future given the past $J$ observations, including the current one.
\begin{equation*}
\label{intro}
\tilde{X}_{t+1}, . . . ,\tilde{ X}_{t+K}  = \underset{X_{t+1},...,X_{t+K}}{arg\;max} \; p(X_{t+1}, . . . , X_{t+K} \big| \hat{X}_{t-J+1}, \hat{X}_{t-J+2}, . . . , \hat{X}_{t})
\end{equation*}

\subsection{Solution proposed}

The underlying idea of predicting future events is drawn from the cognitive neuroscience field \cite{2}: the human mind established complex neural pathways of the physical and casual rules governing the environment primarily through observation and interaction \cite{intro1}-\cite{intro2}. The brain is continuously learning intuitive physics, to name a few: trajectories of a falling ball, flow of fluids, collision physics, and refines the already understood world models from the mismatch between its predictions and what truly occurred. Similarly, the deep learning model is trained on historic data wherein it attempts to find patterns through which it makes a plausible prediction having a physical significance. Although seemingly effortless, the idea is more challenging with the astronomical data.

For the model to make robust predictions from the sequence with spatial and temporal components, it needs to be efficient enough to extract meaningful features, e.g., in our case: how much galaxy is sheared during the merger, view at different angles, the evolution of multiple images due to gravitational lenses and similar such features along with maintaining coherence. Focussing on a small localized part of the input $X_{i}$ for the given time sequence, we could identify myriad visually similar deformations locally due to the temporal coherence, as shown in Fig-\ref{fig1} and Fig-\ref{fig2}. Whereas observing the consecutive frames as a whole would give a visually different but semantically coherent sequence. This inconsistency in the visual appearance of 2D stacked data on various scales is mostly attributed, among other factors, to occlusions and shifts in lighting conditions \cite{2}. Predictive models are able to derive representative spatiotemporal correlations describing the dynamics of the spatiotemporal sequence from this source of temporally organized visual cues.

In our case, the future event prediction task can be associated with a supervised learning approach because the target frame itself acts as a label. Thus, learning by prediction is a self-supervised task, filling the gap between supervised and unsupervised learning. 

\subsection{General problems}
Using an example of our data, we showcase how challenging a prediction problem can be for astronomical data. A single 2D data prediction is difficult primarily due to the lack of contextual information and the infinite possibilities of the latent space. A stacked 2D data at discrete time intervals may assert a temporal sequence, narrowing down to a deterministic outcome and reducing prediction space, but the future prediction at every time step is by nature multimodal.

When there exists an equal possibility of any prediction, the deterministic model tends to output prediction close to the mean value of all the possible outcomes. This randomness is visually expressed as blurriness Fig-\ref{fig3} and Fig-\ref{fig5}. As deterministic models cannot handle real-world settings characterized by chaotic dynamics, incorporating uncertainty into the model becomes crucial.
With an end-to-end model with sufficient data for training, the prediction can be close to the original. One such approach is shown in \cite{3} wherein a considerable amount of radar echo data can be continuously collected so as to reduce the probability space.

\section{Related works}
\label{Related works}
Developments in deep learning, especially in Recurrent Neural Networks (RNNs) and Long-Short Term Memory (LSTM), have greatly tackled this problem. Models \cite{intro2}, \cite{dev1}, \cite{dev2}, \cite{dev3}, \cite{dev4}, \cite{dev5}, \cite{dev10}, present some beneficial insights on how to tackle this problem. Since RNNs face vanishing or exploding gradients, using LSTM and GRUs, these problems were mitigated. Shi \textit{et al.} extended the application of LSTM models to the image \cite{dev6}. Graves \textit{et al.}\cite{dev7} explored multidimensional LSTM and a few stacked recurrent layers to capture abstract spatiotemporal correlations \cite{dev8}. In \cite{3} the authors presented a ConvLSTM model used for Precipitation Nowcasting, showing that the stacked ConvLSTM layers outperform the Fully Connected (FC)-LSTM. FC-LSTM layer has proven effective for handling temporal correlation but shows redundancy with the spatial component.

The use of Autoencoders has further eased the problem. Autoencoders has two parts the encoder where important information is encoded, capturing the significant features of the datasets, and the decoder that reconstructs the output based on those captured features but the predictions generated are low dimensional and blurry. This low-quality feature reconstruction problem is tackled by state-of-art Adversarial networks introduced by Goodfellow \textit{et al.}\cite{dev9}. Drawing inspiration from game theory, the adversarial nets consist of two networks trained simultaneously, both attempting to minimize the loss.

\section{Dataset preparation}
\label{DP}
All our datasets are a spatiotemporal sequence with $10$ 2D images stacked over in discrete time intervals. The datasets were preprocessed to study the model's performance towards the different features that were highlighted. We primarily clustered the datasets in clusters of $2$, $20$, and $60$, done for all the datasets discussed below. We also tested by altering the colormap of the dataset to study and verify a new performance metric: Correctness Factor (CF), which is discussed in detail in Sec-\ref{metric}. The details of the number of datasets and the parameters are given in Table-\ref{dataset}. This section aims to give a brief regarding the datasets we used to train the model.

\begin{table*}
\centering
\caption{CNN model summary}\label{dataset}
\begin{tabular}{ccc}
         Dataset&Input Shape&Total sequence\\
         \hline \hline\\
        Galaxy-mergers&$60 \times 60$&216\\
        Galaxy-mergers&$80 \times 80$&216\\
        Galaxy-mergers&$120 \times 120$&216\\
        Galaxy-mergers ($2$ colormaps)&$60 \times 60$&$2\times 216$\\
        Gravitationally lensed galaxy-mergers ($2$ colormaps)&$60 \times 60$&$ 2 \times 216 $\\
        Gravitationally lensed mono-galaxy ($2$ colormaps)&$60 \times 60$& $2 \times 54 $\\
\hline
\end{tabular}
\end{table*}

\subsection{Dataset 1: Galaxy-mergers}
\label{D1}
We collected pre-simulated galaxy merger datasets available as Galmer dataset\footnote{\url{http://galmer.obspm.fr/}}\cite{gal}. The GalMer database is a library of galaxy merger simulations following a statistical approach by performing and analyzing thousands of simulations of interacting pairs, with initial conditions relevant to all epochs of the universe, at different redshifts.

Although the images need to be manually collected, we developed a web scraping method employed for bulk downloading of images with specific parameters. The code is publicly available at Github\footnote{\url{https://github.com/SSingh087/seq-pred/tree/main/web-scraping}}. Our dataset included $10$ images from $200Myr$ to $700Myr$, restricting our data to orbit-type $1$ only. The sole reason for choosing this orbit was due to the Pericentral distance, which is $8$Kpc, and motion energy is $0$ proving advantageous as giant galaxy usually merges around $700$Myr - $750$Myr. The $0$ motion energy allowed us to capture the cycle with optimum zoom so that the galaxies do not go off the frame the shape evolution (spatial information) is captured. Only Prograde rotation is used, and correspondingly all $4$ available inclinations between the galaxies: $0^\circ$, $45^\circ$, $75^\circ$, and $90^\circ$. Further, in each simulation sequence, data points are captured at $30^\circ$ interval starting from $30^\circ$ and extending to $30^\circ$ for $\phi$ and $\theta$. The sequence is captured at all possible permutations resulting in $9$ sets having $10$ images.

\subsection{Dataset 2: Galaxy-mergers (colormap)}
\label{D2}
Galmer Dataset with two-color maps: hue, saturation, value (HSV), and twilight has all the spatiotemporal features as of the Galmer dataset.

\subsection{Dataset 3: Gravitationally lensed galaxy-mergers}
\label{D3}
Another dataset that we used for training is the gravitationally lensed image of the merging galaxies. We used the Galmer dataset for simulating the lensed image of the galaxy. The aim was to produce a sequence wherein the foreground galaxy has a higher redshift than the background galaxy. We chose foreground galaxy to have a mass of $10^{15} M_{\odot}$ with a redshift, $z$ of $0.05$ and merging galaxy with $z = 0.02$. The lens model was fixed for all simulations, combining the SIE and SHEAR model with the light model as a combination of two SERSIC ELLIPSE and one NIE model. We have carried our simulation in Python3.0 using library lenstronomy \cite{lenstronomy}. Our code is publicly available on Github\footnote{\url{https://github.com/SSingh087/seq-pred/blob/main/lensing/code.py}}.

\subsection{Dataset 4: Gravitationally lensed mono-galaxy}
\label{D4}
Dwarf galaxies that are free from the gravitational influence of other galaxies are gravitationally lensed using the same methodology discussed for Gravitationally lensed galaxy-mergers; also, the central idea of moving foreground galaxy is captured. Simulation is captured at $30^\circ$ interval starting from $30^\circ$  and extending to $90^\circ$  for $\phi$ and $\theta$. The sequence is captured at all possible permutations resulting in $9$ sets having $10$ images.

\section{Proposed Model}
\label{Proposed Model}
\begin{figure*}
\centering
\includegraphics[width=18cm,height=8cm]{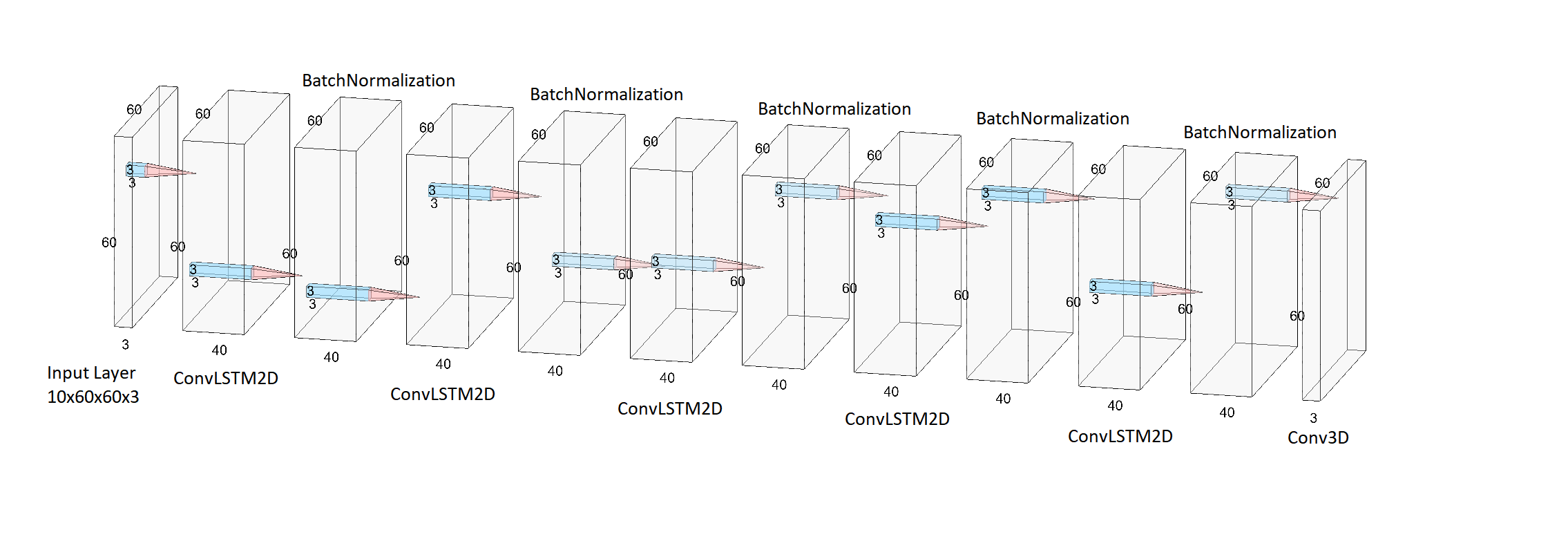}
\caption{Model used for predicting spatiotemporal sequence follows a sequential approach with BatchNormalization layer corresponding to every ConvLSTM layer. The model presented here is for Input shape $60\times 60\times 3$. }
\label{modelfig}
\end{figure*}

In this section, we discuss the DL model that we proposed to overcome the aforementioned problems and predict the spatiotemporal sequence for the datasets discussed in Sec-\ref{DP}. The model architecture is shown in Fig-\ref{modelfig} and summarized in Table-\ref{modeltab}. Although we experimented with various model parameters described in Table-\ref{experiment} and the experimentation results are discussed in Sec-\ref{exp}. This section discusses the fundamental model architecture, which on running for 800 epochs, successfully employed frame prediction with minimum loss. From Fig-\ref{modelfig}, it can be seen that our model follows a sequential approach with tensors passing through various layers sequentially. We avoided the use of Max-pooling or Average-pooling primarily to prevent loss of information since these layers tend to reduce the number of parameters and the computational load. Moreover, the input and output sizes are equivalent. In our model, a Batch Normalization layer follows a ConvLSTM layer, keeping padding as “same” with an argument to return a sequence. The kernel shape and size was fixed for all experimets and activation \texttt{sigmmoid} was used for the Conv3D layer with kernel size of $3 \times 3 \time 3$.

\begin{table*}[t]
\centering
\caption{Model Summary (the model presented here is for Input shape $60\times 60\times 3$)}
\label{modeltab}
\begin{tabular}{ccccc}
         Layer&Input Shape&Output Shape&kernel size&padding\\
         \hline \hline\\
        ConvLSTM&(60, 60, 3)&(60, 60, 40)&(3, 3)&same\\
        BatchNormalization&(60, 60, 40)&(60, 60, 40)&(3, 3)&--\\
        ConvLSTM&(60, 60, 40)&(60, 60, 40)&(3, 3)&same\\
        BatchNormalization&(60, 60, 40)&(60, 60, 40)&(3, 3)&--\\
        ConvLSTM&(60, 60, 40)&(60, 60, 40)&(3, 3)&same\\
        BatchNormalization&(60, 60, 40)&(60, 60, 40)&(3, 3)&--\\
        ConvLSTM&(60, 60, 40)&(60, 60, 40)&(3, 3)&same\\
        BatchNormalization&(60, 60, 40)&(60, 60, 40)&(3, 3)&--\\
        ConvLSTM&(60, 60, 40)&(60, 60, 40)&(3, 3)&same\\
        BatchNormalization&(60, 60, 40)&(60, 60, 40)&(3, 3)&--\\
        Conv3D&(60, 60, 40)&(60, 60, 3)&(3, 3, 3)&same\\
\hline
\end{tabular}
\end{table*}

\section{Metric for model's performance}
\label{metric}
Multiple parameters have to be addressed for a sequence prediction, such as correct temporal prediction, plausible spatial components, brightness, and prediction's sharpness. The review by Oprea \textit{et al.}\cite{2} highlighted the absence of evaluation protocols and metrics that evaluate the predictions by simultaneously fulfilling all these aspects. Evaluating a prediction according to the mismatch between its visual appearance and the ground truth is not always reliable because the metrics penalize the predictions that deviate from the ground truth. This results in the predictions being blurry, rather than sharp and bright, as the model attempt to predict values close to the ground truth \cite{met1}, \cite{met2}, \cite{met3}.

Addressing the same issue, we propose a novel performance metric for 2D spatiotemporal sequences,  tested for our datasets, and has performed notably. We do expect to have similar performance for other 2D spatiotemporal datasets as well. In our case, a 2D frame is a weighted combination of three color channels - R, G, B. At the end of 800 epochs, model performance was evaluated not solely based on the loss function but a metric that we define as \texttt{Correctness Factor (CF)}. CF is a combination of two functions- \texttt{Root mean square pixel-wise difference (RMSPWD)} and \texttt{Channel-wise Standard deviation (CSD)}. RMSPWD is the square root difference of the squared pixel values between the ground truth and the predicted sequence for each frame individually for each color channel, and CSD is the difference between the standard deviations of the pixel values of the ground truth and predicted frame corresponding to each channel. In Eq-\ref{eqr}, we define the mathematical approach we used towards the newly defined metric CF.

\label{eqr}
\begin{multline*}
    z_{o}^{C_1} = f(x,y) = \tilde{X}_{single\;color\; channel} \\
    z_{p}^{C_1} = g(x,y) ={X}_{single\;color\; channel}  \\
    RMSPWD = \alpha = \sqrt{\lvert z_{o}^{2} - z_{p}^{2}} \rvert \\
    CSD = \gamma= \sqrt{\frac{\sum_{i=0}^{M \times N} \big(z_{{o}_i}^{C_1}- \frac{\sum_{i=0}^{M \times N} z^{C_1}_{{o}_i}}{M \times N}\big)^{2}}{M \times N}} - \\ \sqrt{\frac{\sum_{i=0}^{M \times N} \big(z^{C_1}_{{p}_i}- \frac{\sum_{i=0}^{M \times N}z^{C_1}_{{p}_i}}{M \times N}\big)^{2}}{M \times N}}
\end{multline*}
In Eq-\ref{eqr} $z_{o}^{C_1}$ and $z^{C_1}_{p}$ are 3D tensors with symbols as explained in Eq-\ref{intro} wherein $C_1$ represents single color channel. Tensor $z$ can be written as a function of $x$ representing the frame's width and $y$ the length for an individual frame. 

\begin{figure}[h]
\centering
    \subfigure{\includegraphics[width=8cm,height=5cm]{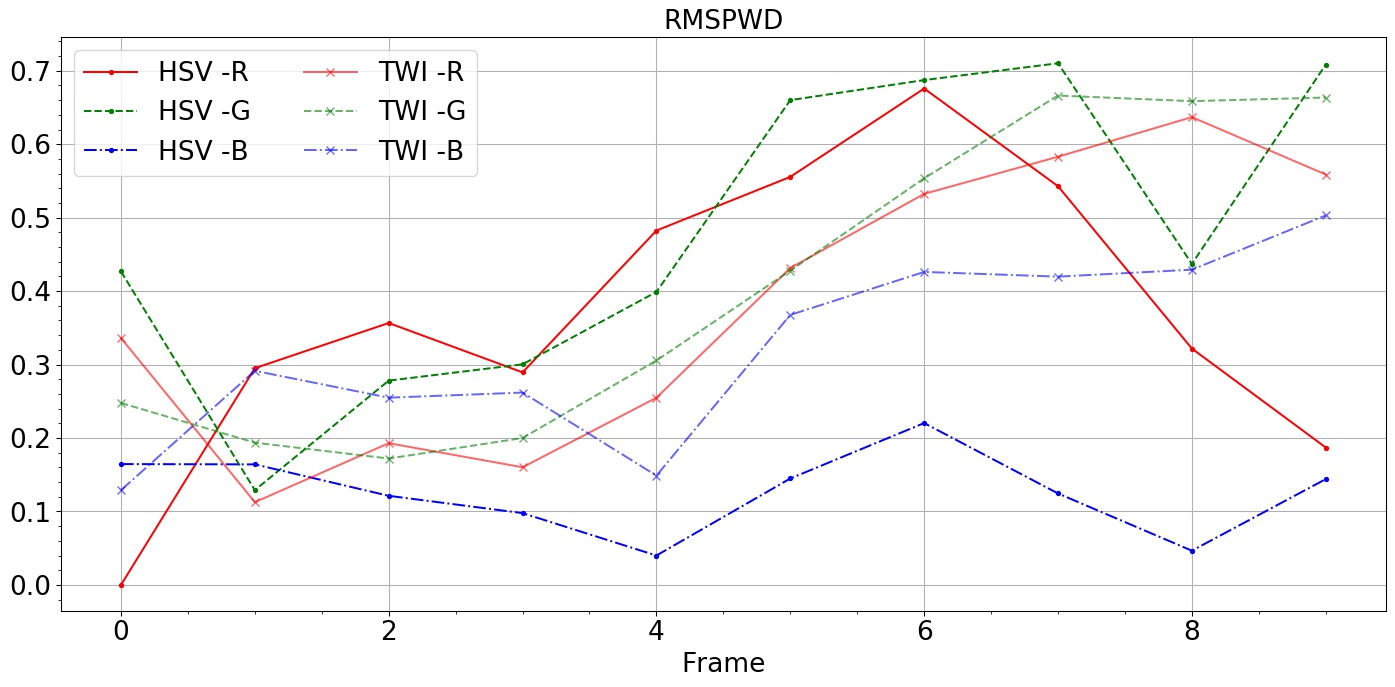}}
    \subfigure{\includegraphics[width=8cm,height=5cm]{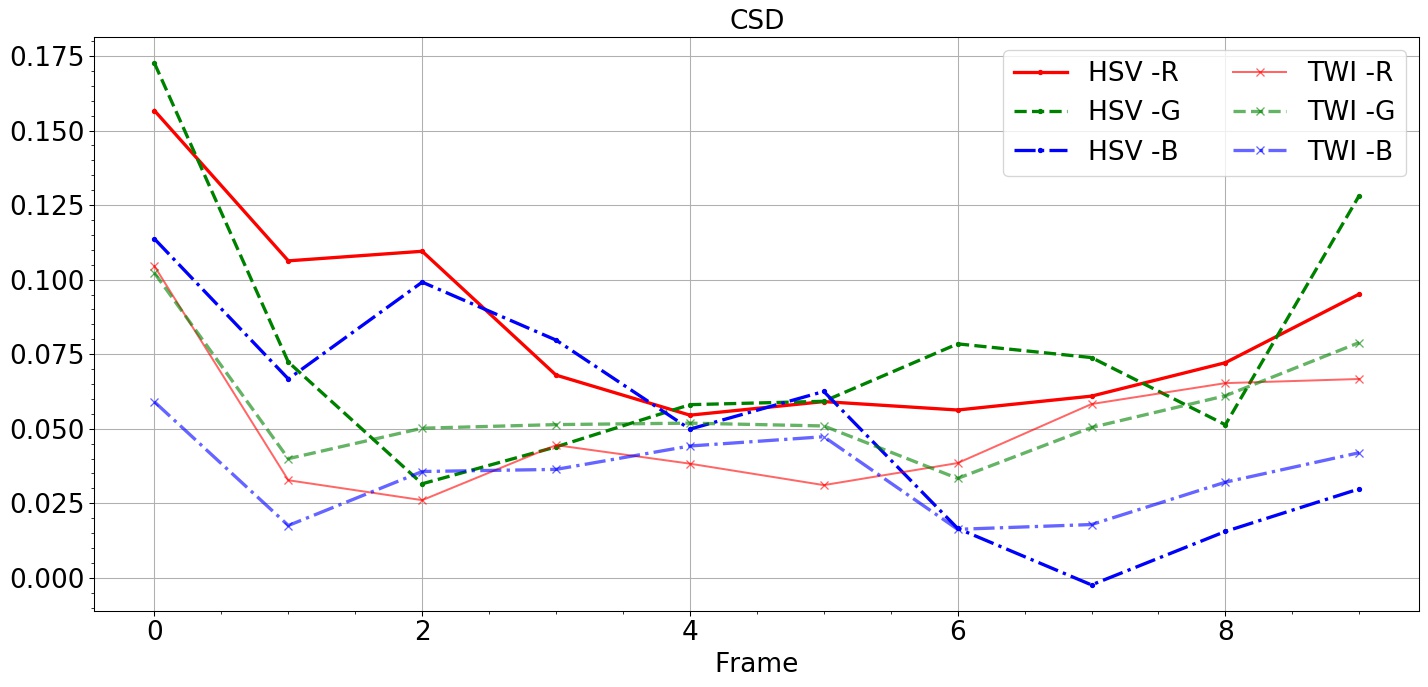}}
    \caption{The left column shows the RMSPWD and the right column shows the CSD. This is evaluated in reference to Fig-\ref{fig3} and Fig-\ref{fig5}}
    \label{rmspwd}
\end{figure}

\texttt{RMSPWD} returns the distance between the predicted frame and the ground truth, giving an idea of how far the prediction is from the ground-truth capturing the idea of frame brightness. Using Fig-\ref{fig3} and Fig-\ref{fig5}\footnote{The resulting sequence can be viewed as 60 frames per second at: \url{https://github.com/SSingh087/seq-pred}} we demonstrate the RMSPWD in Fig-\ref{rmspwd} for $M, N=(75,75)$ evaluated for $10$ frames\footnote{$M$ and $N$ refers to same terminology as in Eq-\ref{intro}}. It can be seen that channels of HSV show extreme scale differences for $M, N=(75,75)$. R-channel initially has an exact prediction, but CSD increased for higher frames, especially the frames where the foreground galaxy almost crosses over the background galaxy. RMSPWD records such similar trends for all $M, N \in M \times N$ signifying the predicted frame's sharpness. Also, in Fig-\ref{rmspwd}, we show how the \texttt{CSD} varies for each frame. CSD measures the amount of dispersion of pixel values; the lower the CSD lower the pixel's deviation from the mean value. For our example, physical significance can be inferred in terms of mass accretion. The higher deviation represents scattered mass throughout the frame, whereas, lower deviation represents accreted mass in a localized for fixed angles (i, $\phi, \theta$). CSD directly measures how blurry the predicted image is by reporting the difference, thereby giving a measure of image sharpness.

\begin{figure}[h]
\centering
    \subfigure{\includegraphics[width=8cm,height=5cm]{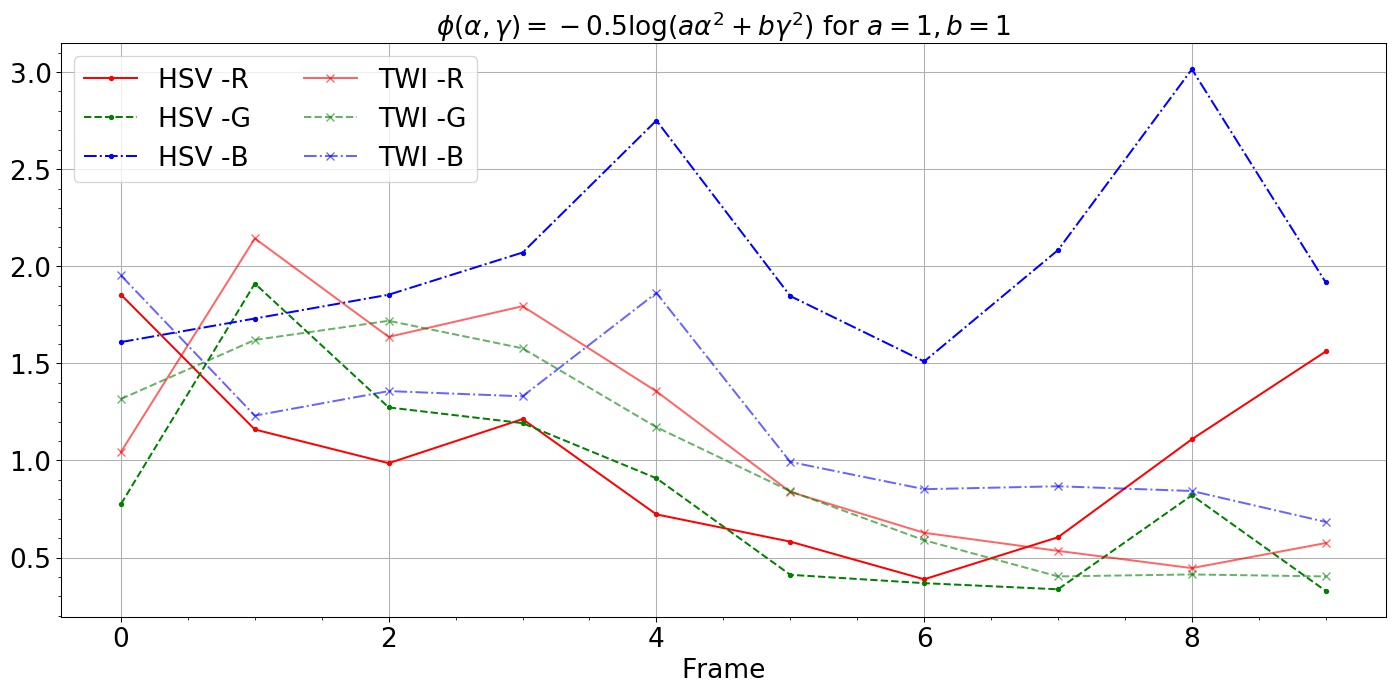}}
    \subfigure{\includegraphics[width=8cm,height=5cm]{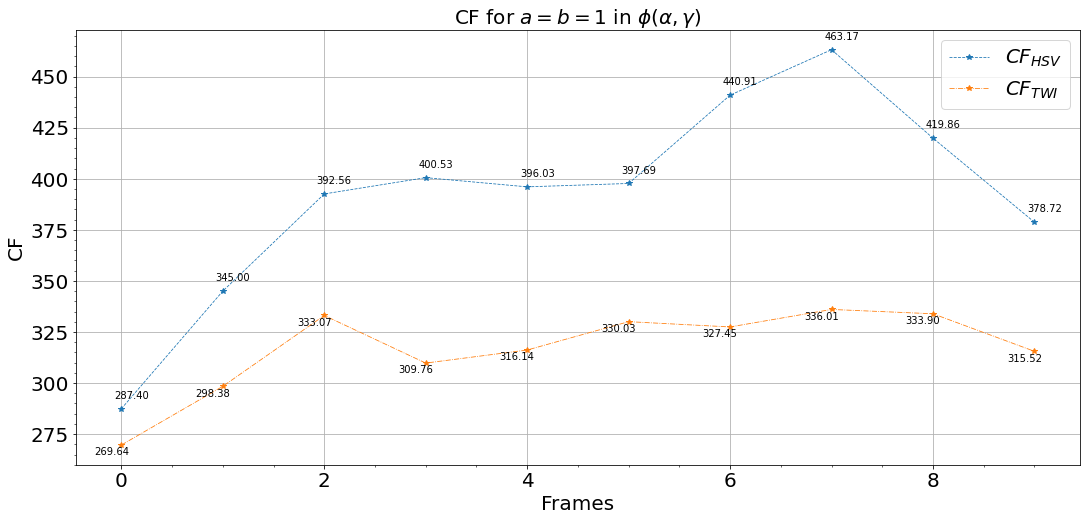}}
    \caption{The left column shows the $\phi(\alpha,\gamma)$ for $(a, b)=1$ and the right column shows the CF whose formulation is given in Eq-\ref{eqcf}. This is evaluated in reference to Fig-\ref{fig3} and Fig-\ref{fig5}}
    \label{SF_show}
\end{figure}

We define a function $\phi(\alpha,\gamma)$ in Eq-\ref{phi} which outputs a negative $\log$ of the combination of the two metrics evaluated per color channel. A minimum of the function will result in the prediction as close to the ground truth. The advantage of the above formalism allows us to establish a metric that combines two fundamental properties of the spatiotemporal prediction: spatial coherence governed by the brightness and temporal components measured by the frame's sharpness. In Fig-\ref{rmspwd}, \ref{SF_show} we highlight how the the performance is. Choosing $a$ and $b$ as unity the range of $\phi(\alpha,\gamma)$ is $\kappa \in [-0.5\log{2}, \infty)$. The maximum value of $\alpha$ and $\gamma$ can be 1, which shows the maximum difference (deviation from the ground truth). And the minimum value of $\phi(\alpha,\gamma)$ can be $0$, which is the perfect prediction. In Eq-\ref{proof} we show the condition for the minimum error for a \texttt{close-to-perfect} prediction. 

\begin{equation}
    \phi(\alpha,\gamma) = \log{\frac{1}{\sqrt{a \alpha^2+b \gamma^2}}}
    \label{phi}
\end{equation}

All of the above metrics are defined for individual color channels. We define CF in Eq-\ref{eqcf}, wherein $\phi^R_i$ represents the $R$ color channel corresponding to the intersection of $M^{th}$ row and $R^{th}$ column resulting in a single pixel. Similarly for $B$ and $G$ color channels. The idea is that each color channel represents a latent space of its own, and all the functions that we have defined earlier are evaluated in their individual latent space (Fig-\ref{phif}). CF combines those latent spaces to output a single-valued metric for an individual frame. The higher the CF more \texttt{close-to-perfect} will be the prediction. The result can be concluded from Fig-\ref{rmspwd},\ref{SF_show}.
\begin{equation}
\label{eqcf}
CF = \sqrt{\sum_{i=0}^{M\times N}[\phi^R_i(\alpha,\gamma)]^2+\sum_{i=0}^{M\times N}[\phi^G_i(\alpha,\gamma)]^2+\sum_{i=0}^{M\times N}[\phi^B_i(\alpha,\gamma)]^2}
\end{equation}

\section{Training and testing}
\label{tt}
We trained our model for 800 epochs using Binary crossentropy\footnote{\url{https://keras.io/api/losses/probabilistic_losses/}} as the suitable loss function, with optimizer Adadelta \footnote{\url{https://ruder.io/optimizing-gradient-descent/index.html}}.

\section{Experimentation}
\label{exp}
In this section we discuss our experimentation results on the datasets mentioned in Sec-\ref{DP} and study its effect on the loss function and CF. The experimentation results are summarized in Table-\ref{experiment}.

\begin{table}[!h]
\caption{Dataset experimentation parameters}
\label{experiment}
\begin{center}
\begin{tabular}{ |c|c|c|c|c } 
\hline
Dataset & Varying parameters & Loss function & CF \\
\hline 
\multirow{3}{2em}{1}& Image shape & Fig-\ref{figd1AC0}, Fig-\ref{figd1AC00}  & Fig-\ref{figd1AC000}\\ 
                    & Number of Filters & Fig-\ref{figd1FS1}, Fig-\ref{figd1FS2} & Fig-\ref{figd1FS3}\\
                    & Number of Layers & Fig-\ref{figd1NL1},Fig-\ref{figd1NL2} & Fig-\ref{figd1NL31}, Fig-\ref{figd1NL32}\\
\hline
\multirow{1}{2em}{2} & Colormap & Fig-\ref{figd21}, Fig-\ref{figd22}  & Fig-\ref{figd23}\\
\hline
\multirow{1}{2em}{3} & Colormap & Fig-\ref{figd31}, Fig-\ref{figd32}  & Fig-\ref{figd33}\\
\hline
\multirow{1}{2em}{4} & Colormap & Fig-\ref{figd41}, Fig-\ref{figd42}  & Fig-\ref{figd43}\\
\hline
\end{tabular}
\end{center}
\end{table}

\section{Discussion and Conclusion}
\label{conclude}
In this section, we explain our results and the notable observations for the datasets, as mentioned in Sec-\ref{DP} and the varying parameters as compiled in Table-\ref{experiment}. In dataset-1 (\ref{D1}) from Fig-\ref{figd1AC0} we can observe that the loss function decreases rapidly for all the image-size once number of epochs are close to $100$ and converges after crossing $600-700$ epochs. We also observed that the clustering has a significant effect, and a more stable solution was achieved when the number of clusters was 60 or 0. Thus, from our experiments, we concluded that the reduction in the number of clusters did not assisted towards accelerating the accuracy. A plausible explanation can be that the DL learning model predicts various pixel values corresponding to the input frames and attempts to mimic them, reduction in the number of clusters, which is another way of reducing the spread of the pixel values, results in a larger difference between the ground-truth-frame and the predicted frame. We can also observe a general trend with validation loss, which tends to increase and peaks between $30-100$ and then maintains a similar gradient as that of prediction loss. Although less prominent for image-size 60, it is clearly visible for the other two image-sizes: 80 and 120.

From Fig-\ref{figd1AC00} we can conclude the similar points with additional information of image-size 120 has a more stable solution with the least loss. Fig-\ref{figd1AC000} also echoes the same features. CF highlights individually for each frame and their prediction performance. We observe that for all the cases, image-size 120 has a higher CF score. Although for the case where the number of clusters is $2$, a lower CF is observed but regains a higher score around $4^{th}-6^{th}$ frame. A general trend is also observed wherein the CF score decreases rapidly after $4^{th}$ frame for image-size 60 and 80. A reason that can be accounted for this general trend is after the $4^{th}$ frame (Fig-\ref{fig2}, Fig-\ref{fig1}) the galaxy merger cycle reaches a stage wherein the gravitational potential begins shifting so quickly that star orbits are greatly altered and lose trace of their prior orbit. This process is called “violent relaxation” \cite{conc1}. At this stage, image-size 60 cannot completely capture the stars that deviate from the orbit and resulting in blurred or \texttt{confused} prediction by the model.

We also observed that the activation function does not assist in prediction rather negatively affects the performance. The same keypoint can be highlighted from the Fig\ref{figd1AC10}. Since the image-size, we chose for this experiment set was 60, a decrease in CF score post $4^{th}$ frame is observed as for the possible reason explained above. Fig-\ref{figd1FS1} highlights the idea that a higher number of filters improve performance. However, we have not tested for the convergence limit where on further increasing the number of filters, the performance stops increasing. Fig-\ref{figd1FS2} bolsters this idea as the rapid fall in loss function for $10$ filters is close to 400-500 epochs. One of the advantages of using CF as the metric can be highlighted in Fig-\ref{figd1FS3}, which highlights that $10$ filters outperform 30 filters for image-size 60. Thus, the loss function is not the sole measure of performance. Fig-\ref{figd1NL2} validates the idea that was stacking ConvLSTM layers improve the performance with Fig-\ref{figd1NL1} validating the idea that 0 and 60 clusters outperform lower clusters. Fig-\ref{figd1NL31}, Fig-\ref{figd1NL32} bolsters the same key point.

Further moving on to dataset-2 (Sec-\ref{D2}) from Fig-\ref{figd21} frames with no colormap outperforms others. Also from Fig-\ref{figd22} and Fig-\ref{figd23} the same idea can be captured. Twilight colormap does not perform well and has a lower CF, and the same trend is observed with other datasets. The clustering trend is still maintained here. Dataset-3 (Sec-\ref{D3}) and dataset-4 (Sec-\ref{D4}) highlights the same trend of HSV performing better than the Twilight colormap. For datasets-3, 4, these results can be verified with human observation. 

The complete animation for $10$ frames is available at Github\footnote{\url{https://github.com/SSingh087/seq-pred/blob/main/README.md}} wherein each result is verified with human observation. CF is a new metric that we propose for dealing with the spatiotemporal sequence. The formulation has been tested on our dataset only, and a further extension and study is proposed and is underway.
\newpage
\begin{acknowledgements}
 We would like to thank authors of \cite{2} for compiling the useful models and giving an in-depth insights of the current frame prediction algorithms. We wish to show our gratitude to many people without whom this work would not have been complete. We would also like to acknowledge Project Horizon\footnote{\url{http://www.projet-horizon.fr/rubrique3.html}} for maintaining the Galmer database. We also wish to express our gratitude towards all the reviews that were crucial for this work.
\end{acknowledgements}

\bibliographystyle{plain}
\bibliography{auth} %
\nocite{*}

\begin{appendix} 

\begin{figure*}[p]
\centering
    \subfigure{\includegraphics[width=6cm,height=3.5cm]{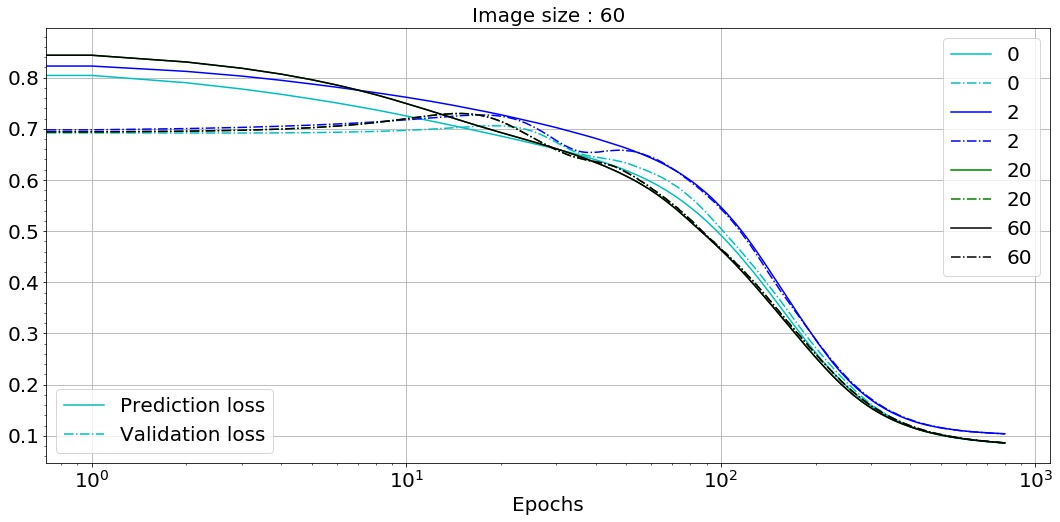}}
    \subfigure{\includegraphics[width=6cm,height=3.5cm]{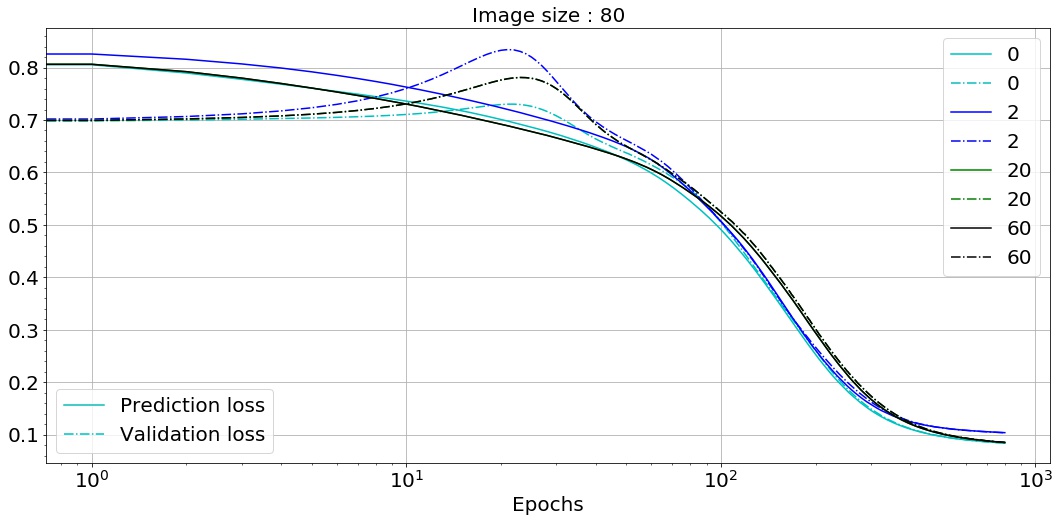}}
    \subfigure{\includegraphics[width=6cm,height=3.5cm]{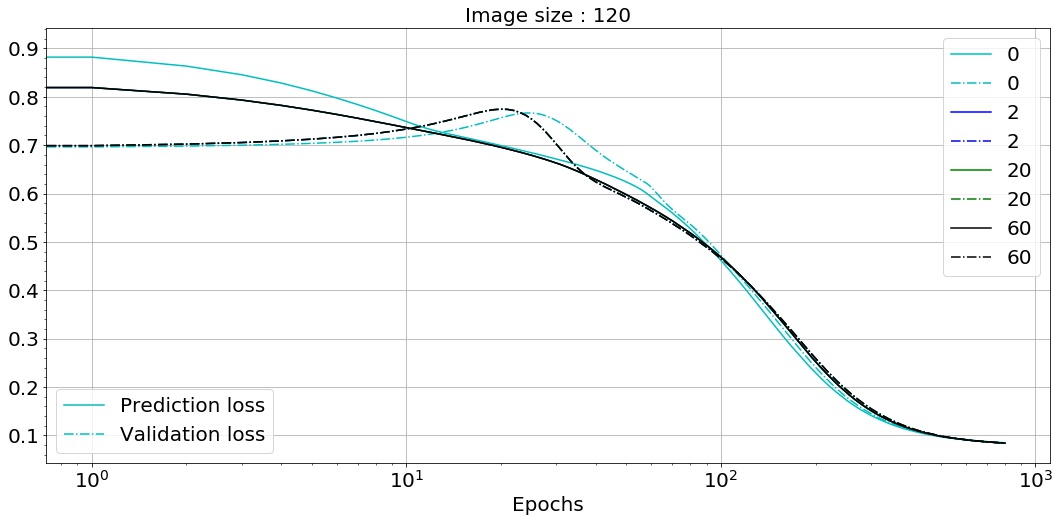}}
    \caption{Variation of loss function with number of clusters for for each image size}
    \label{figd1AC0}
\end{figure*}

\begin{figure*}[p]
\centering
    \subfigure{\includegraphics[width=4.5cm,height=3.5cm]{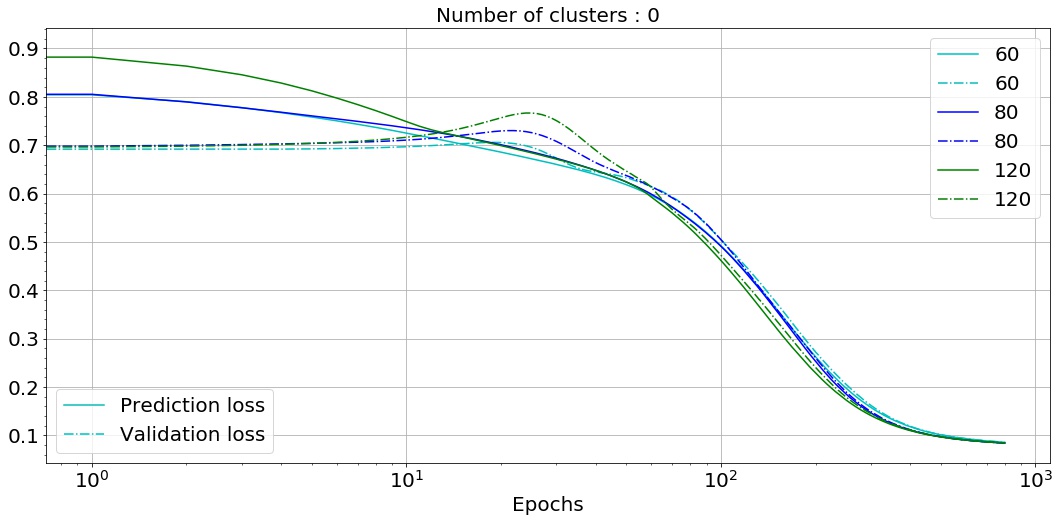}}
    \subfigure{\includegraphics[width=4.5cm,height=3.5cm]{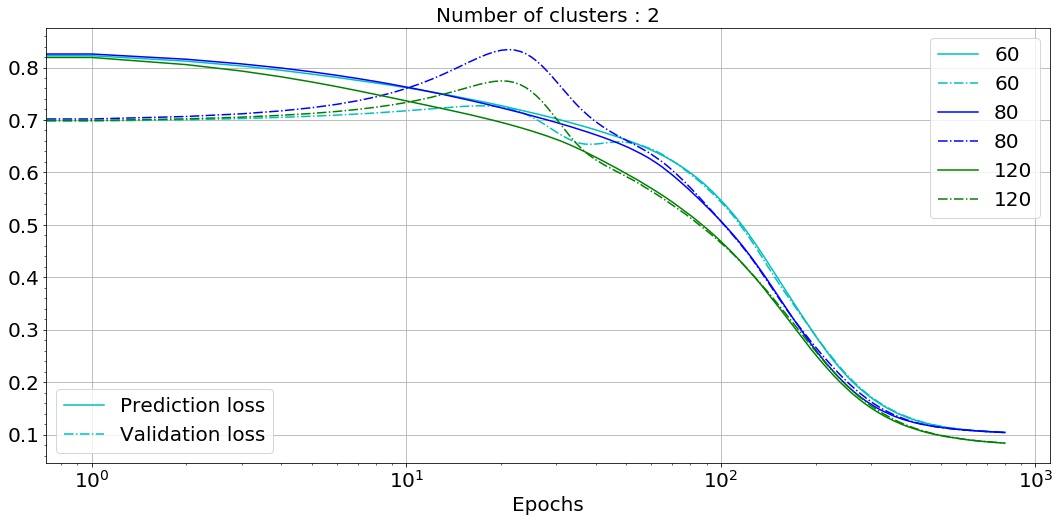}}
    \subfigure{\includegraphics[width=4.5cm,height=3.5cm]{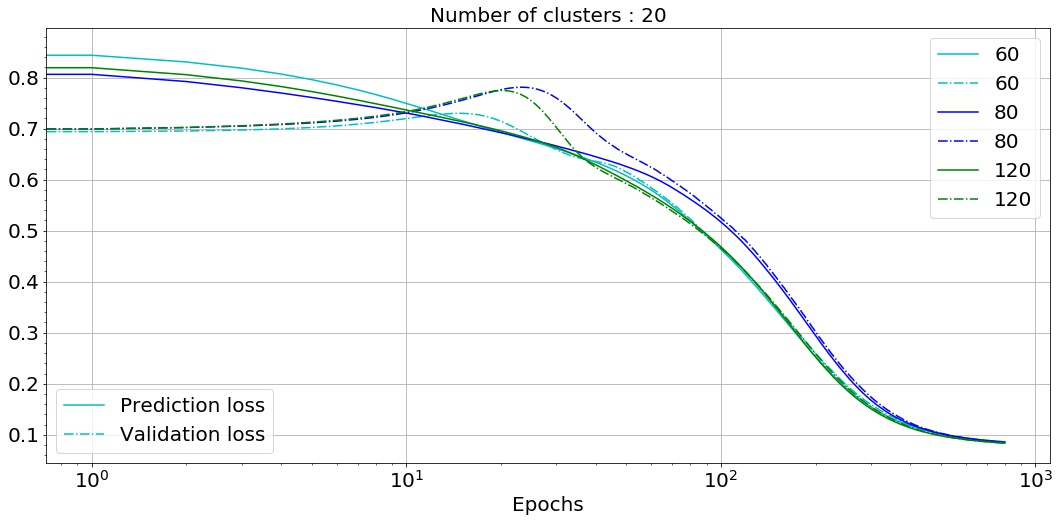}}
    \subfigure{\includegraphics[width=4.5cm,height=3.5cm]{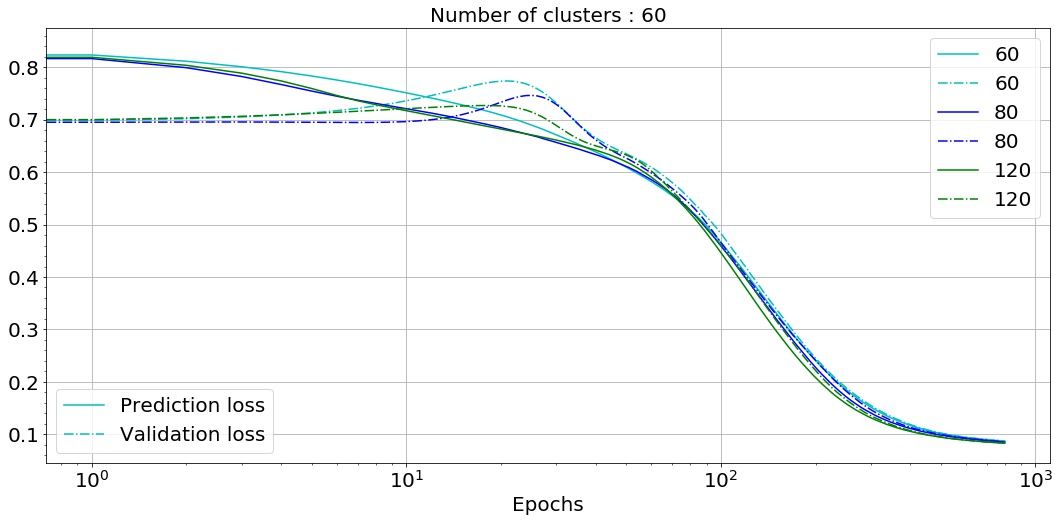}}
    \caption{Variation of loss function with image size corresponding to number of clusters}
    \label{figd1AC00}
\end{figure*}

\begin{figure*}[p]
\centering
    \subfigure{\includegraphics[width=8cm,height=5cm]{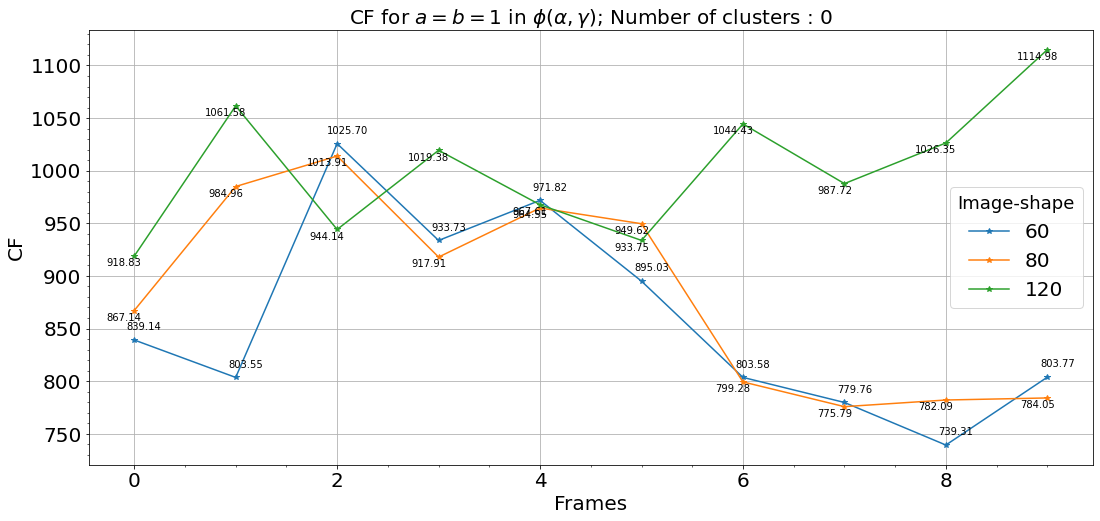}}
    \subfigure{\includegraphics[width=8cm,height=5cm]{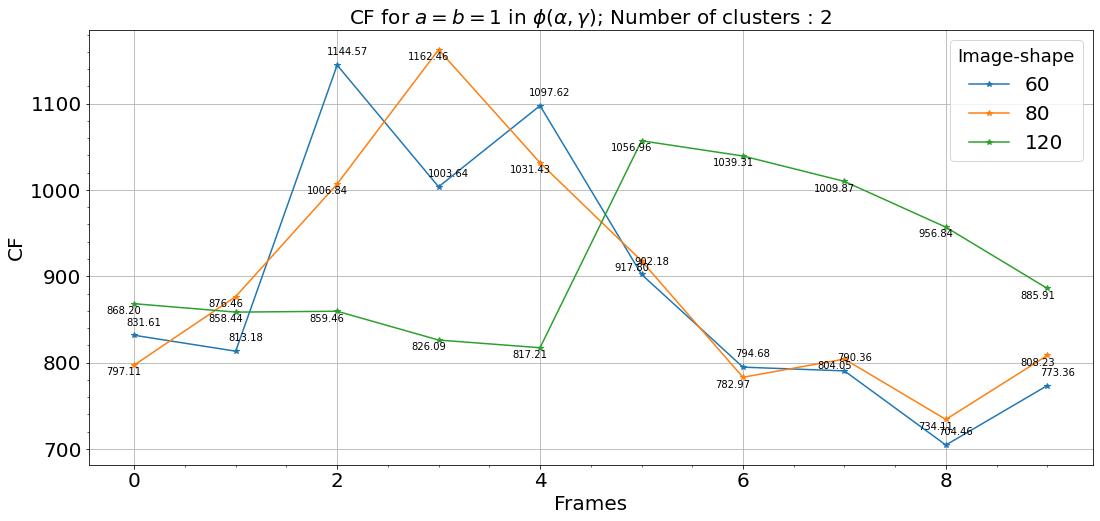}}
    \subfigure{\includegraphics[width=8cm,height=5cm]{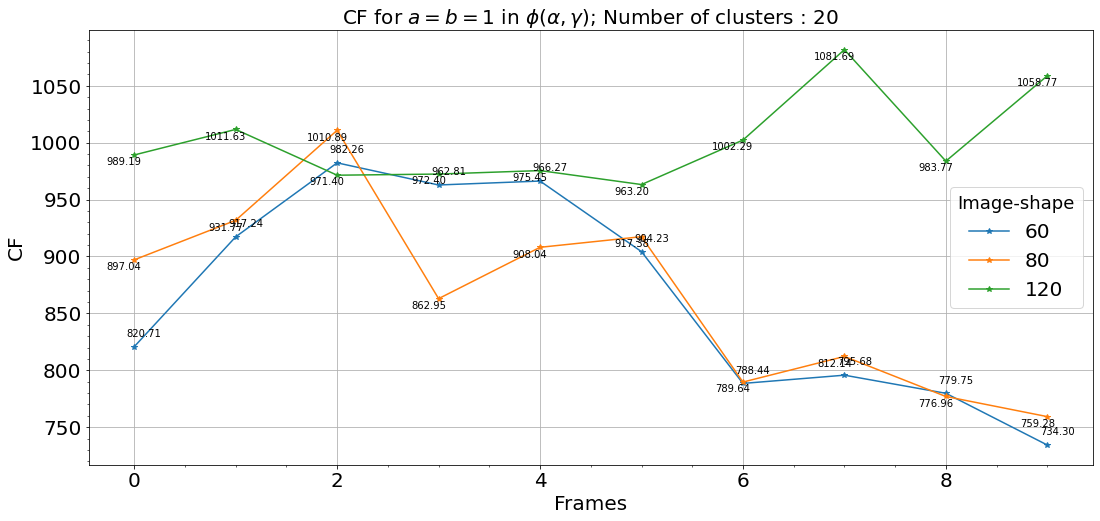}}
    \subfigure{\includegraphics[width=8cm,height=5cm]{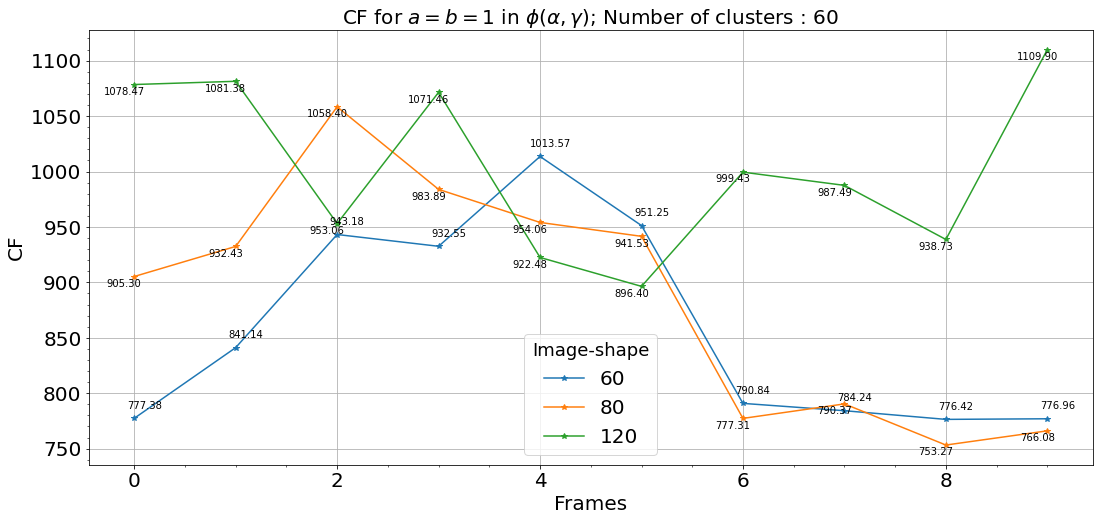}}
    \caption{Variation of CF with varying Image-size corresponding to each cluster}
    \label{figd1AC000}
\end{figure*}

\begin{figure*}[p]
\centering
    \subfigure{\includegraphics[width=8cm,height=5cm]{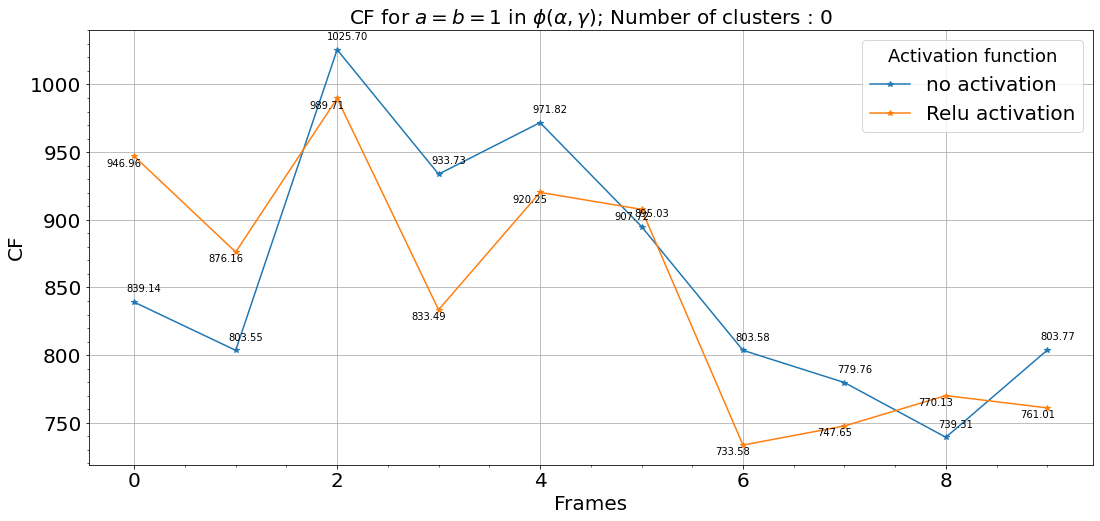}}
    \subfigure{\includegraphics[width=8cm,height=5cm]{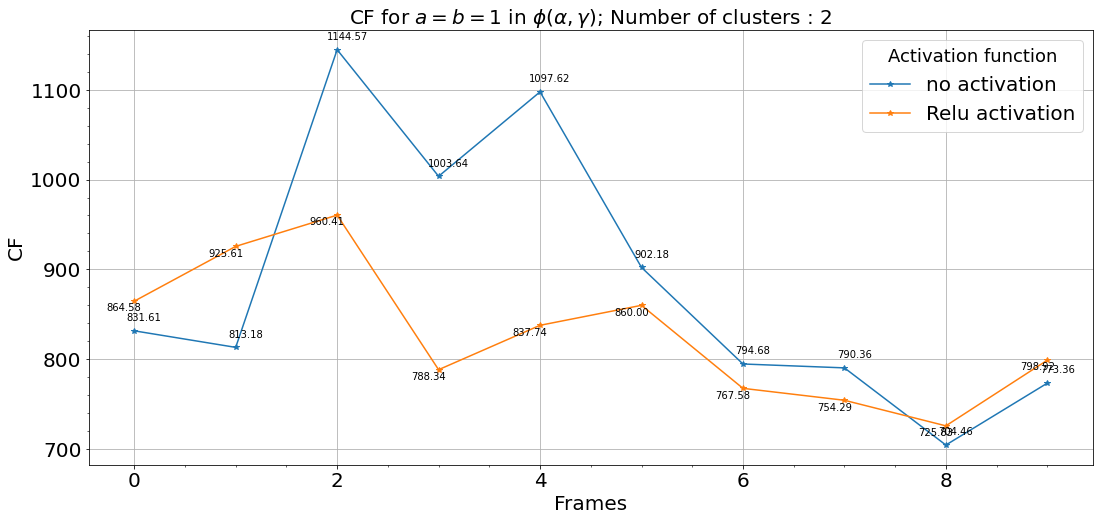}}
    \subfigure{\includegraphics[width=8cm,height=5cm]{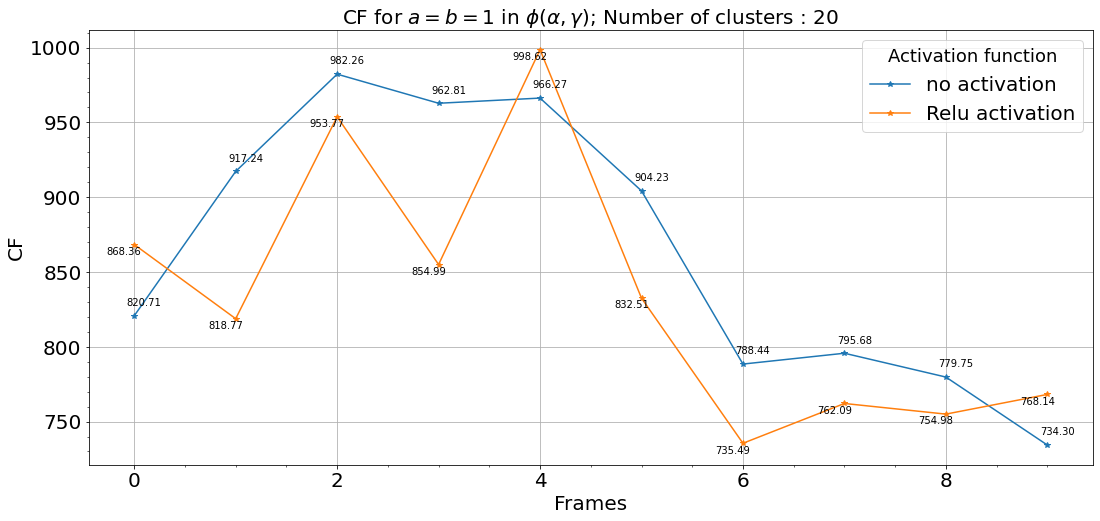}}
    \subfigure{\includegraphics[width=8cm,height=5cm]{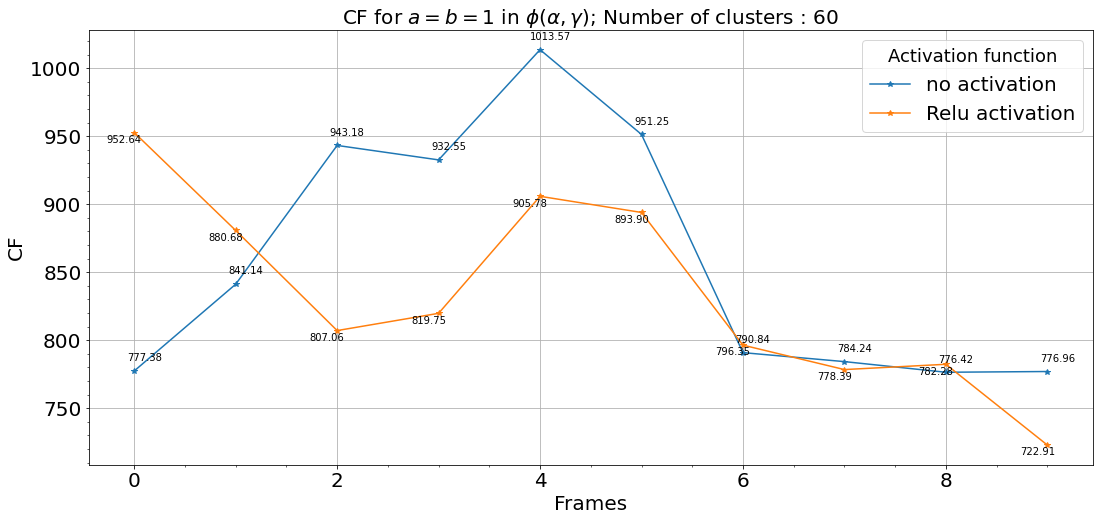}}
    \caption{Variation of CF with activation function (Relu) for each clusters}
    \label{figd1AC10}
\end{figure*}

\begin{figure*}[p]
\centering
    \subfigure{\includegraphics[width=6cm,height=3.5cm]{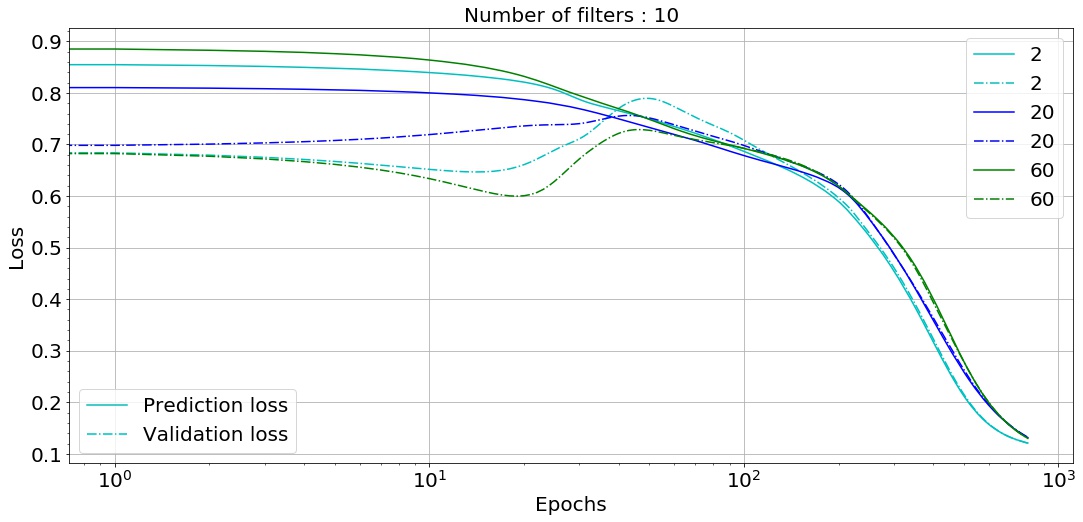}}
    \subfigure{\includegraphics[width=6cm,height=3.5cm]{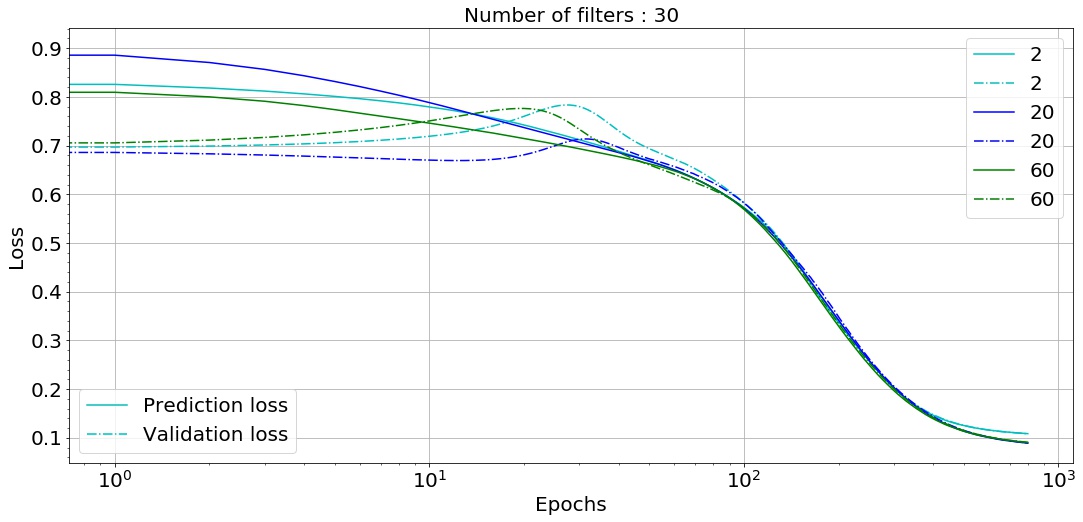}}
    \subfigure{\includegraphics[width=6cm,height=3.5cm]{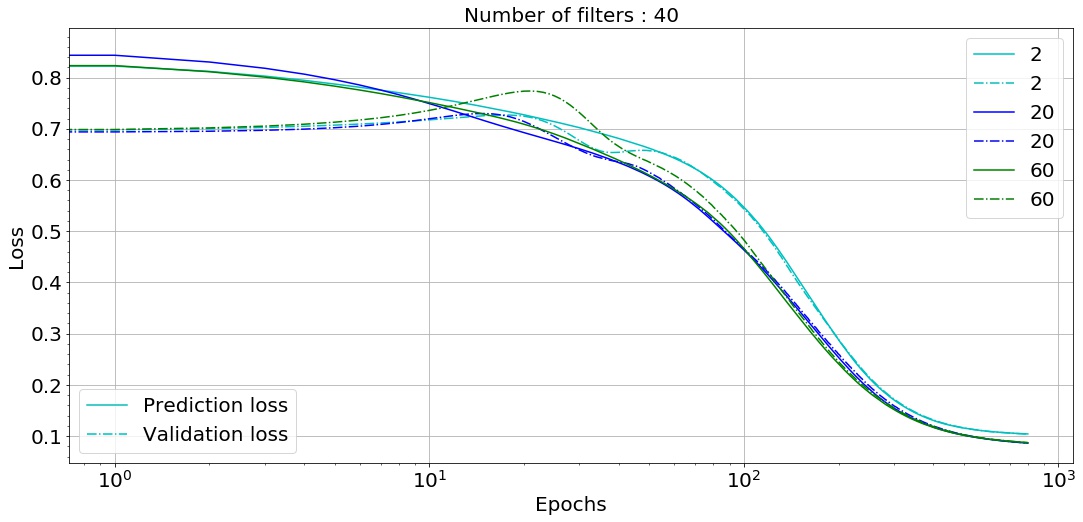}}
    \caption{Variation of loss function with varying number of filters corresponding to each cluster}
    \label{figd1FS1}
\end{figure*}

\begin{figure*}[p]
\centering
    \subfigure{\includegraphics[width=4.5cm,height=3.5cm]{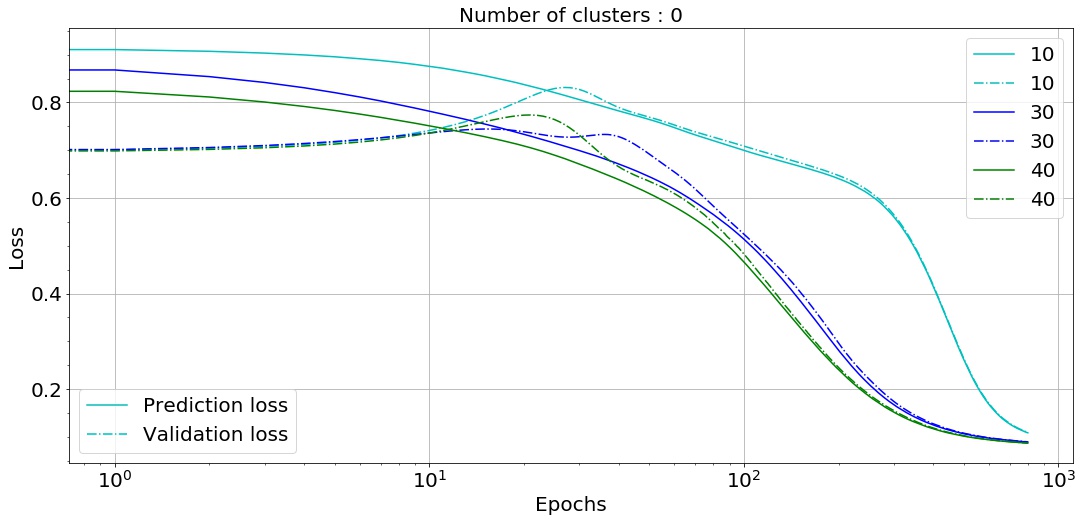}}
    \subfigure{\includegraphics[width=4.5cm,height=3.5cm]{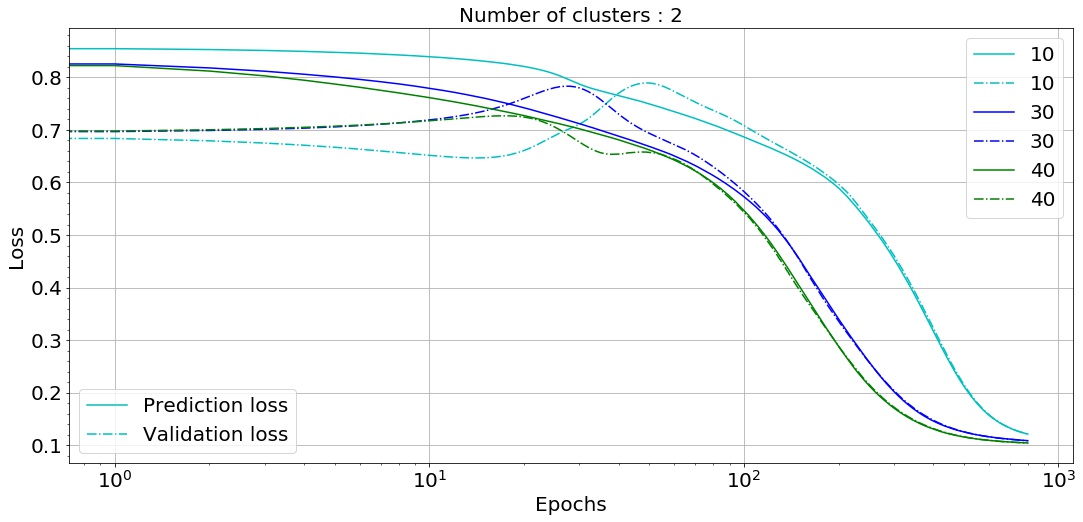}}
    \subfigure{\includegraphics[width=4.5cm,height=3.5cm]{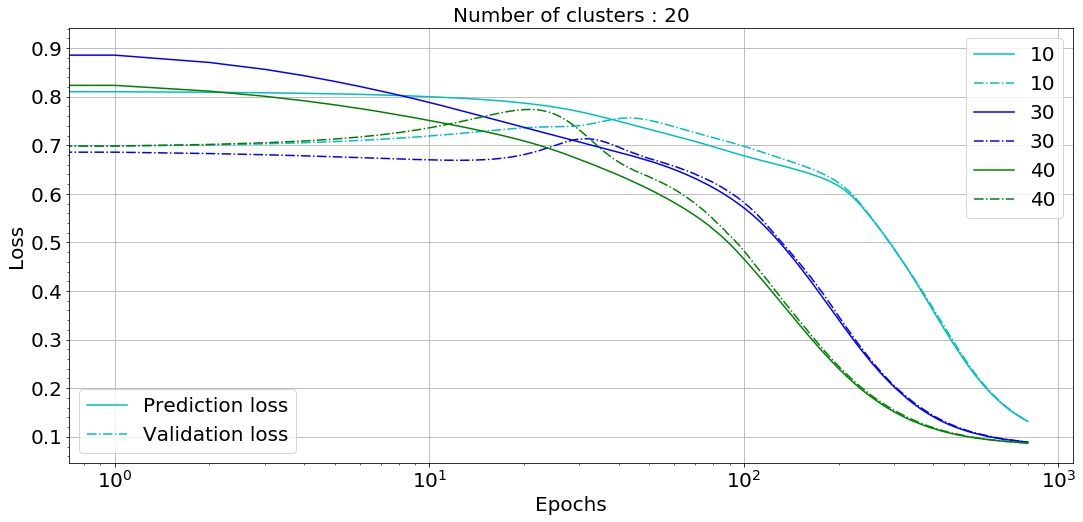}}
    \subfigure{\includegraphics[width=4.5cm,height=3.5cm]{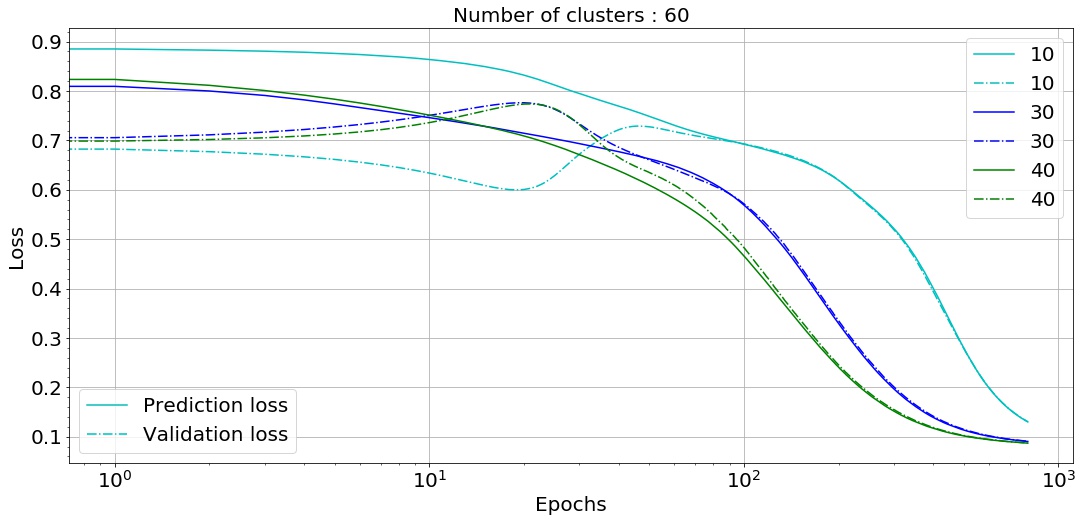}}
    \caption{Variation of loss function with each cluster corresponding to number of filters}
    \label{figd1FS2}
\end{figure*}

\begin{figure}[p]
\centering
    \subfigure{\includegraphics[width=8cm,height=5cm]{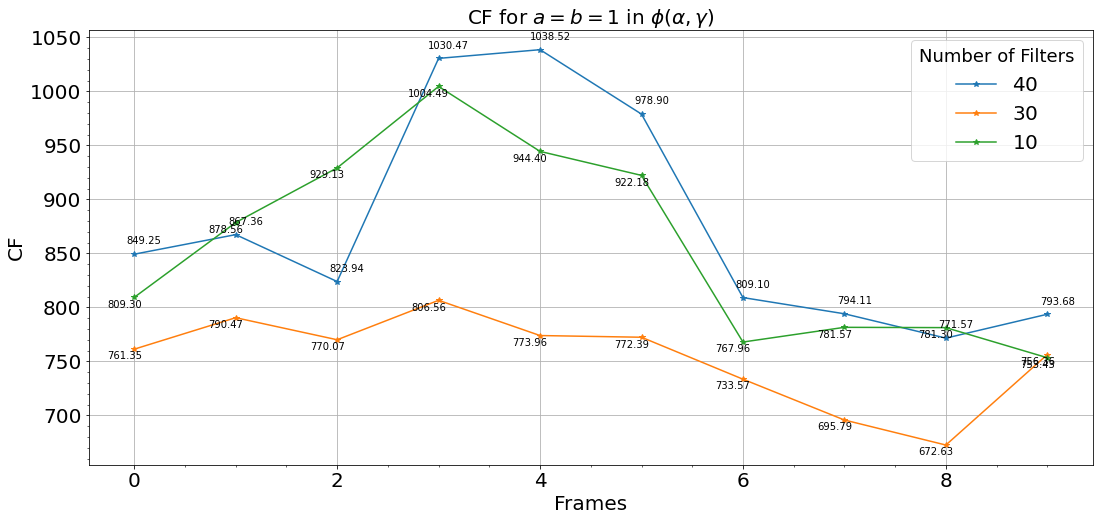}}
    \caption{Variation of CF with number of filters}
    \label{figd1FS3}
\end{figure}

\begin{figure*}[p]
\centering
    \subfigure{\includegraphics[width=6cm,height=3.5cm]{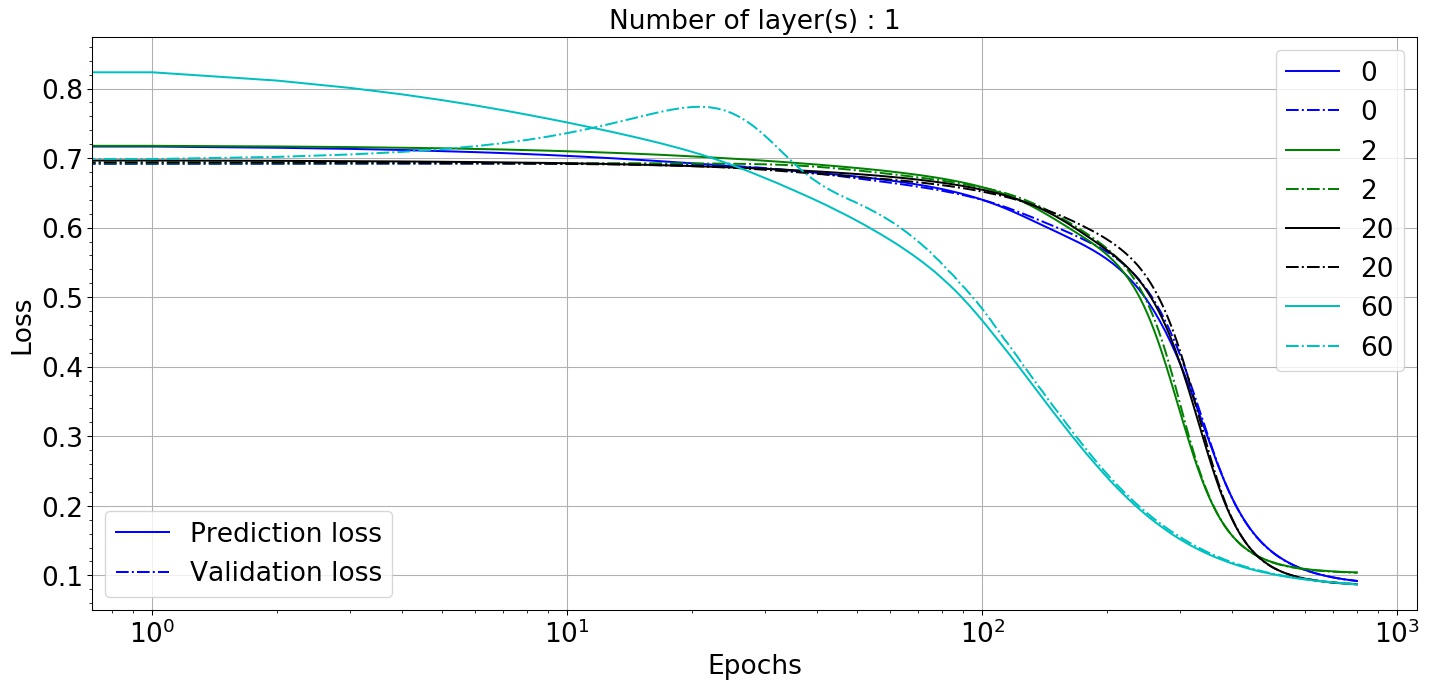}}
    \subfigure{\includegraphics[width=6cm,height=3.5cm]{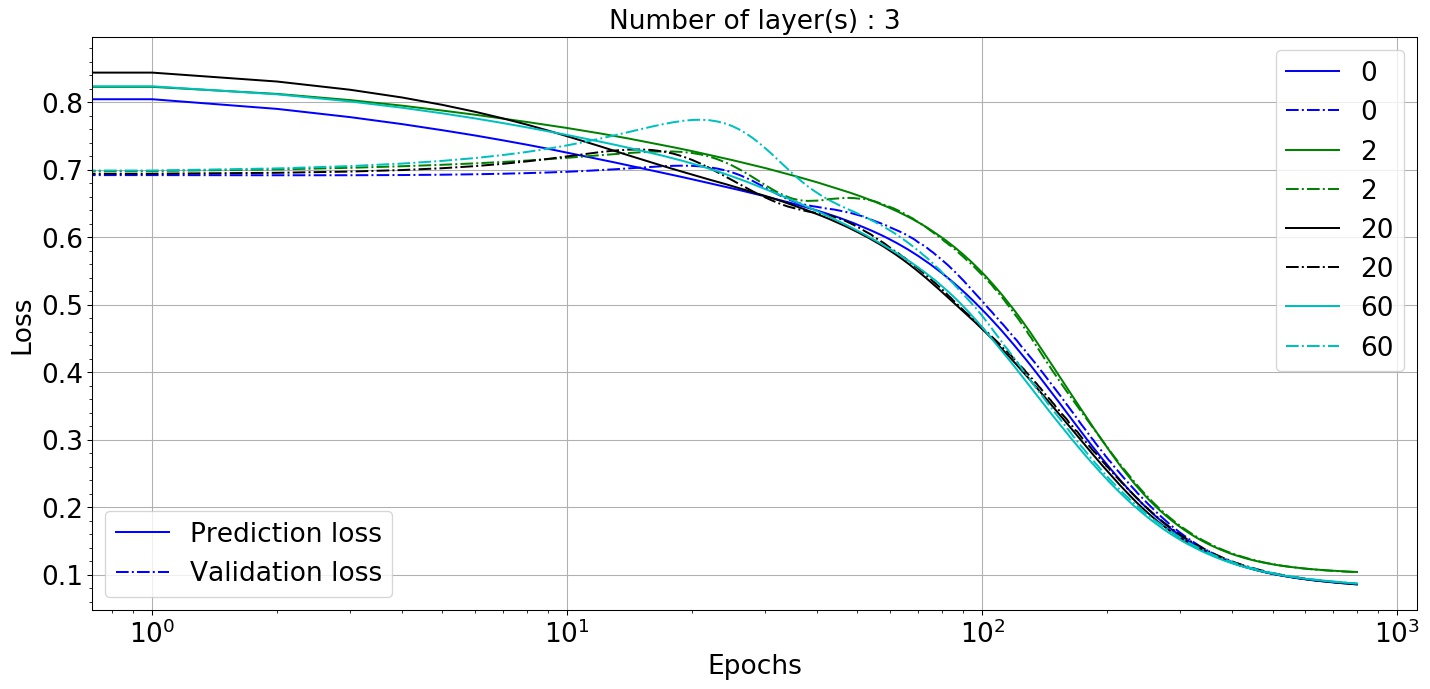}}
    \subfigure{\includegraphics[width=6cm,height=3.5cm]{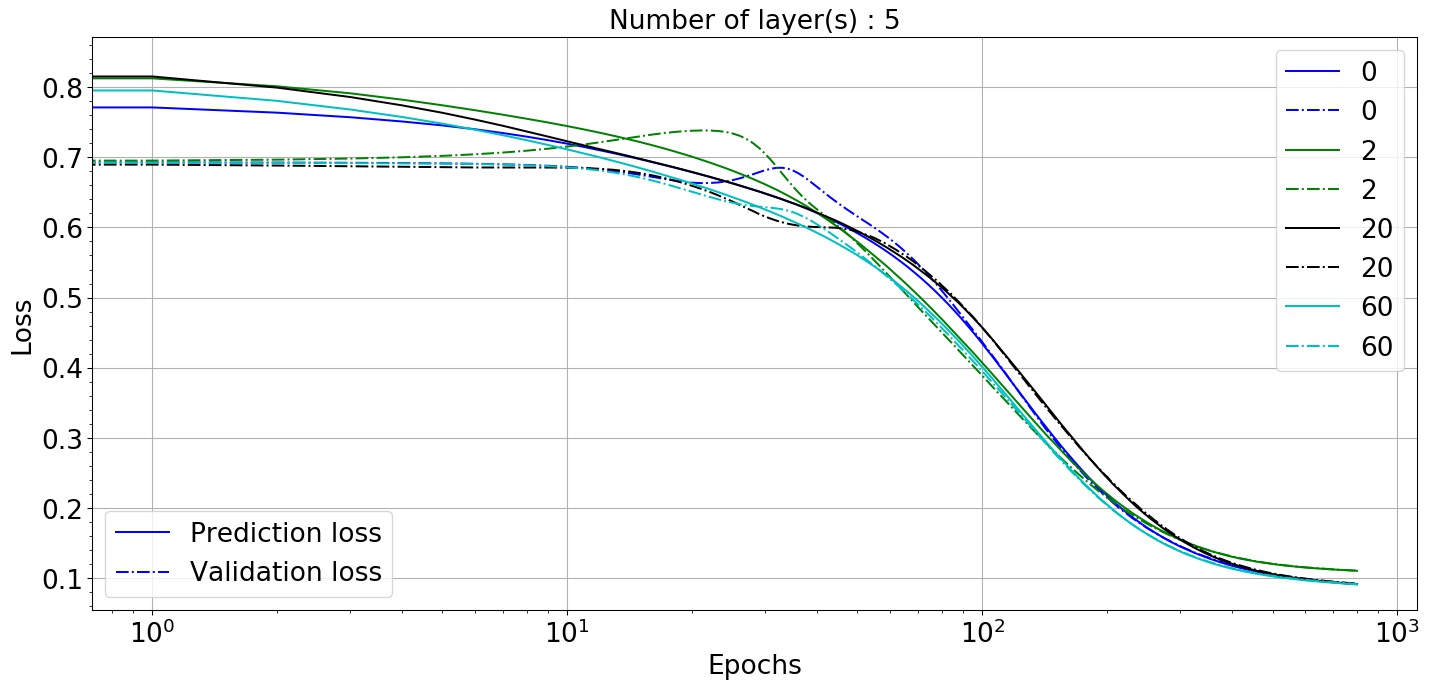}}
    \subfigure{\includegraphics[width=6cm,height=3.5cm]{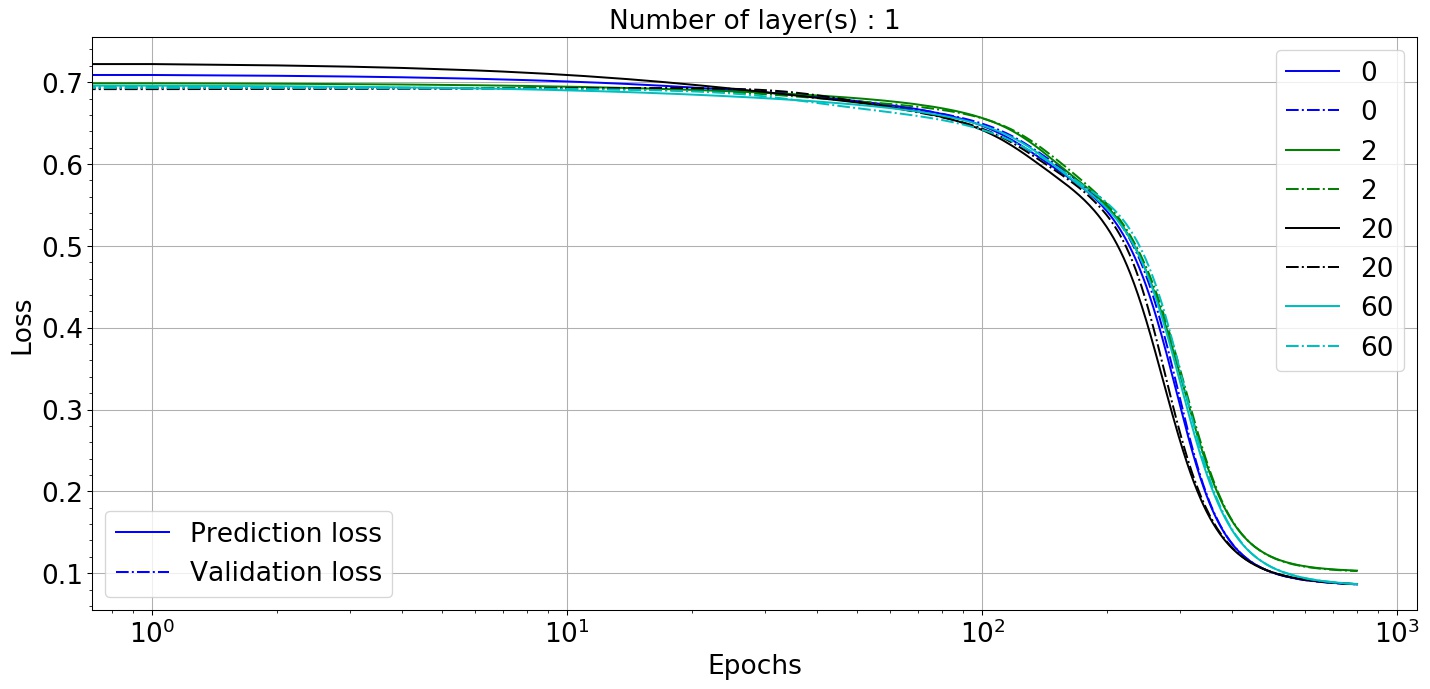}}
    \subfigure{\includegraphics[width=6cm,height=3.5cm]{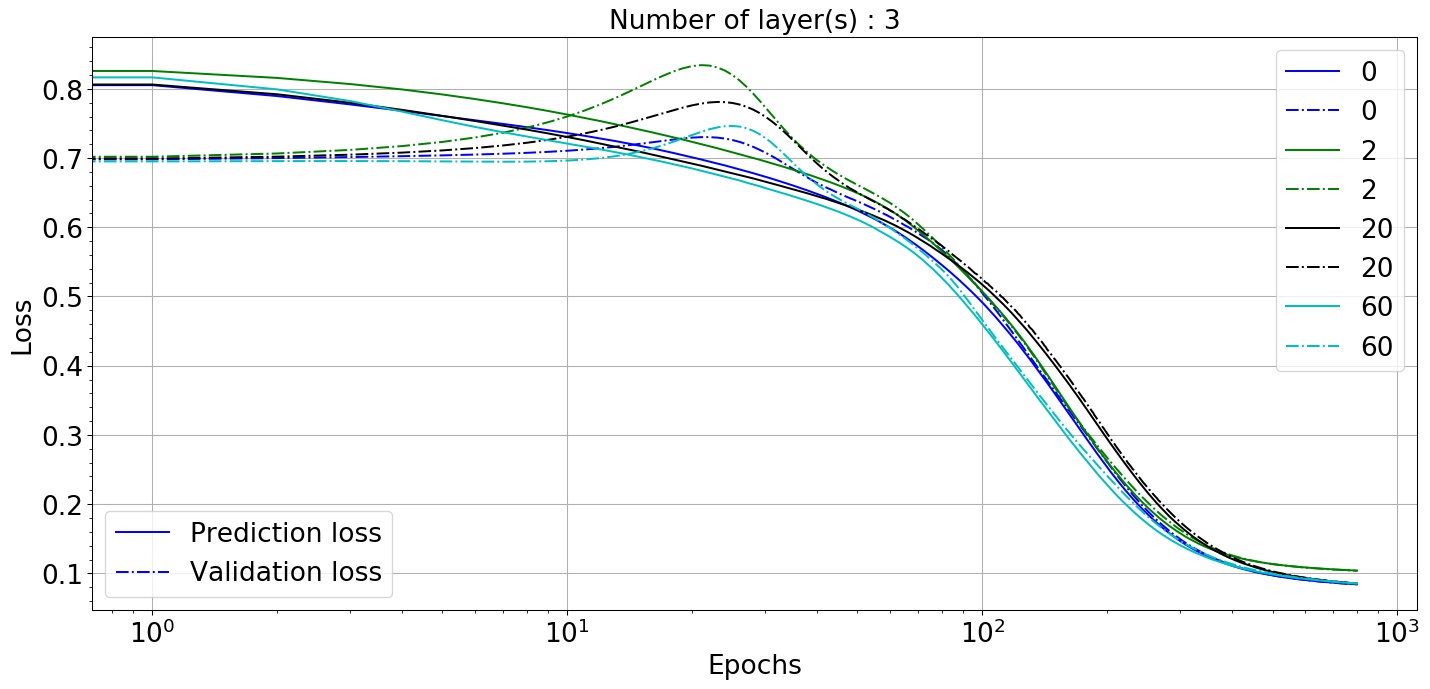}}
    \subfigure{\includegraphics[width=6cm,height=3.5cm]{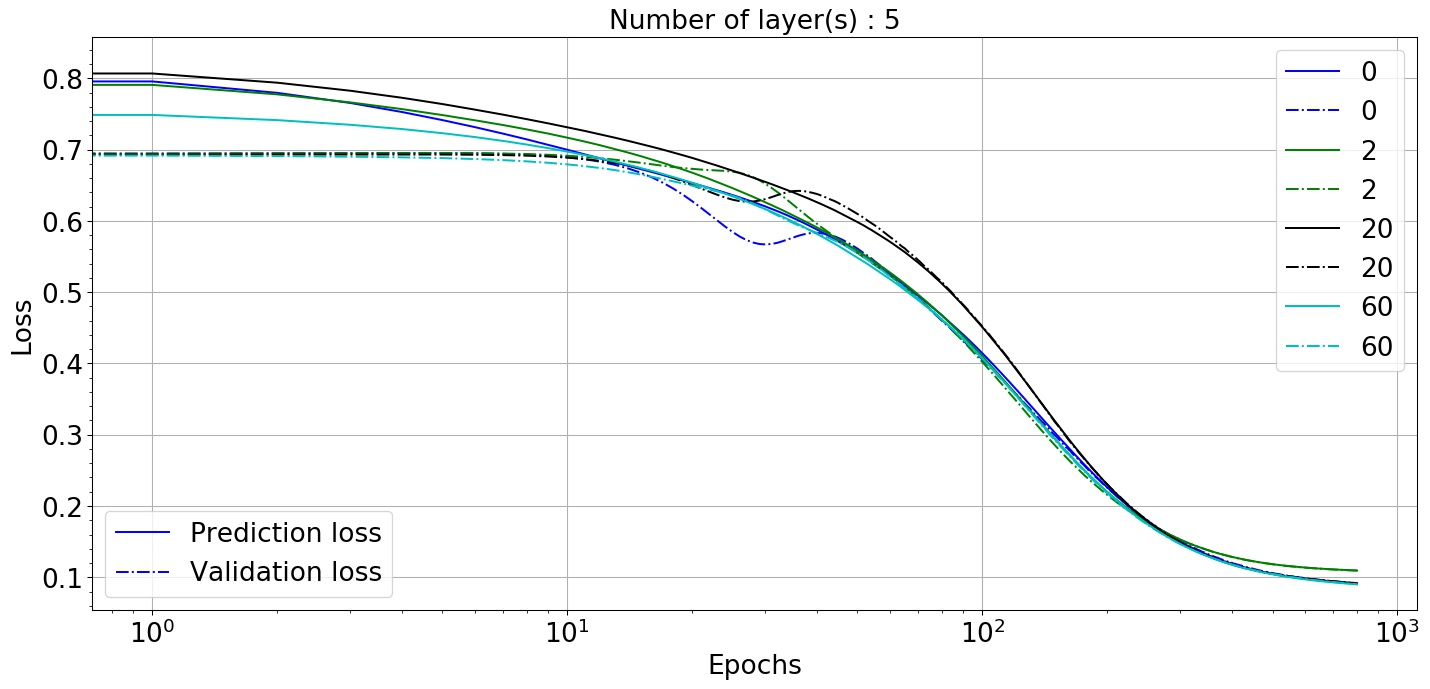}}
    \caption{Variation of loss function with number of clusters corresponding to number of layers. The upper row represents the variation for Image-size $60$ and the lower row for image size $80$.}
    \label{figd1NL1}
\end{figure*}

\begin{figure*}[p]
\centering
    \subfigure{\includegraphics[width=4.5cm,height=3.5cm]{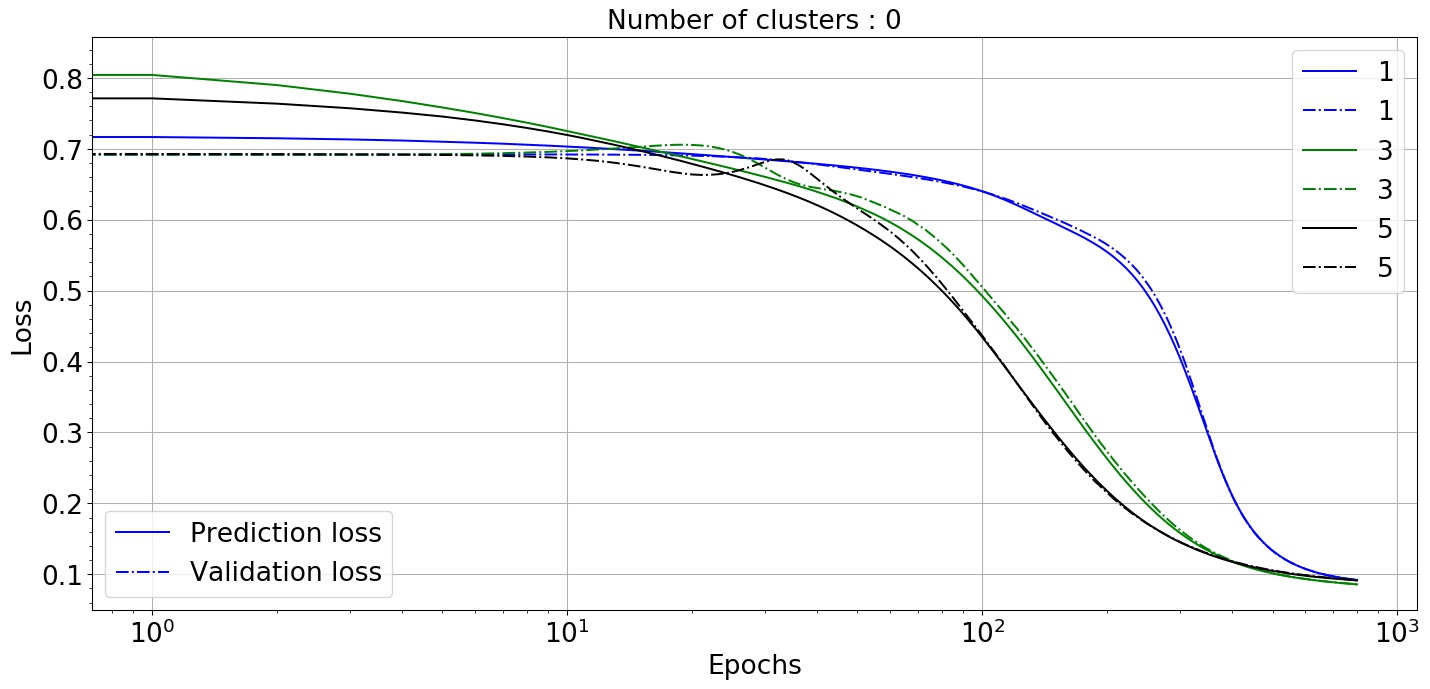}}
    \subfigure{\includegraphics[width=4.5cm,height=3.5cm]{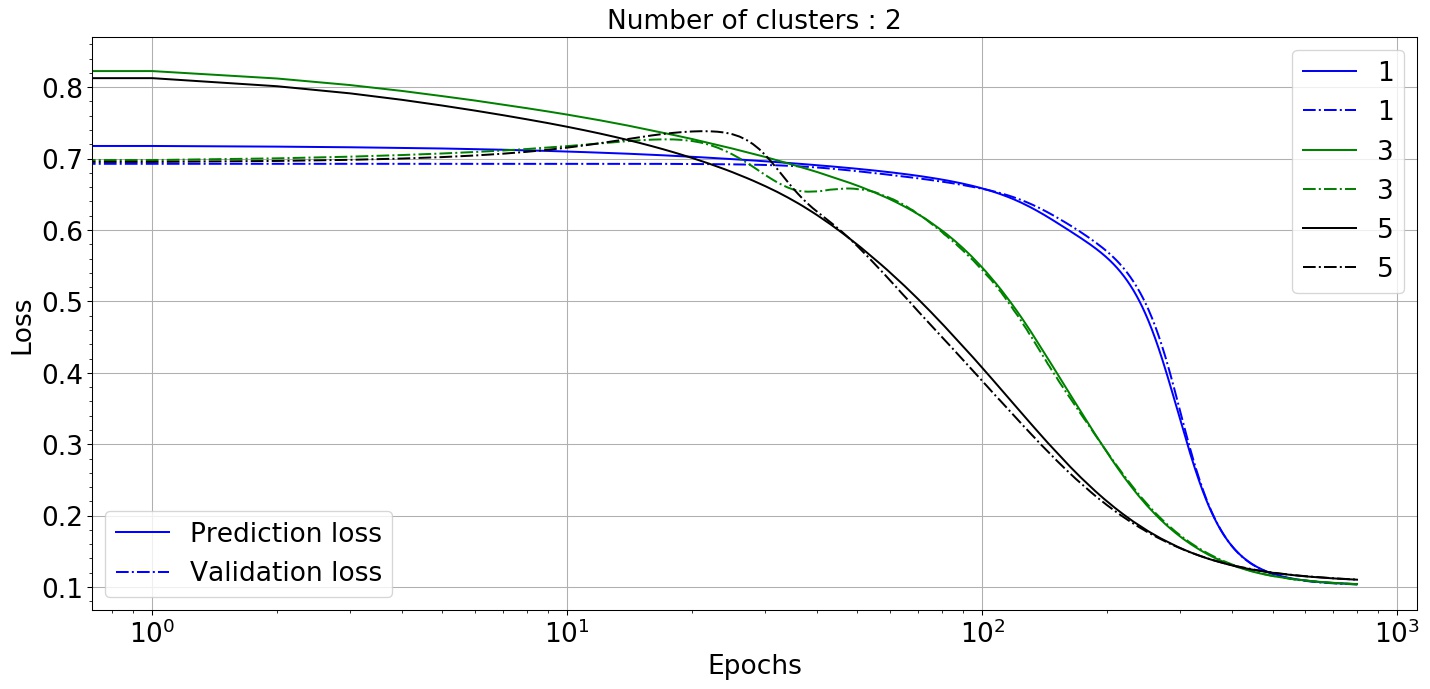}}
    \subfigure{\includegraphics[width=4.5cm,height=3.5cm]{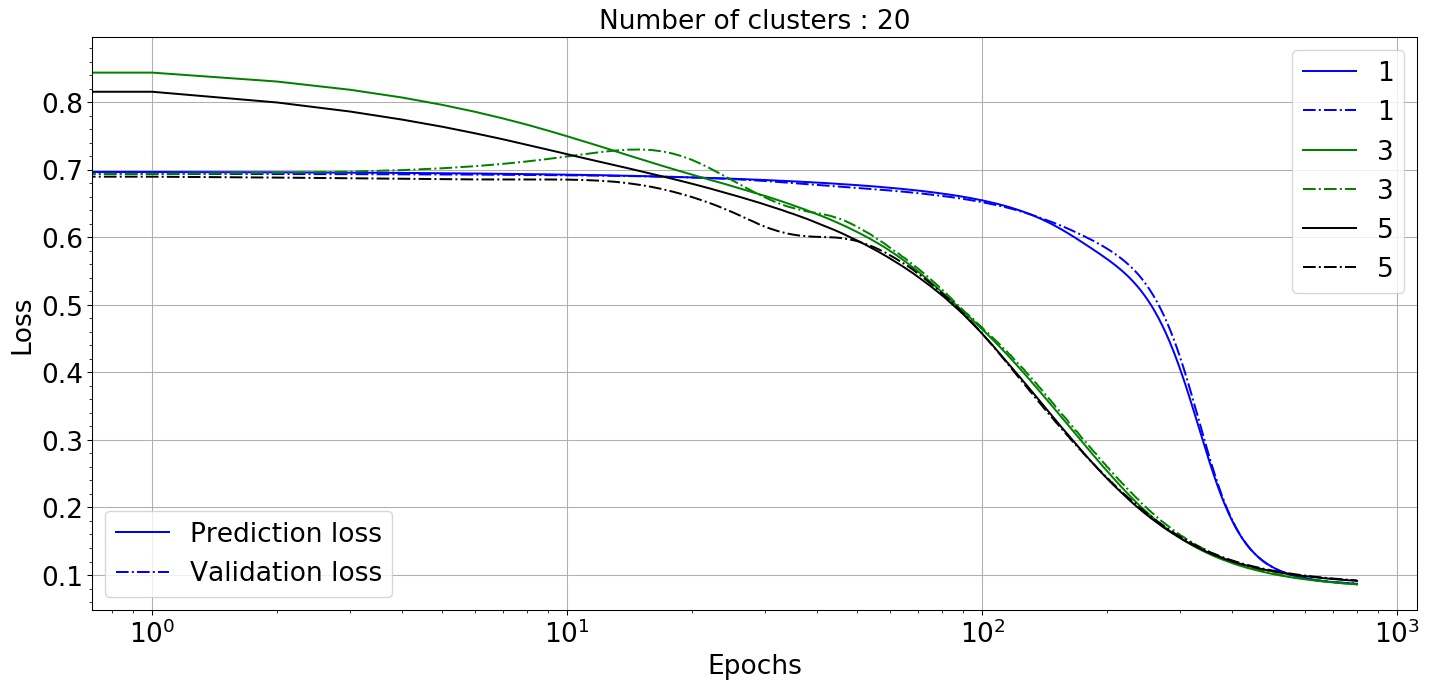}}
    \subfigure{\includegraphics[width=4.5cm,height=3.5cm]{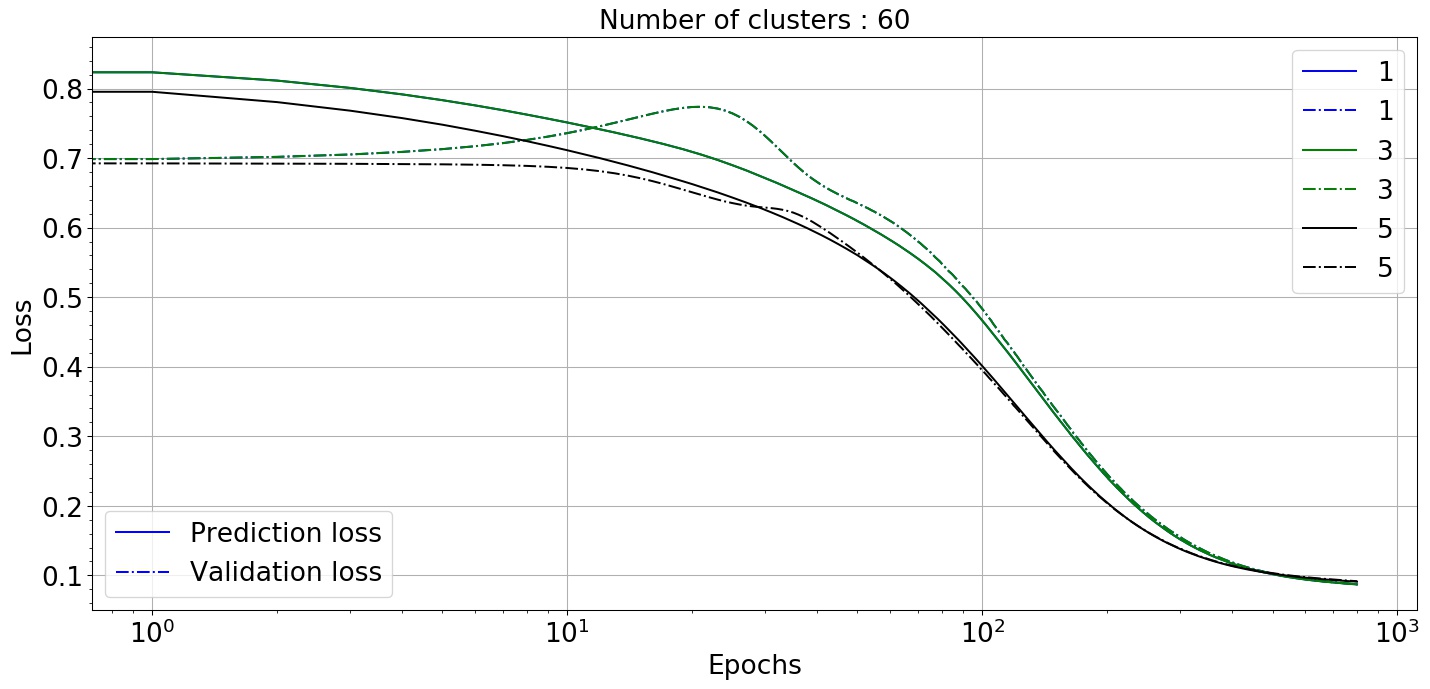}}
    \caption{Variation of loss function with number of layers corresponding to each clusters}
    \label{figd1NL2}
\end{figure*}

\begin{figure*}[p]
\centering
    \subfigure{\includegraphics[width=8cm,height=5cm]{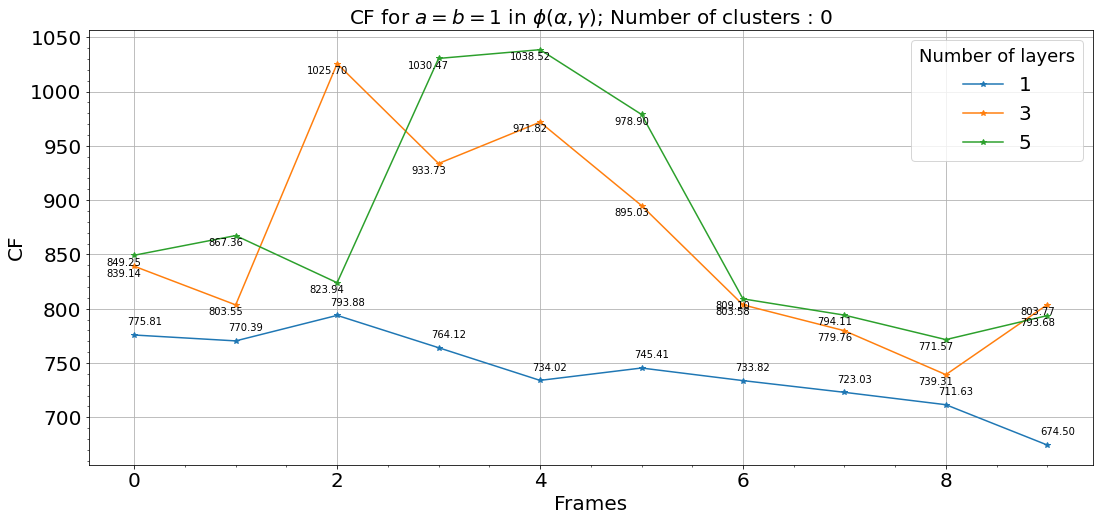}}
    \subfigure{\includegraphics[width=8cm,height=5cm]{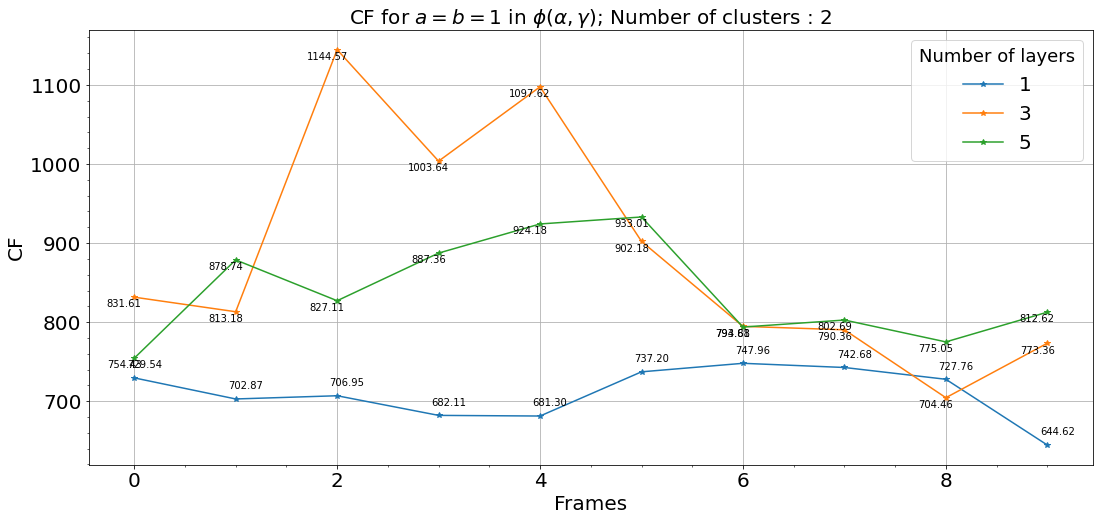}}
    \subfigure{\includegraphics[width=8cm,height=5cm]{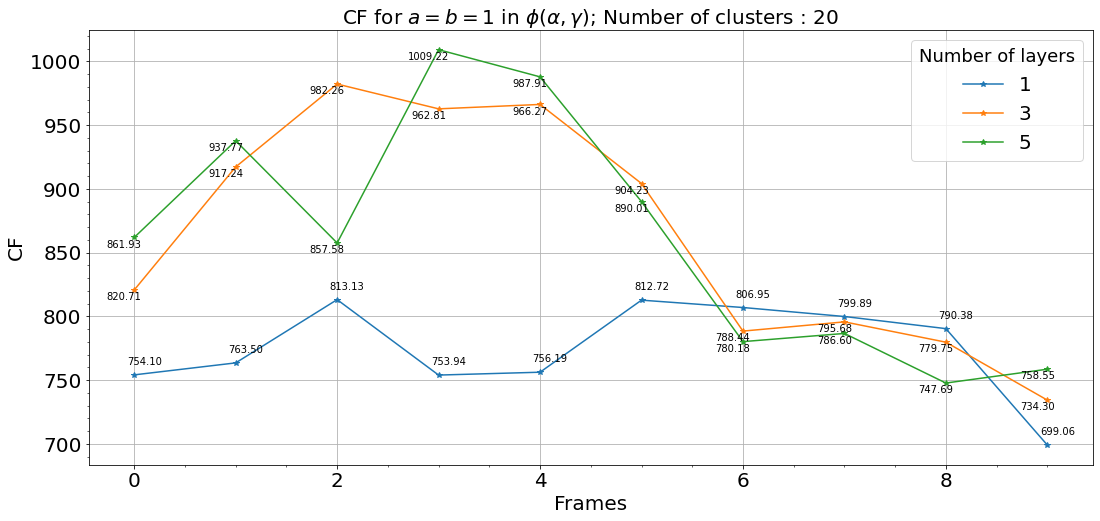}}
    \subfigure{\includegraphics[width=8cm,height=5cm]{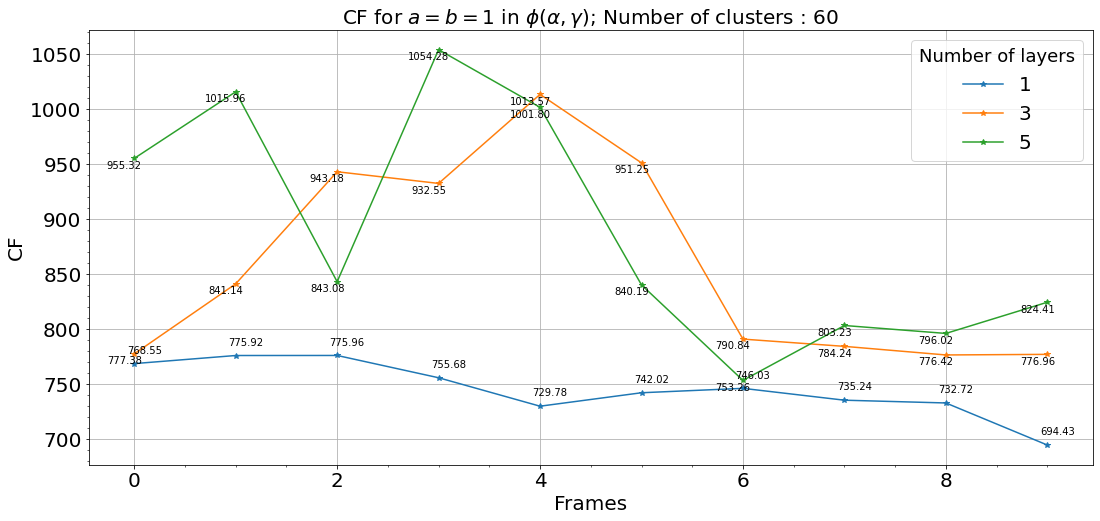}}
    \caption{Variation of CF with number of layers corresponding to each clusters}
    \label{figd1NL31}
\end{figure*}

\begin{figure*}[p]
\centering
    \subfigure{\includegraphics[width=8cm,height=5cm]{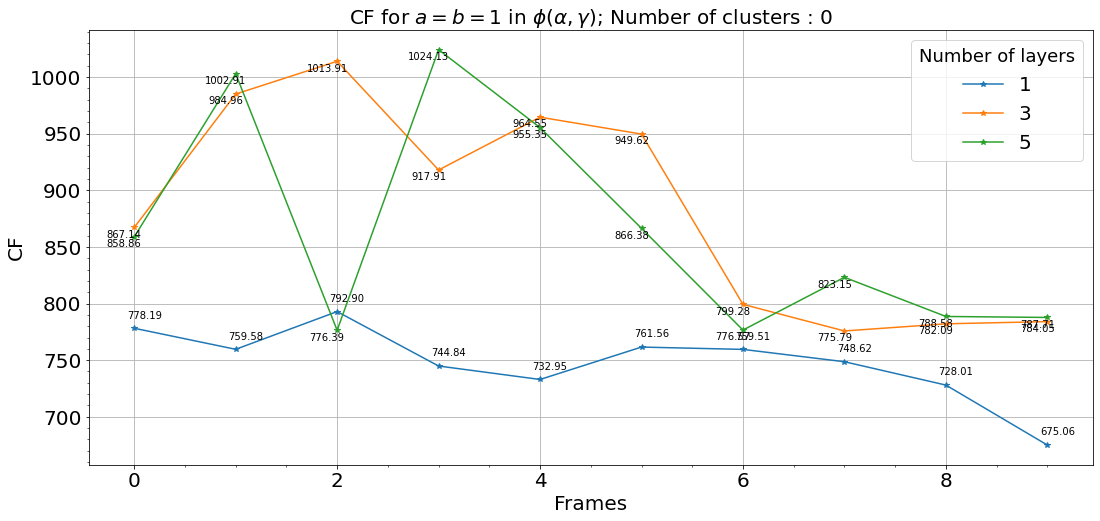}}
    \subfigure{\includegraphics[width=8cm,height=5cm]{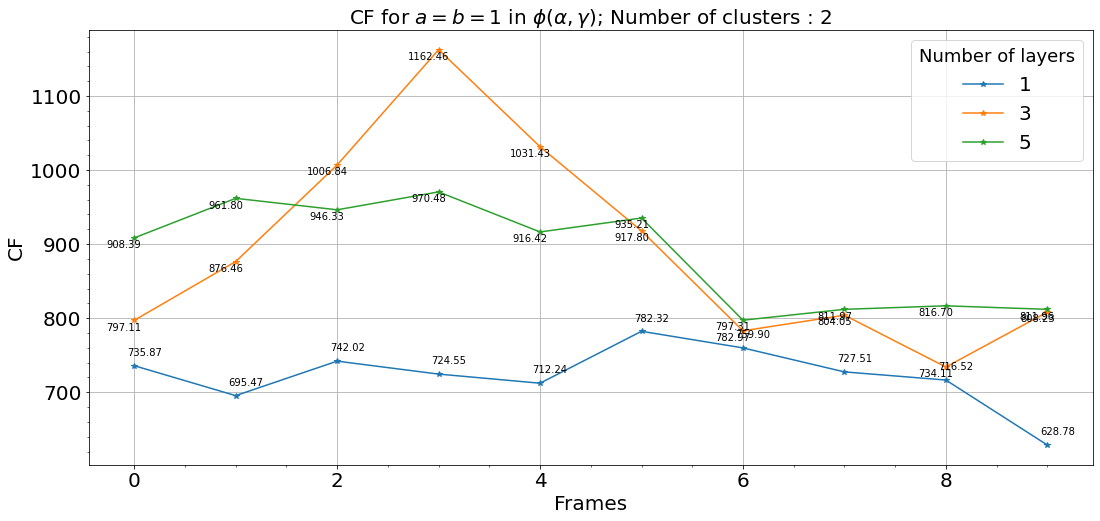}}
    \subfigure{\includegraphics[width=8cm,height=5cm]{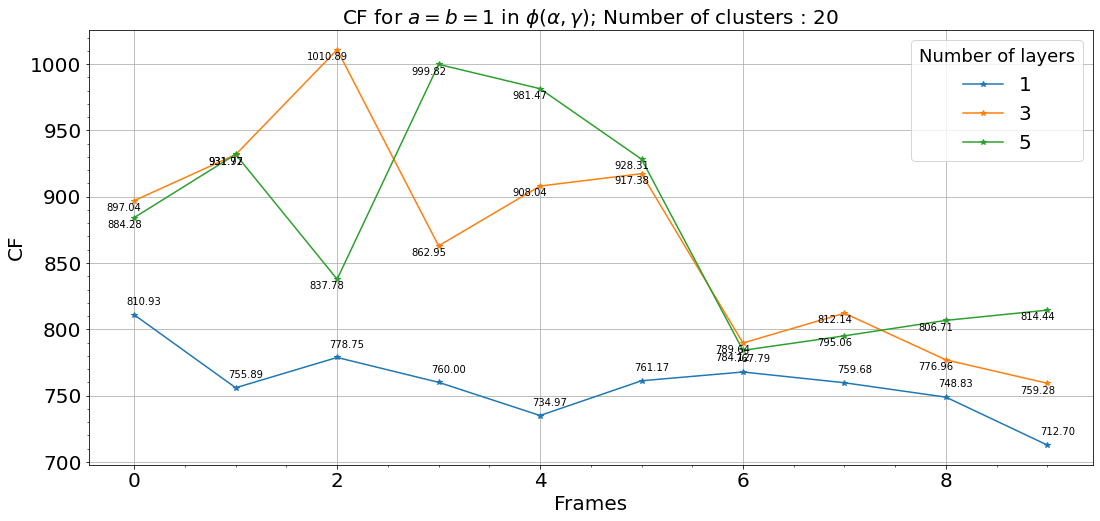}}
    \subfigure{\includegraphics[width=8cm,height=5cm]{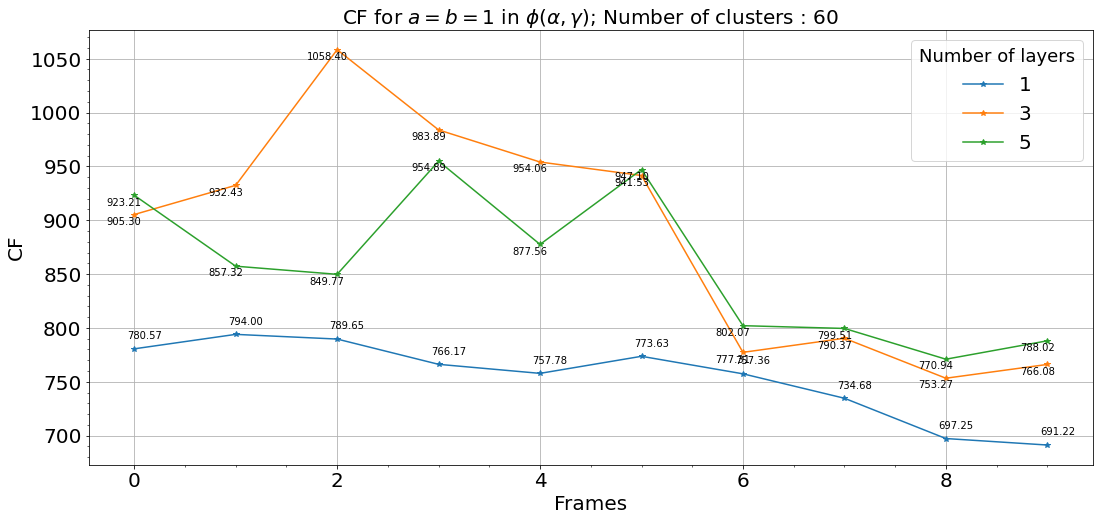}}
    \caption{Variation of CF with number of layers corresponding to each clusters}
    \label{figd1NL32}
\end{figure*}

\begin{figure*}[p]
\centering
    \subfigure{\includegraphics[width=4.5cm,height=3.5cm]{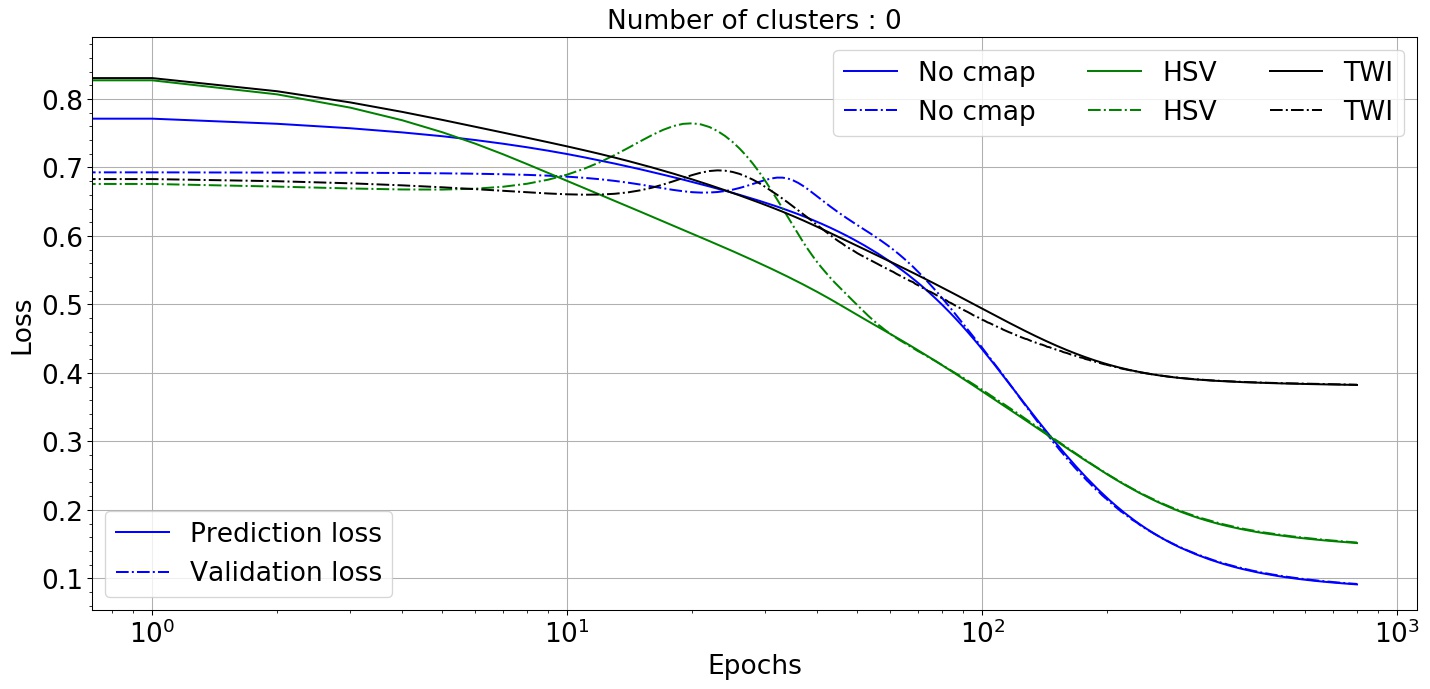}}
    \subfigure{\includegraphics[width=4.5cm,height=3.5cm]{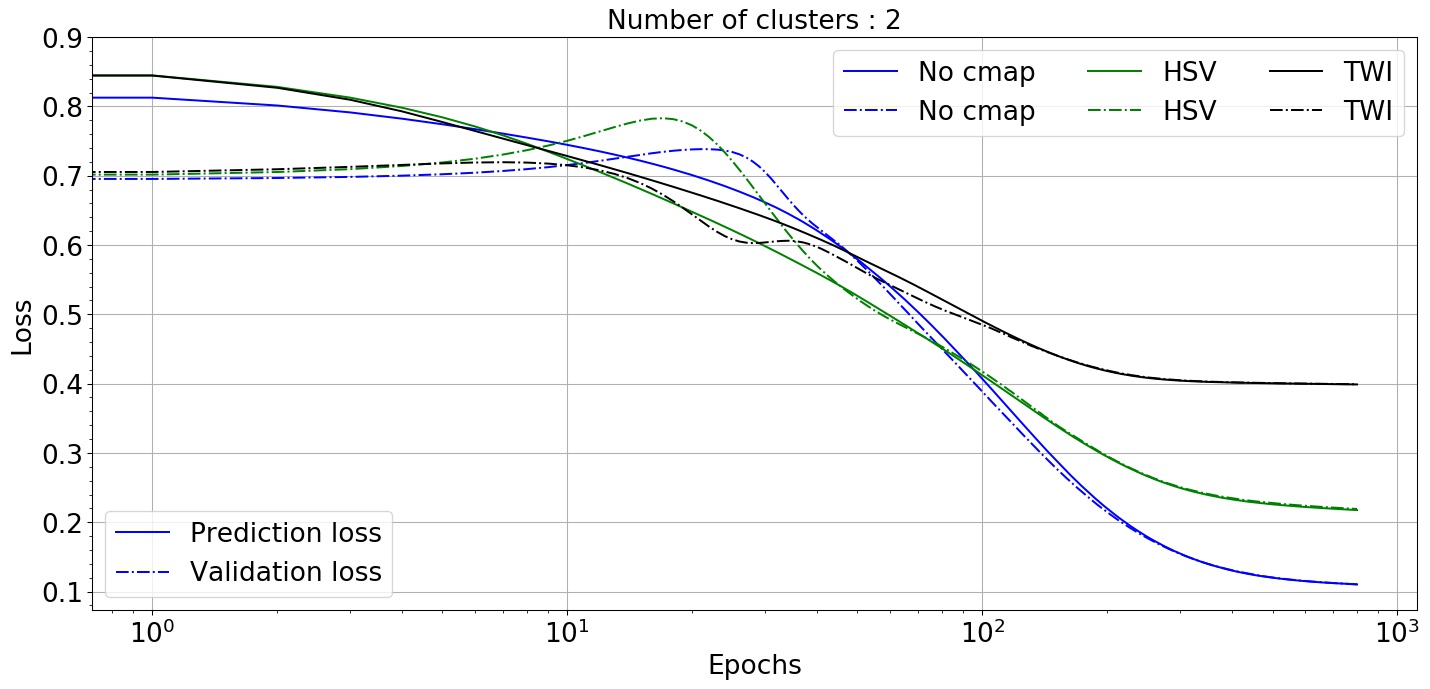}}
    \subfigure{\includegraphics[width=4.5cm,height=3.5cm]{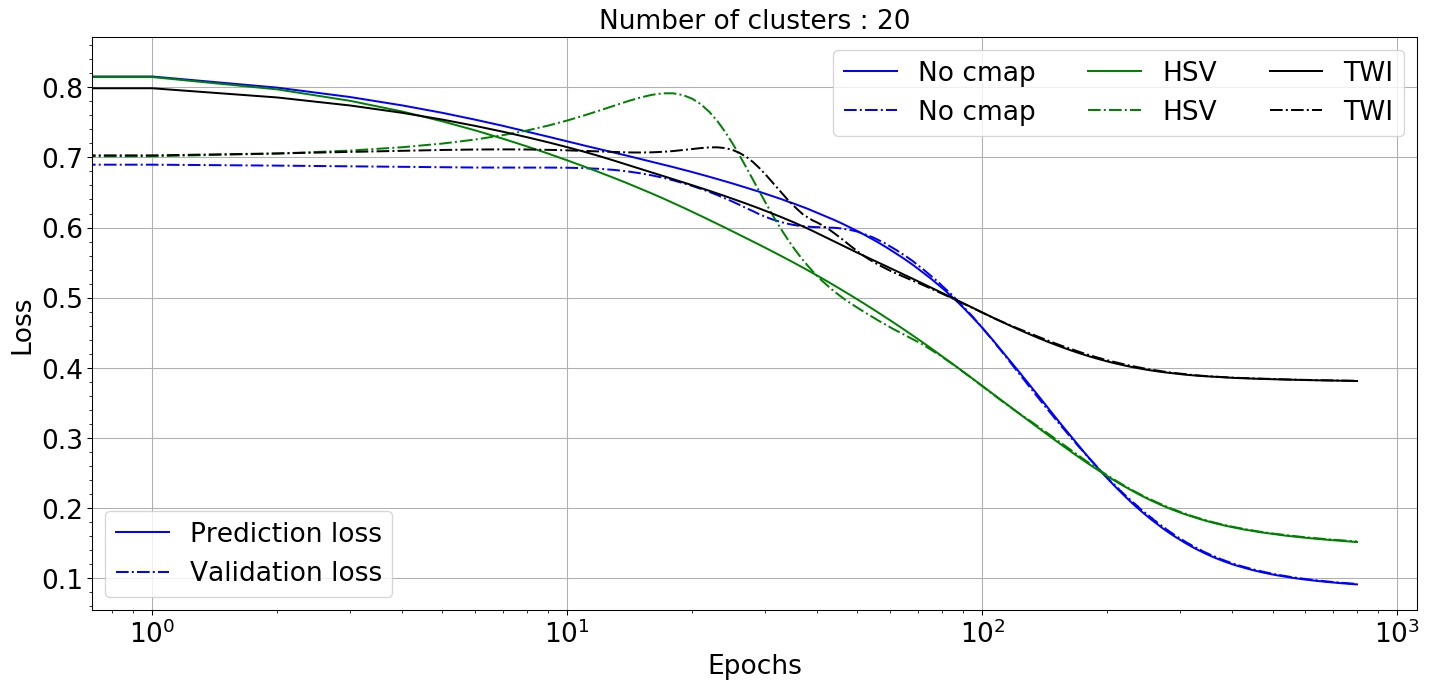}}
    \subfigure{\includegraphics[width=4.5cm,height=3.5cm]{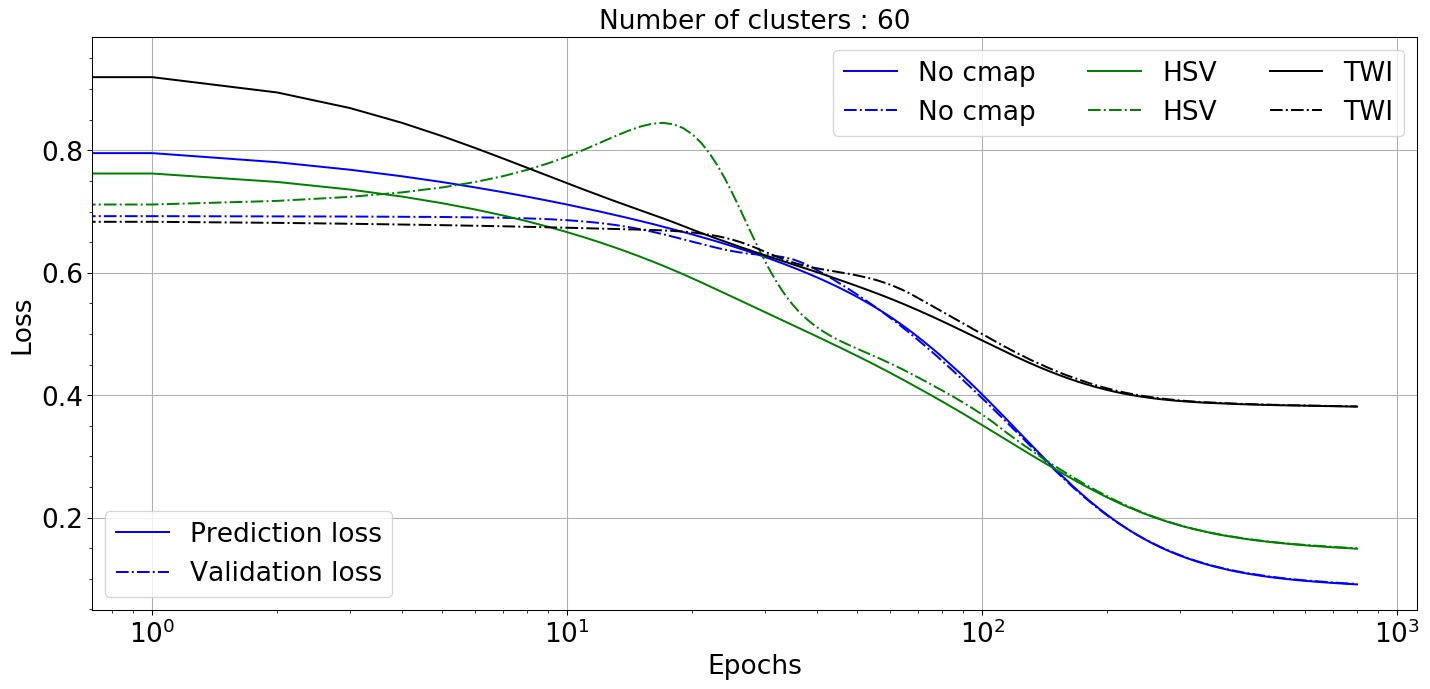}}
    \caption{Variation of loss function with colormap corresponding to each clusters}
    \label{figd21}
\end{figure*}

\begin{figure*}[p]
\centering
    \subfigure{\includegraphics[width=6cm,height=3.5cm]{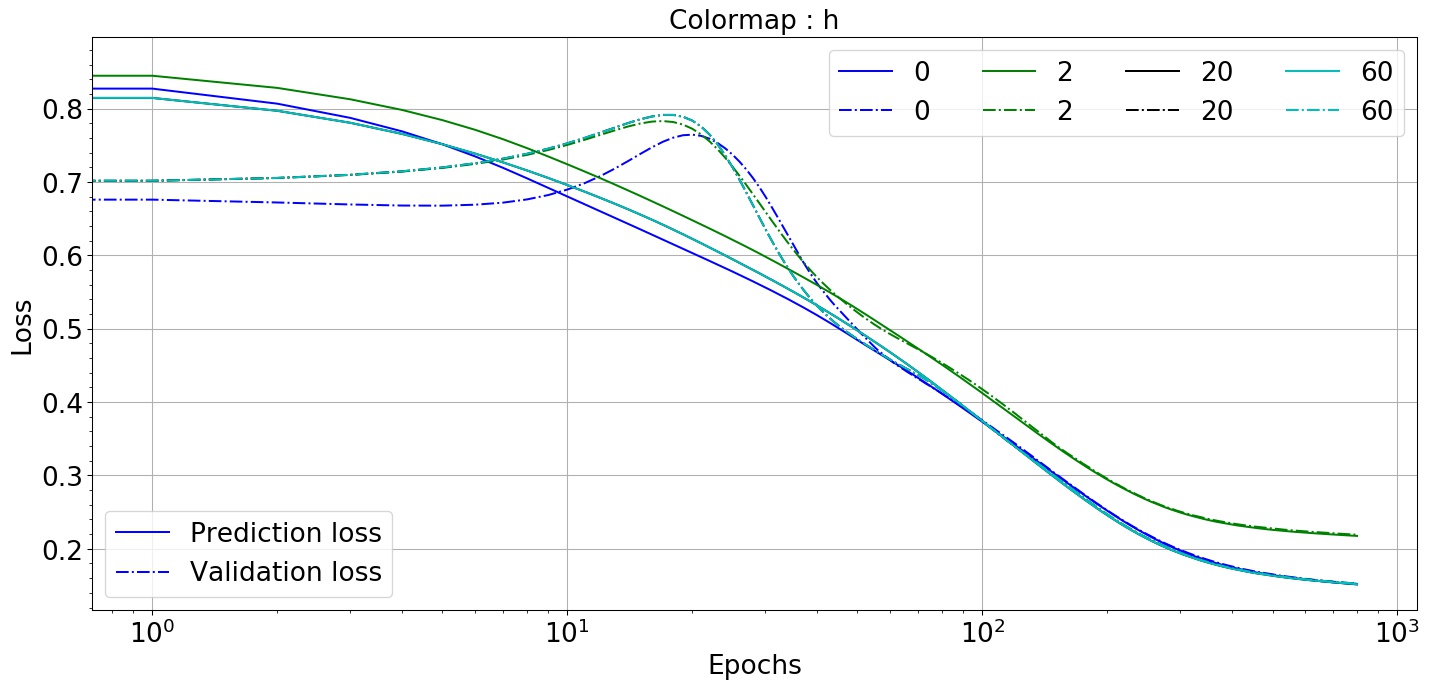}}
    \subfigure{\includegraphics[width=6cm,height=3.5cm]{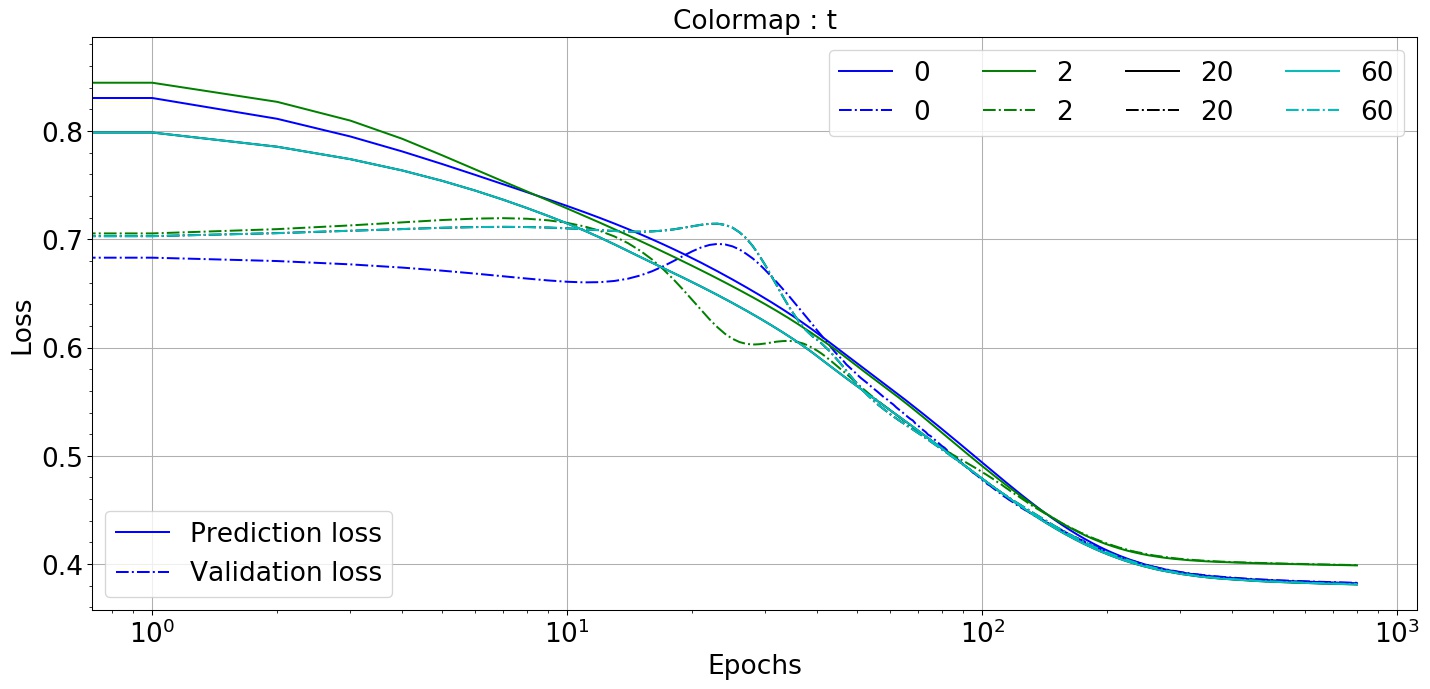}}
    \caption{Variation of loss function with each clusters corresponding to the two colormaps}
    \label{figd22}
\end{figure*}

\begin{figure*}[p]
\centering
    \subfigure{\includegraphics[width=8cm,height=5cm]{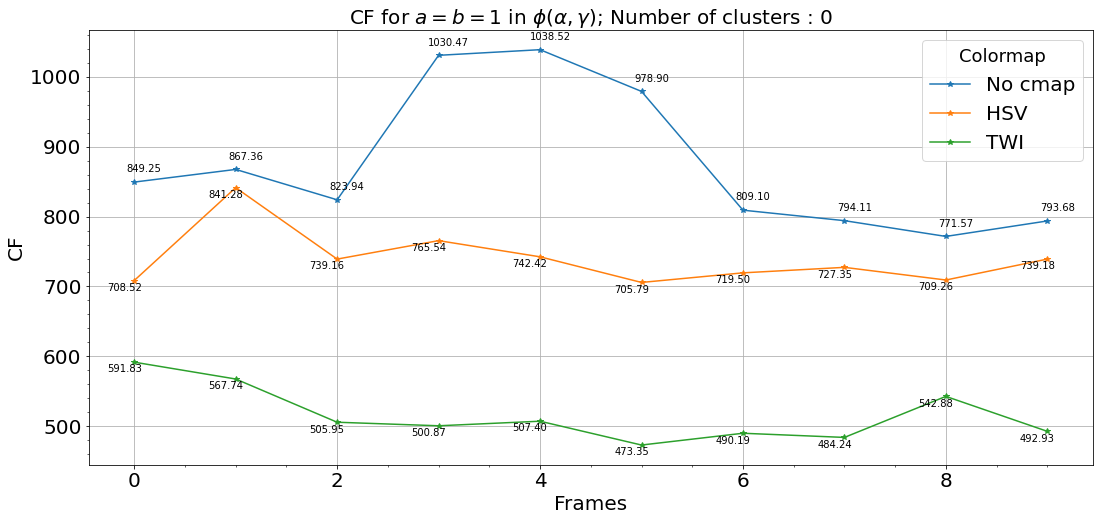}}
    \subfigure{\includegraphics[width=8cm,height=5cm]{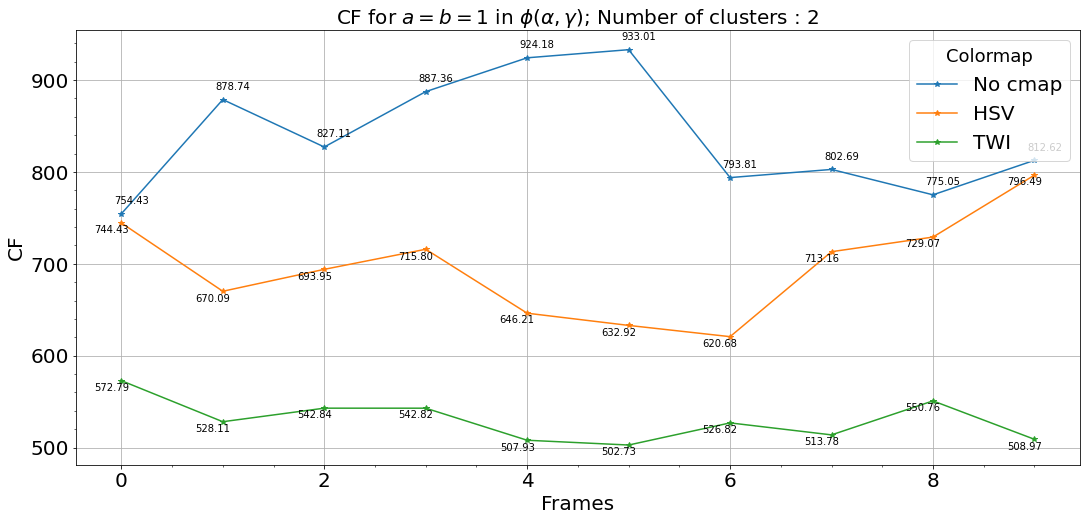}}
    \subfigure{\includegraphics[width=8cm,height=5cm]{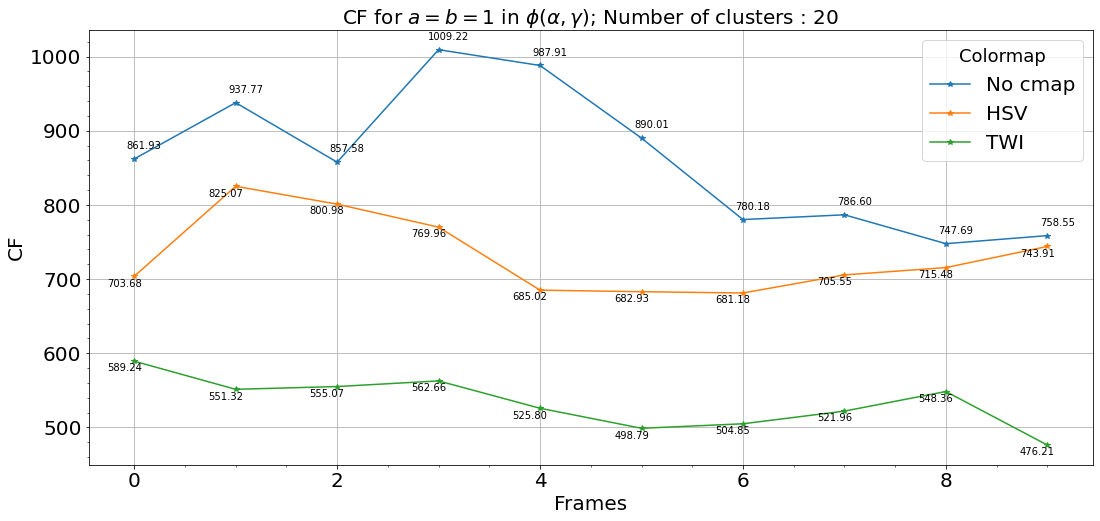}}
    \subfigure{\includegraphics[width=8cm,height=5cm]{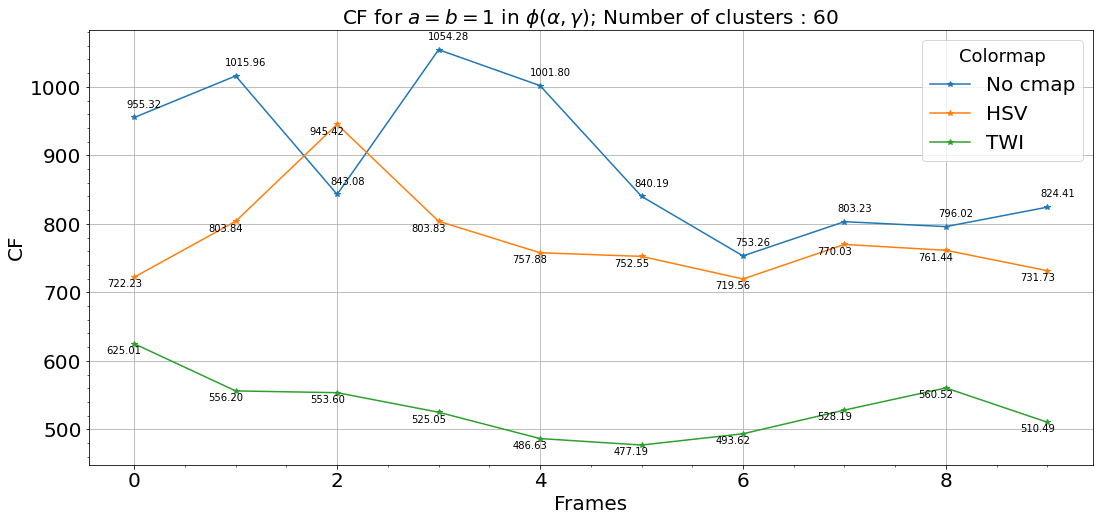}}
    \caption{Variation of CF with colormap corresponding to each clusters}
    \label{figd23}
\end{figure*}
\begin{figure*}[p]
\centering
    \subfigure{\includegraphics[width=6cm,height=3.5cm]{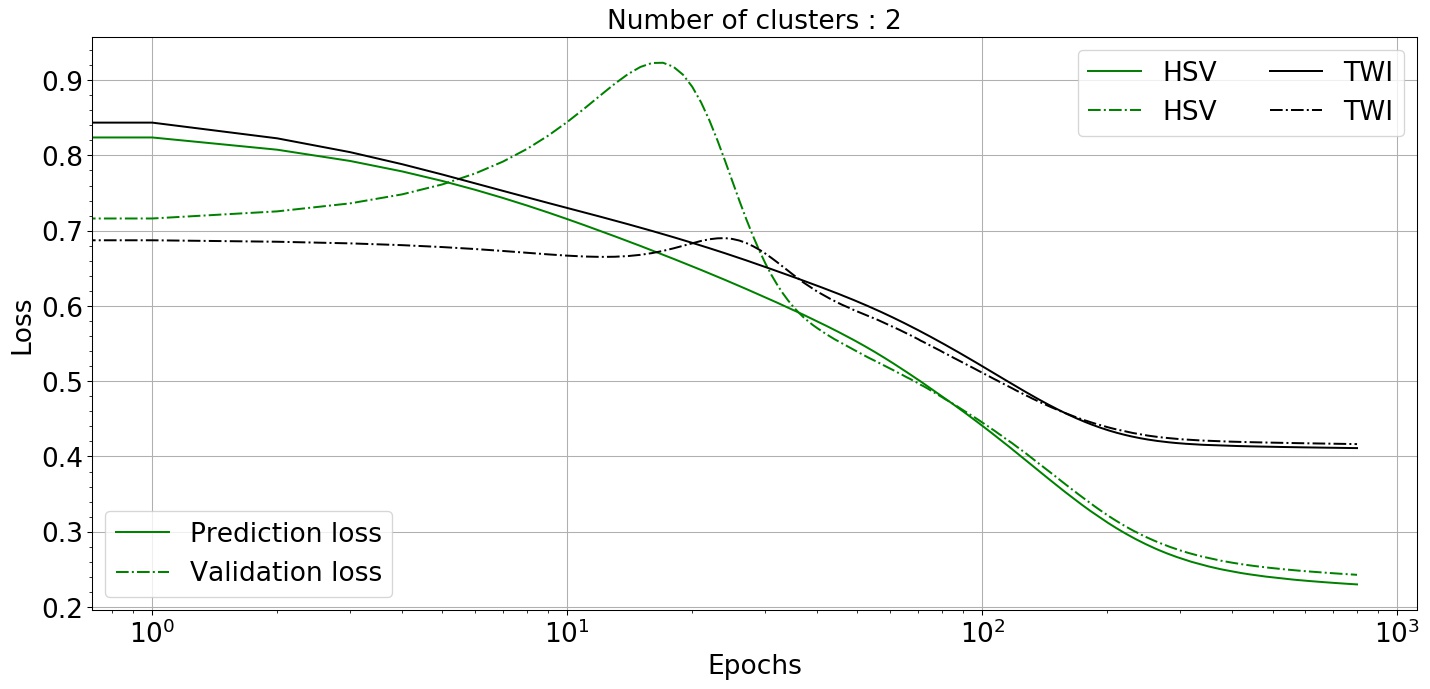}}
    \subfigure{\includegraphics[width=6cm,height=3.5cm]{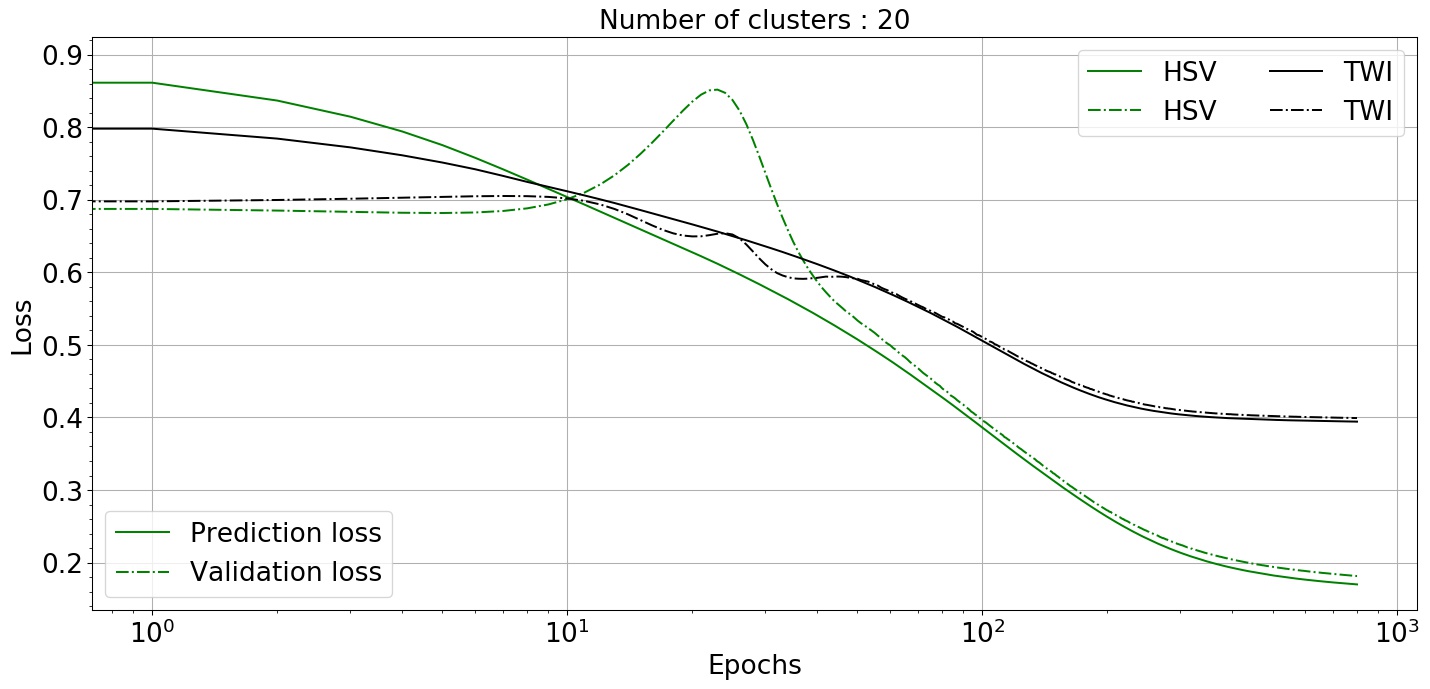}}
    \subfigure{\includegraphics[width=6cm,height=3.5cm]{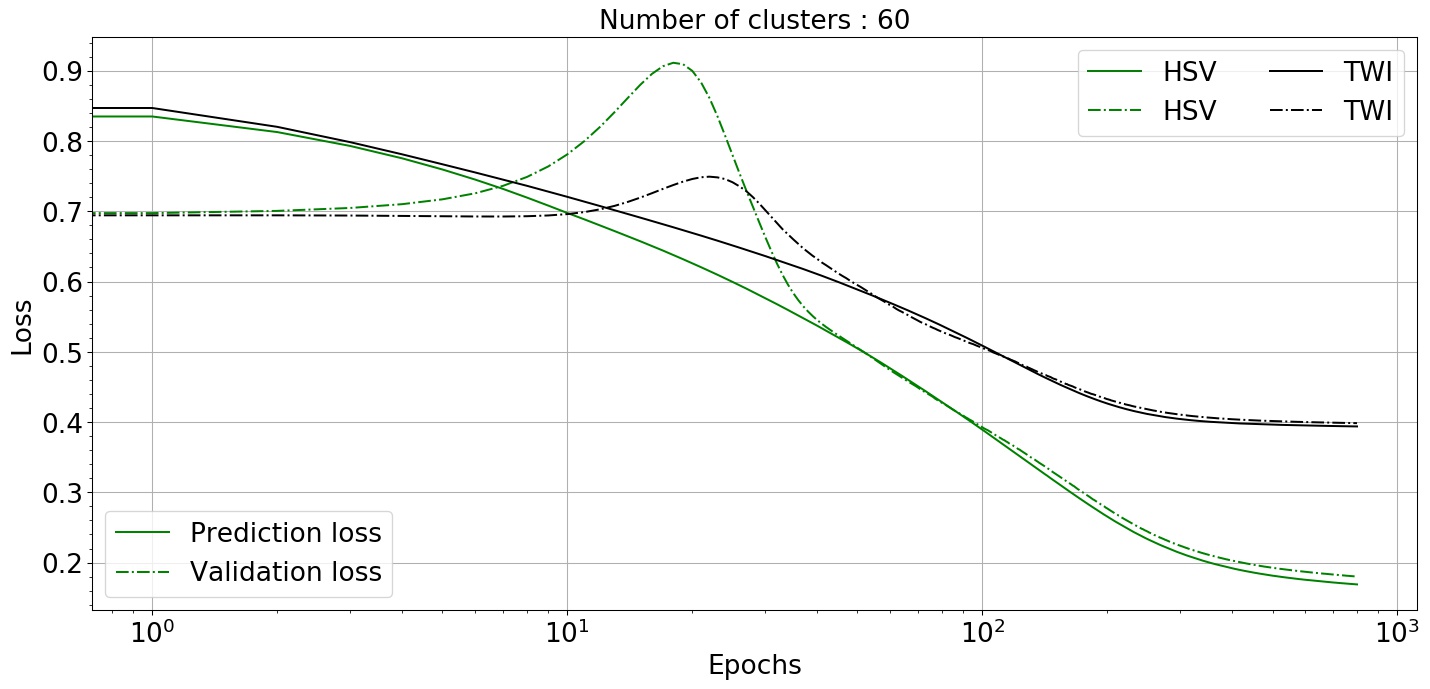}}
    \caption{Variation of loss function with colormap corresponding to each clusters}
    \label{figd31}
\end{figure*}

\begin{figure*}[p]
\centering
    \subfigure{\includegraphics[width=6cm,height=3.5cm]{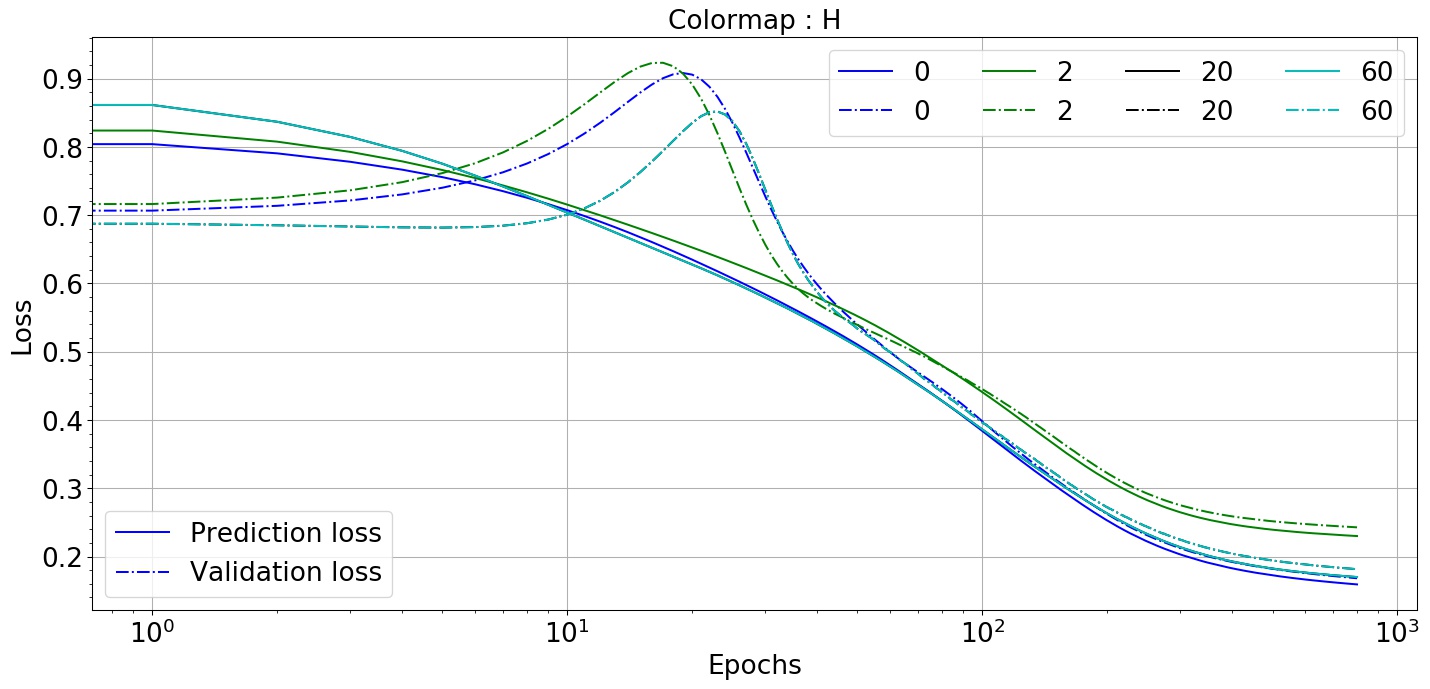}}
    \subfigure{\includegraphics[width=6cm,height=3.5cm]{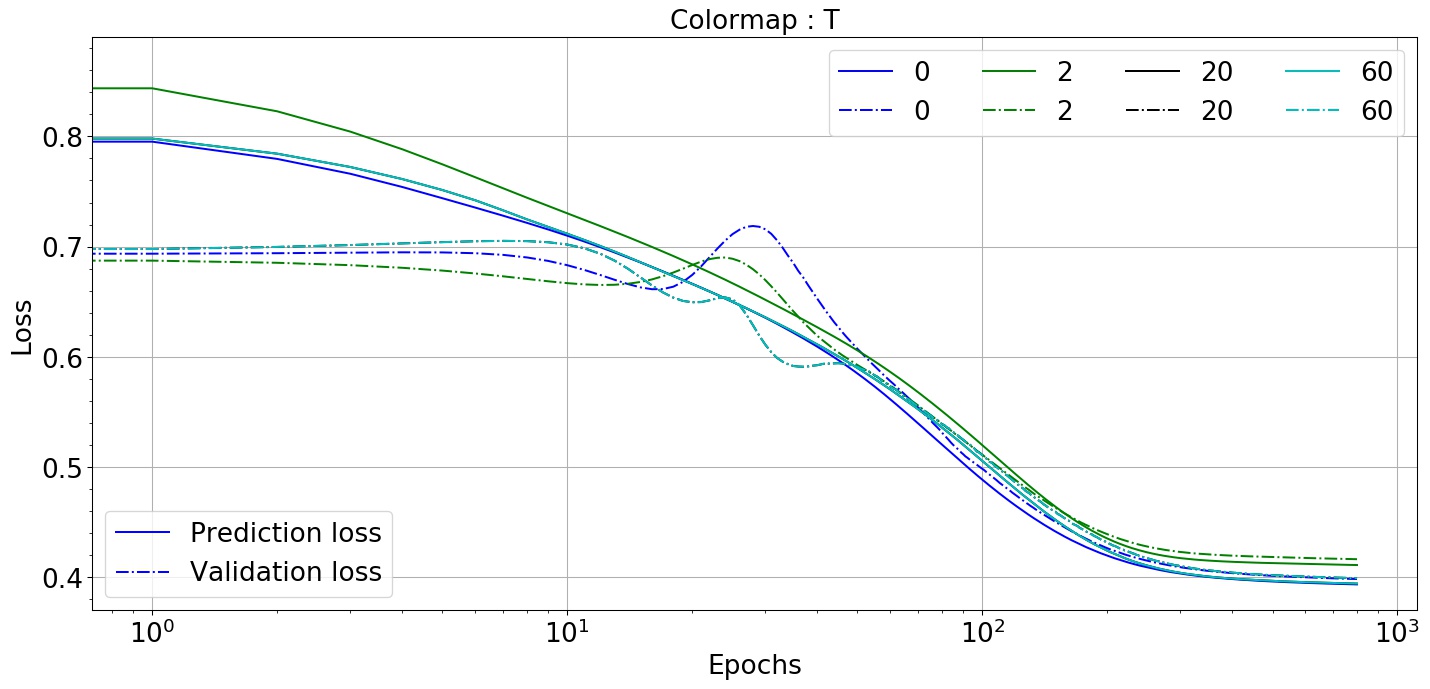}}
    \caption{Variation of loss function with each clusters corresponding to the two colormaps}
    \label{figd32}
\end{figure*}

\begin{figure*}[p]
\centering
    \subfigure{\includegraphics[width=8cm,height=5cm]{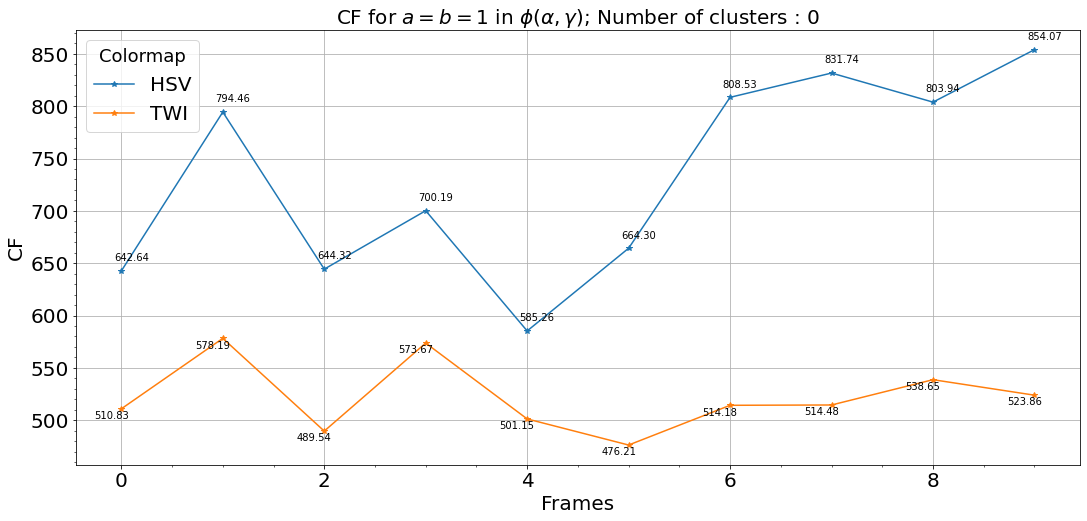}}
    \subfigure{\includegraphics[width=8cm,height=5cm]{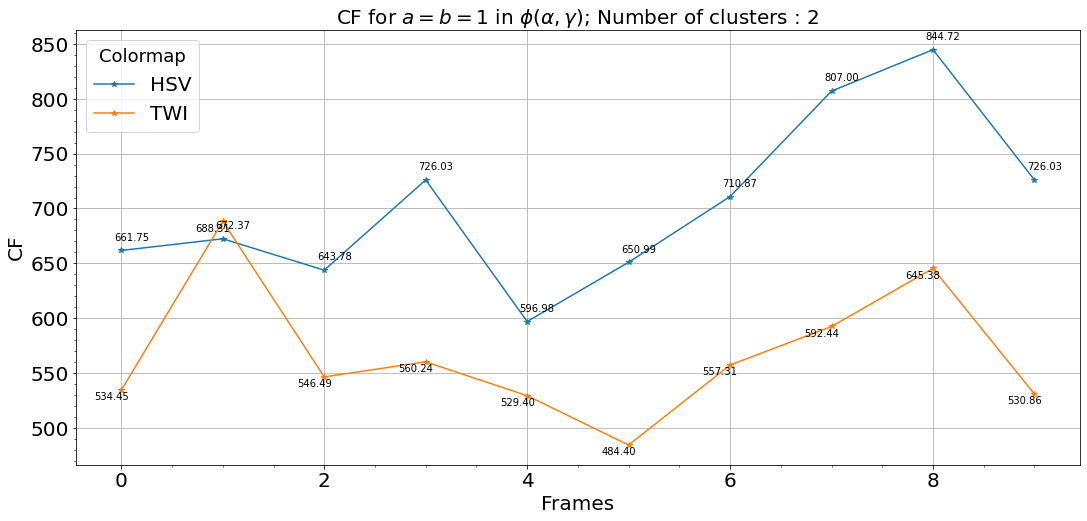}}
    \subfigure{\includegraphics[width=8cm,height=5cm]{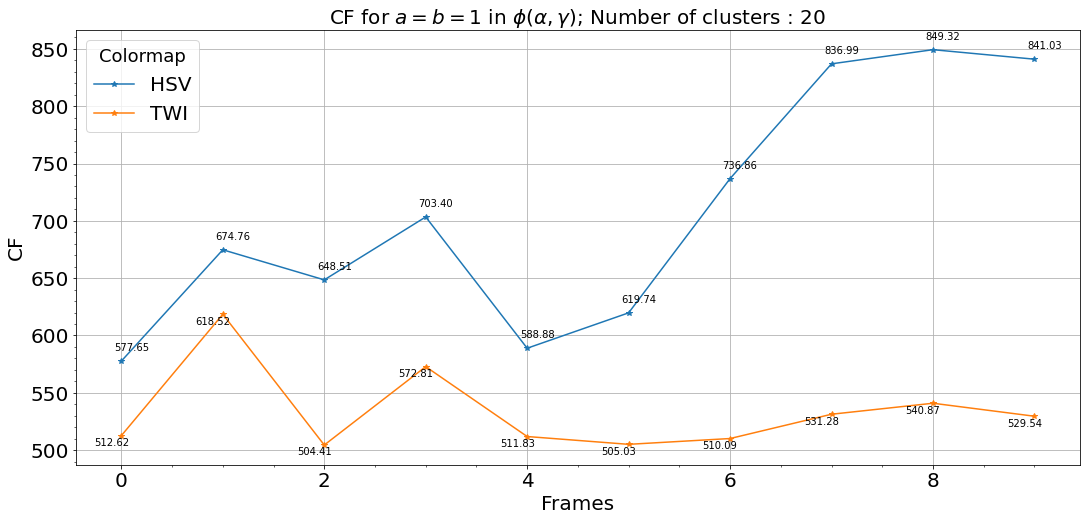}}
    \subfigure{\includegraphics[width=8cm,height=5cm]{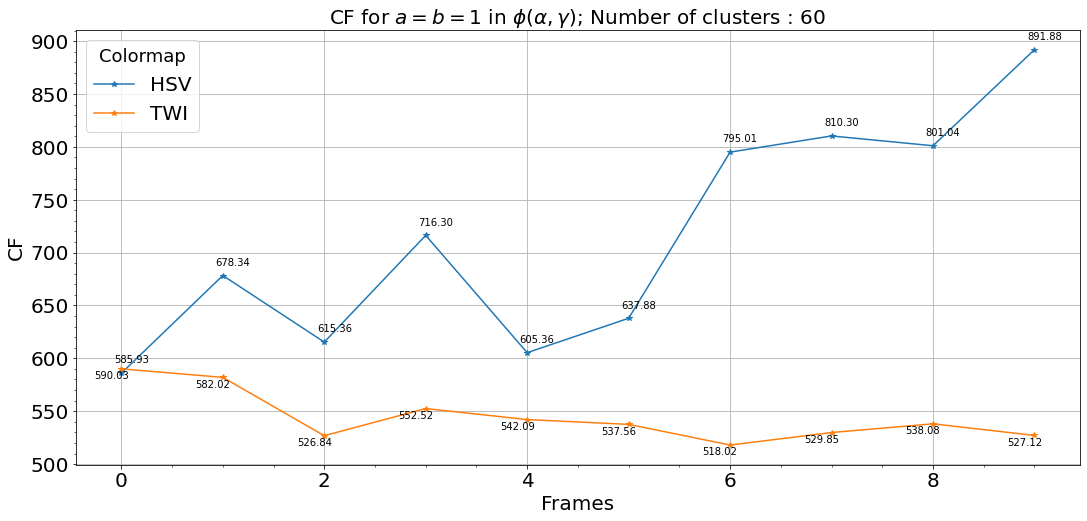}}
    \caption{Variation of CF with colormap corresponding to each clusters}
    \label{figd33}
\end{figure*}

\begin{figure*}[p]
\centering
    \subfigure{\includegraphics[width=6cm,height=3.5cm]{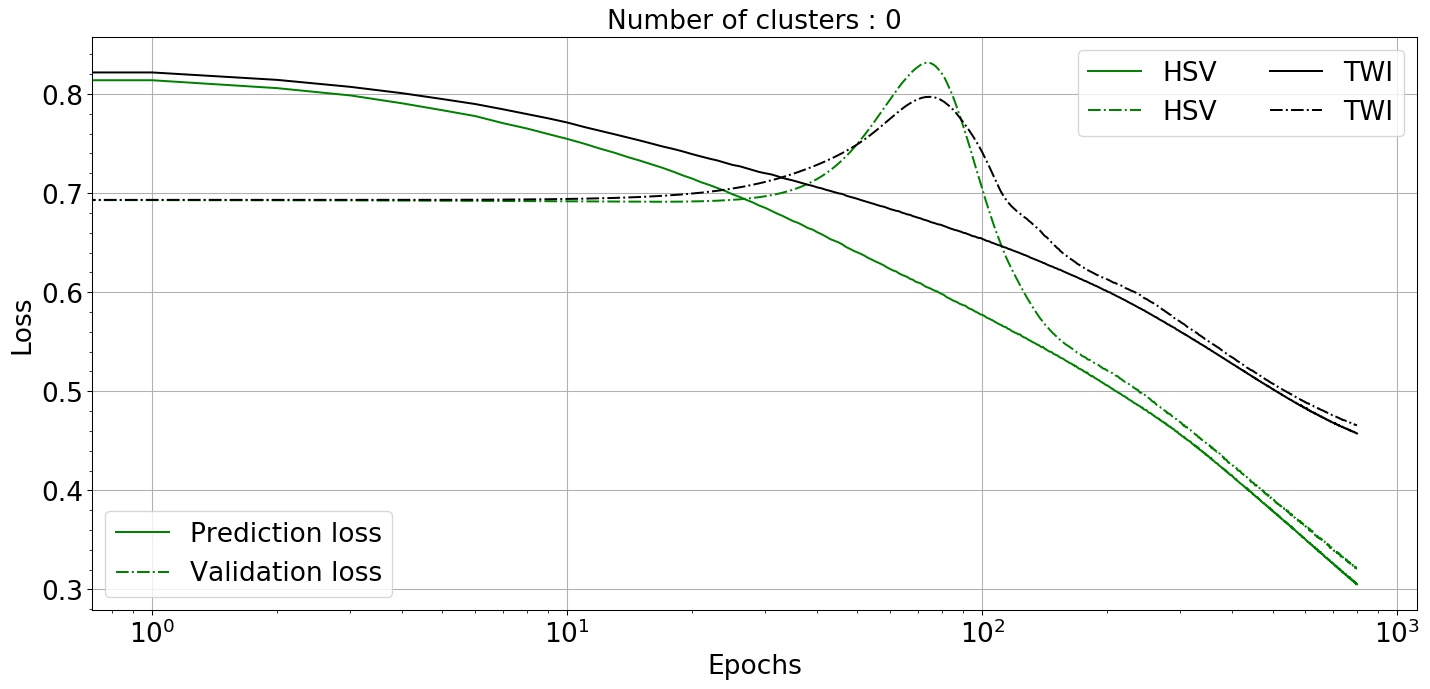}}
    \subfigure{\includegraphics[width=6cm,height=3.5cm]{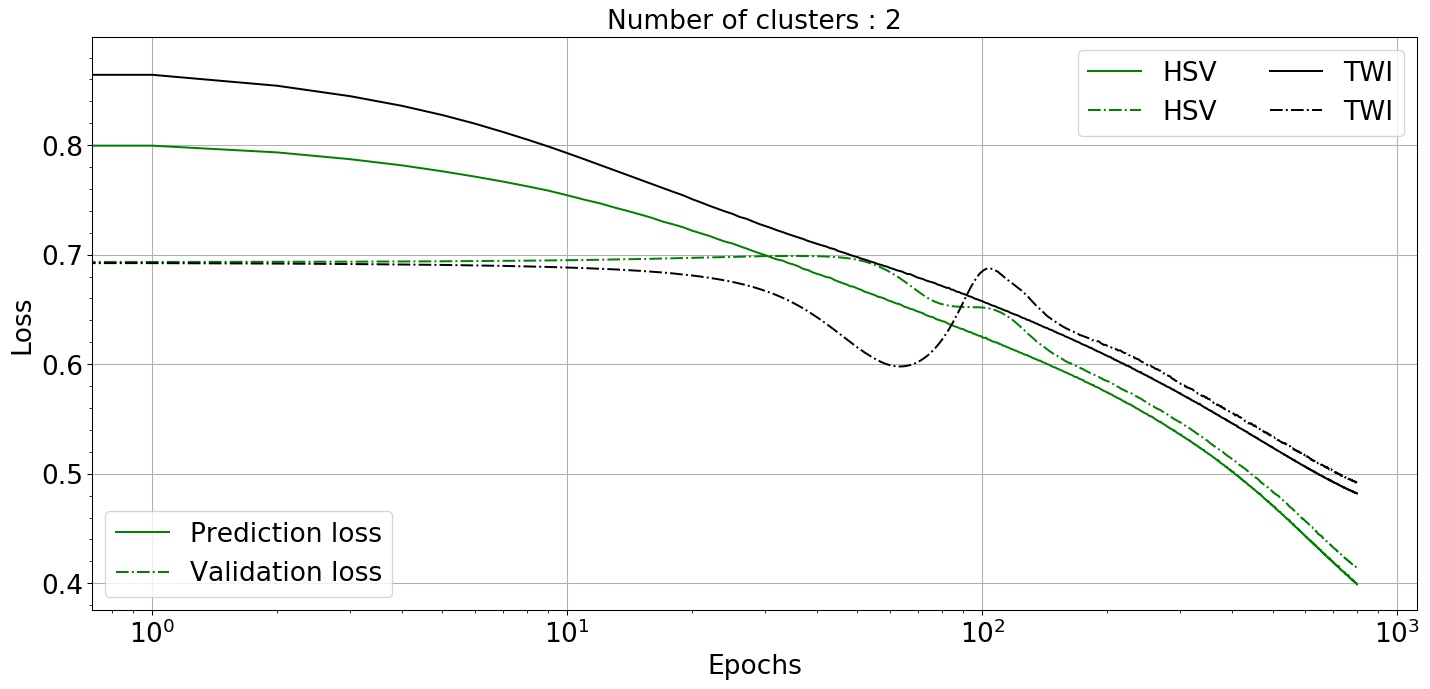}}
    \subfigure{\includegraphics[width=6cm,height=3.5cm]{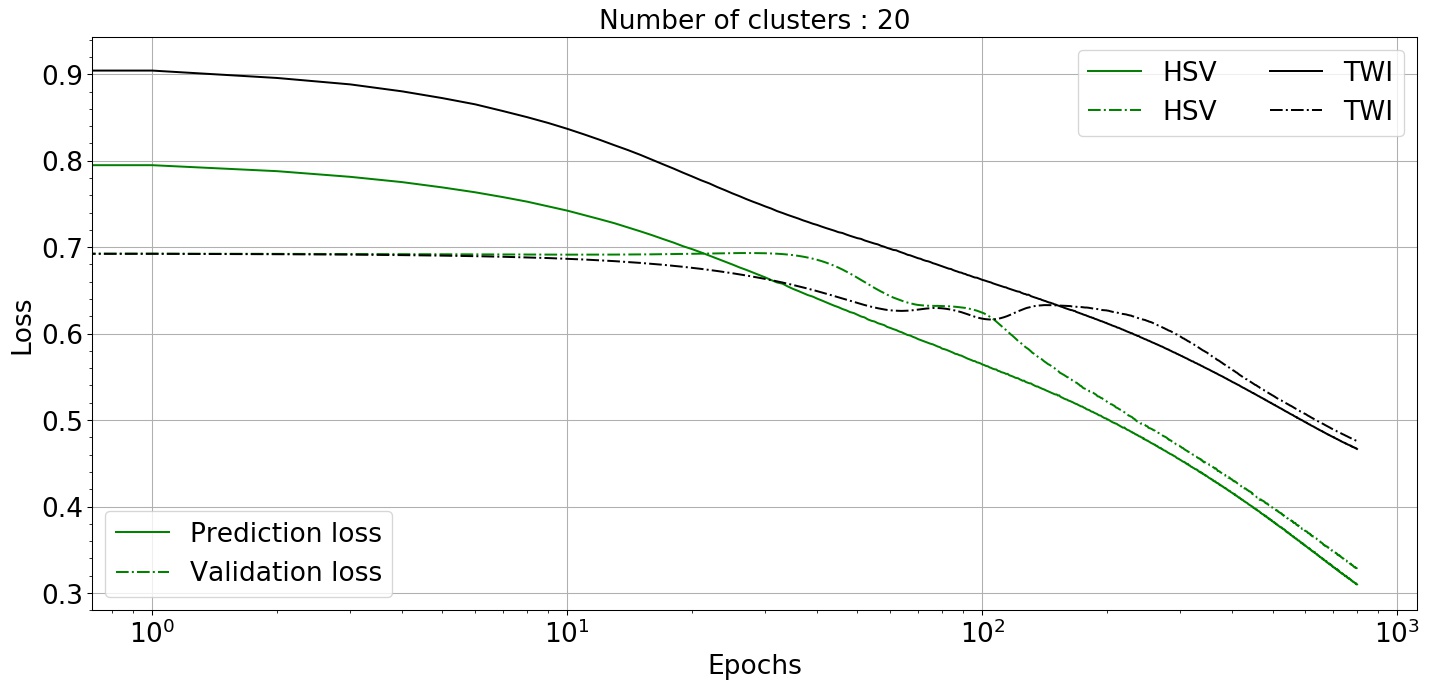}}
    \caption{Variation of loss function with colormap corresponding to each clusters}
    \label{figd41}
\end{figure*}

\begin{figure*}[p]
\centering
    \subfigure{\includegraphics[width=6cm,height=3.5cm]{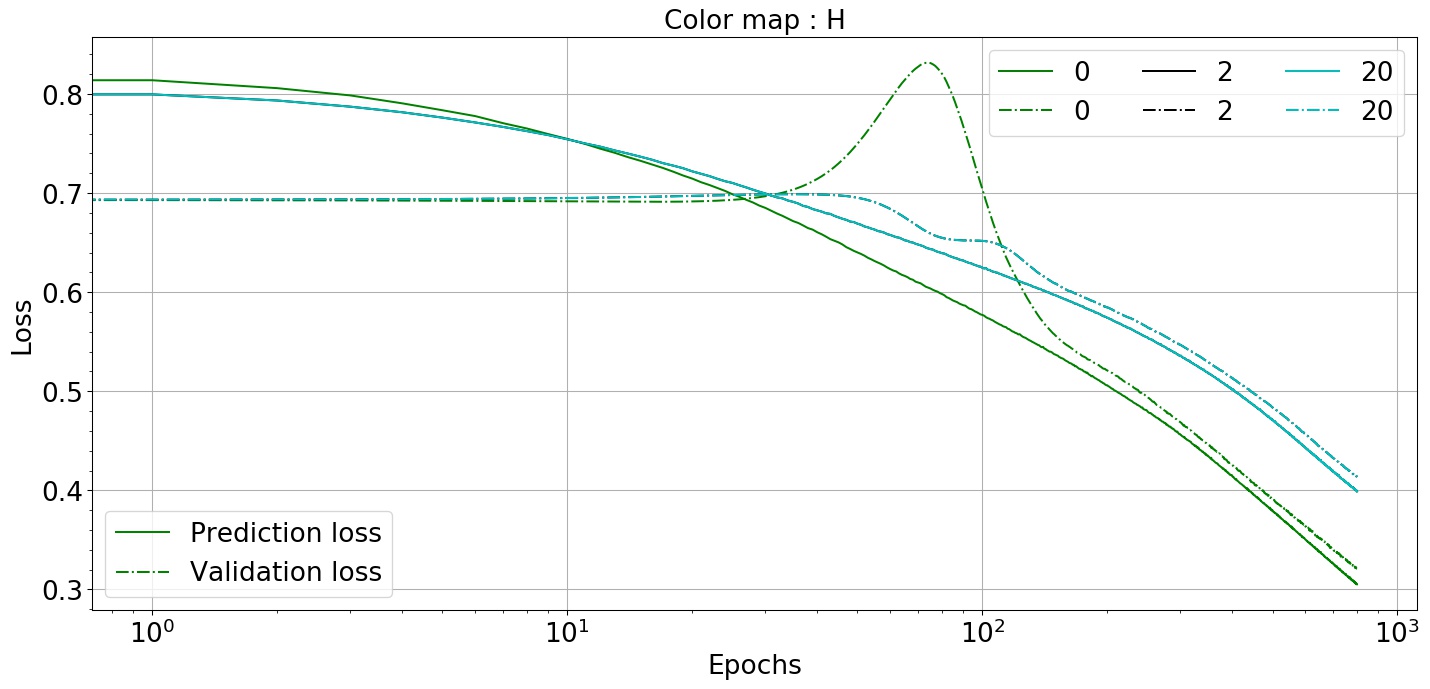}}
    \subfigure{\includegraphics[width=6cm,height=3.5cm]{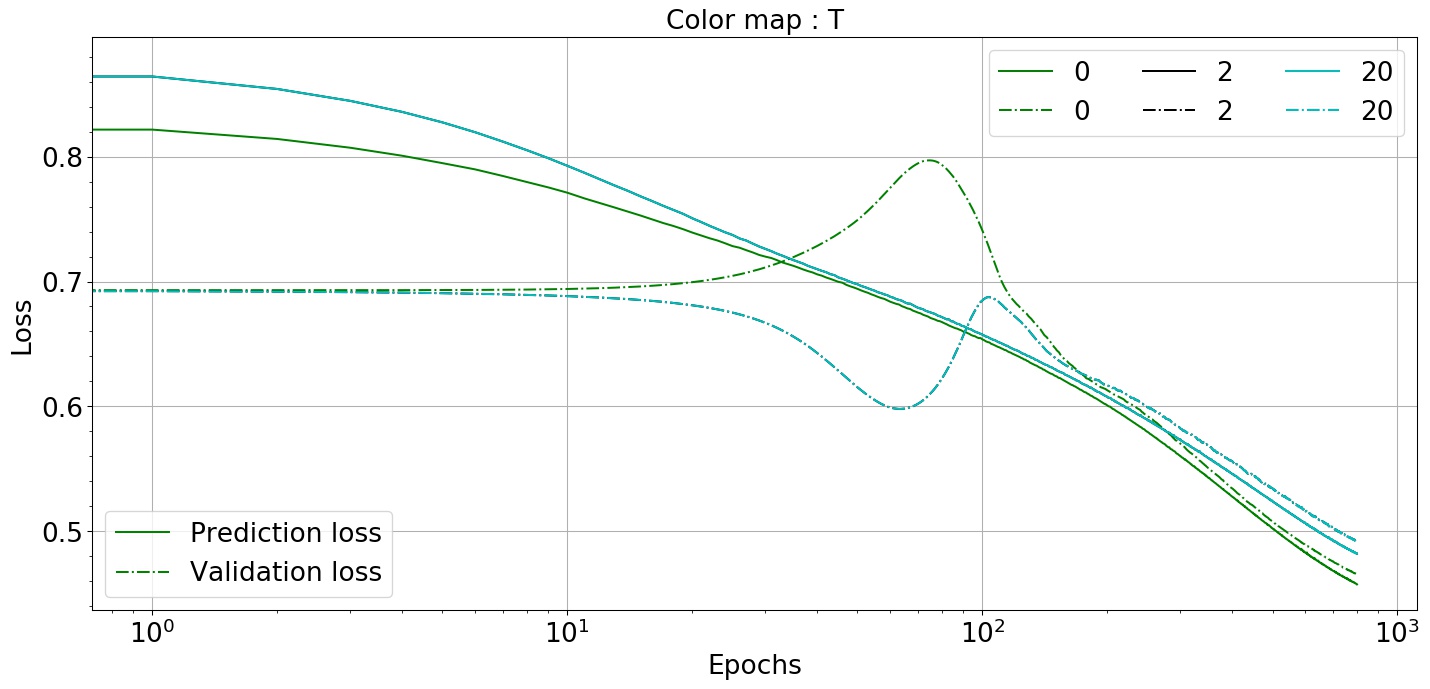}}
    \caption{Variation of loss function with each clusters corresponding to the two colormaps}
    \label{figd42}
\end{figure*}

\begin{figure*}[p]
\centering
    \subfigure{\includegraphics[width=8cm,height=5cm]{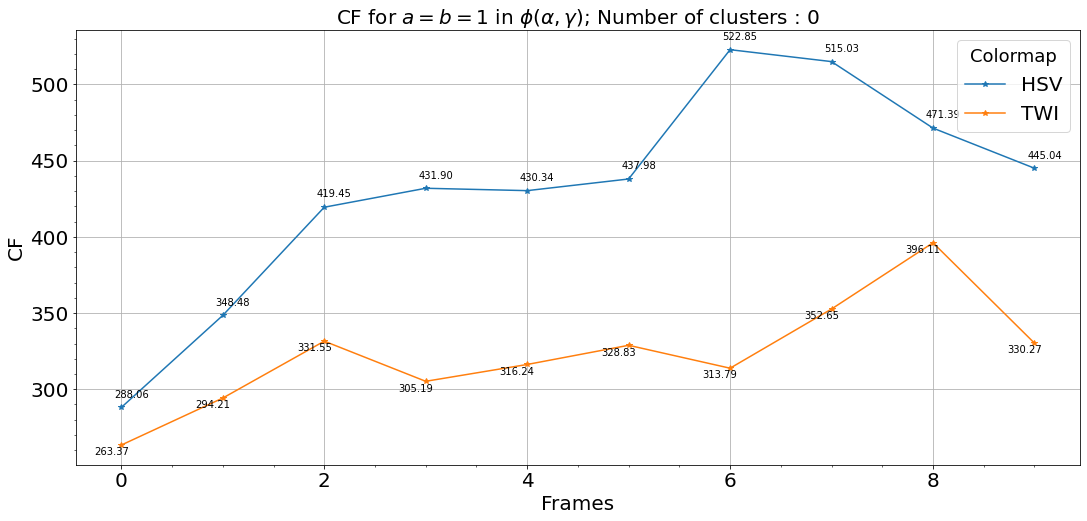}}
    \subfigure{\includegraphics[width=8cm,height=5cm]{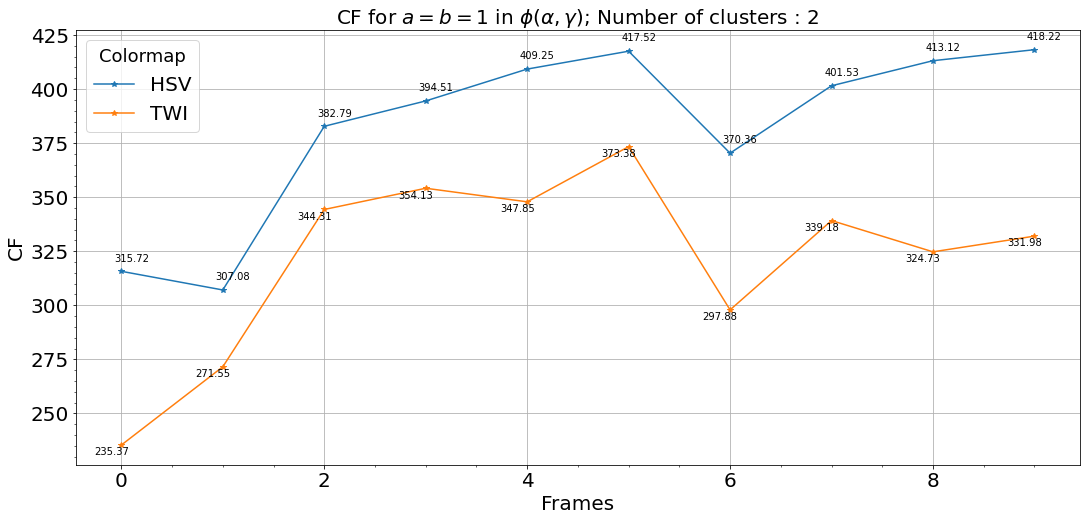}}
    \subfigure{\includegraphics[width=8cm,height=5cm]{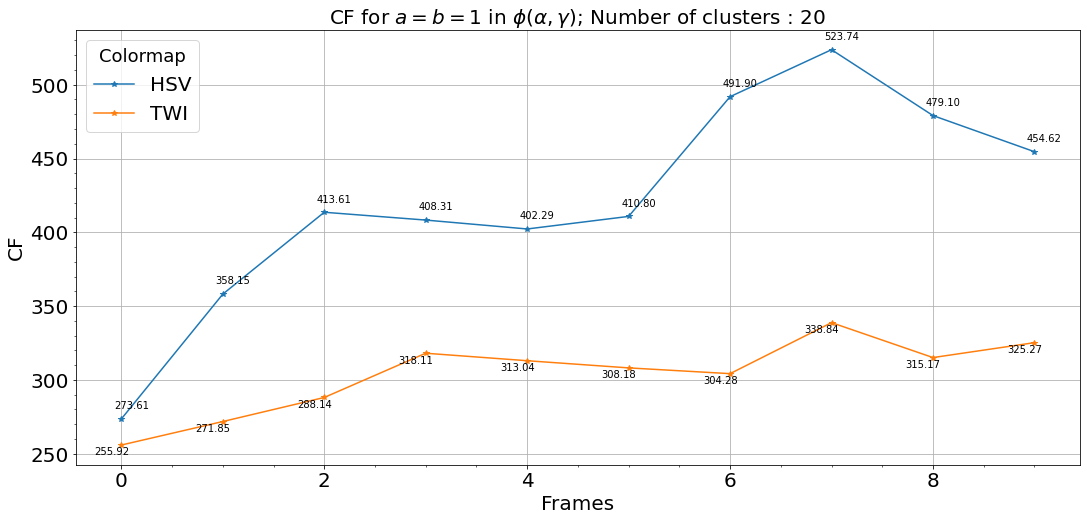}}
    \caption{Variation of CF with colormap corresponding to each clusters}
    \label{figd43}
\end{figure*}

\begin{figure*}
\centering
    \subfigure{\includegraphics[width=3cm,height=2cm]{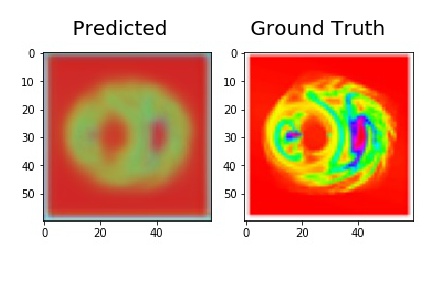}}
    \subfigure{\includegraphics[width=3cm,height=2cm]{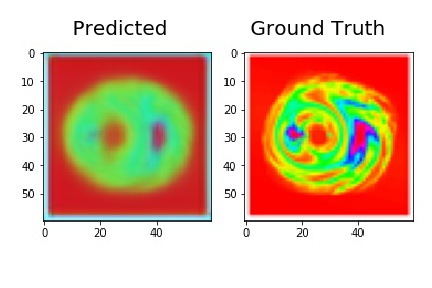}}
    \subfigure{\includegraphics[width=3cm,height=2cm]{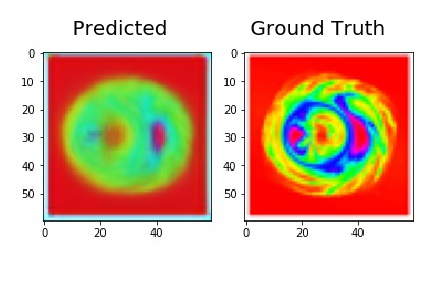}}
    \subfigure{\includegraphics[width=3cm,height=2cm]{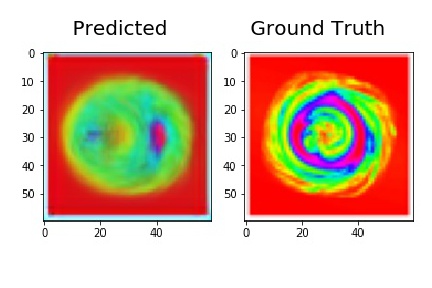}}
    \subfigure{\includegraphics[width=3cm,height=2cm]{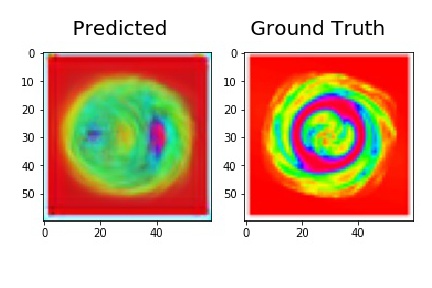}}
    \subfigure{\includegraphics[width=3cm,height=2cm]{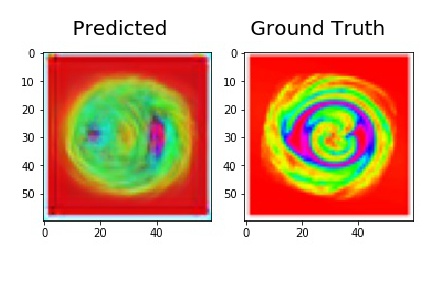}}
    \subfigure{\includegraphics[width=3cm,height=2cm]{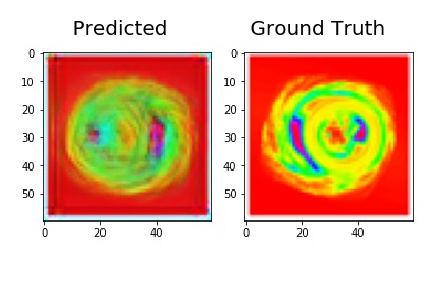}}
    \subfigure{\includegraphics[width=3cm,height=2cm]{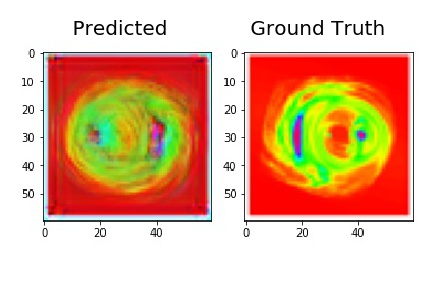}}
    \subfigure{\includegraphics[width=3cm,height=2cm]{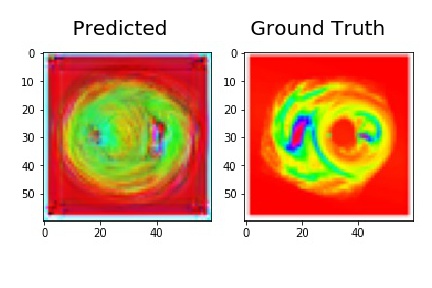}}
    \subfigure{\includegraphics[width=3cm,height=2cm]{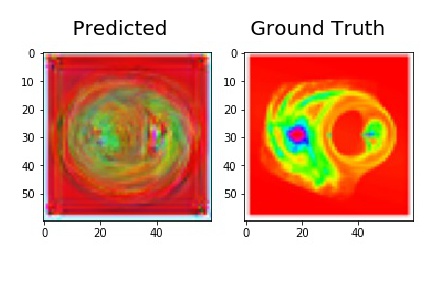}}
    \caption{Predicted frame and Ground-truth frame plotted side-by-side for all $10$ framed for HSV colormap}
    \label{fig3}
\end{figure*}

\begin{figure*}
\centering
    \subfigure{\includegraphics[width=3cm,height=2cm]{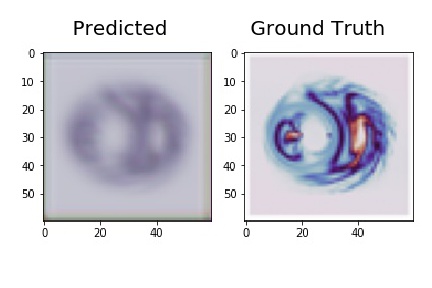}}
    \subfigure{\includegraphics[width=3cm,height=2cm]{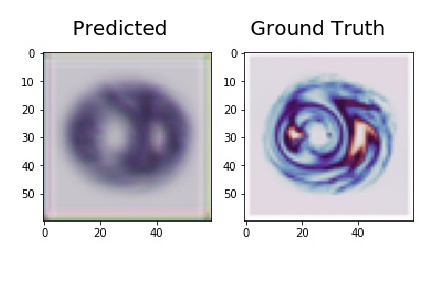}}
    \subfigure{\includegraphics[width=3cm,height=2cm]{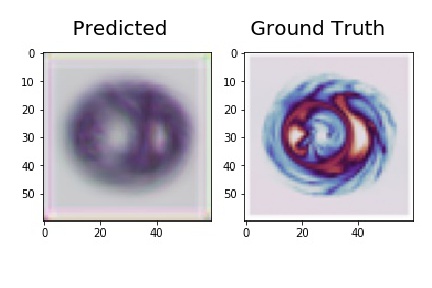}}
    \subfigure{\includegraphics[width=3cm,height=2cm]{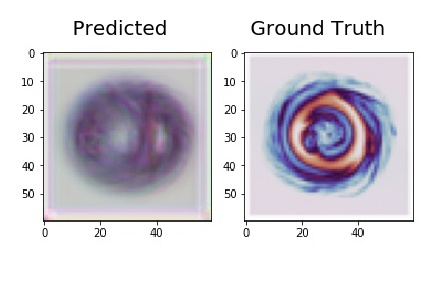}}
    \subfigure{\includegraphics[width=3cm,height=2cm]{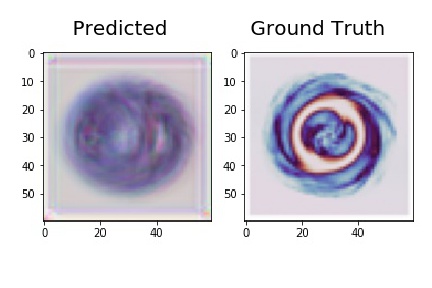}}
    \subfigure{\includegraphics[width=3cm,height=2cm]{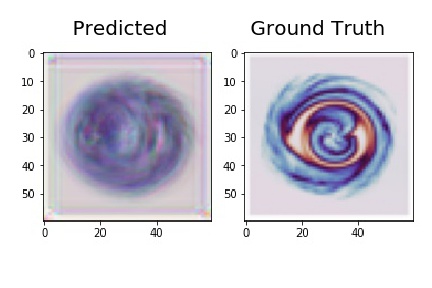}}
    \subfigure{\includegraphics[width=3cm,height=2cm]{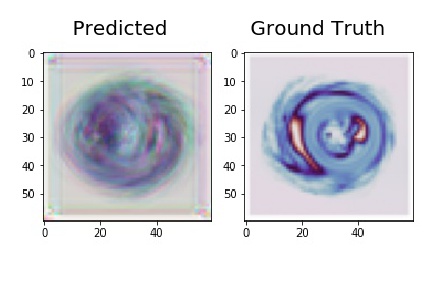}}
    \subfigure{\includegraphics[width=3cm,height=2cm]{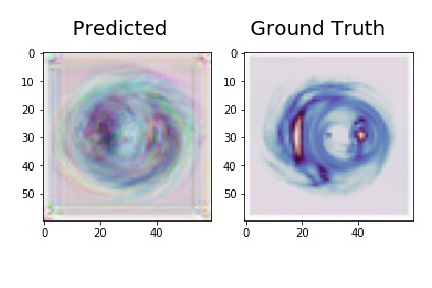}}
    \subfigure{\includegraphics[width=3cm,height=2cm]{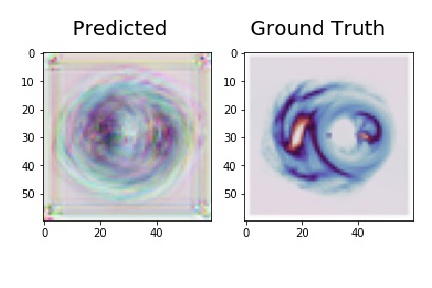}}
    \subfigure{\includegraphics[width=3cm,height=2cm]{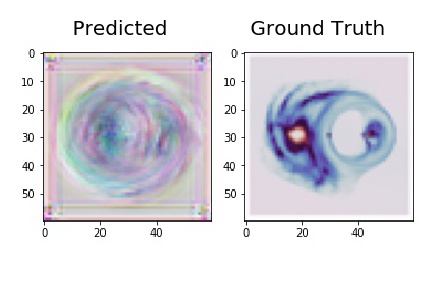}}
    \caption{Predicted frame and Ground-truth frame plotted side-by-side for all $10$ framed for TWILIGHT colormap}
    \label{fig5}
\end{figure*}

\begin{figure*}
\centering
    \subfigure{\includegraphics[width=1.5cm,height=1.5cm]{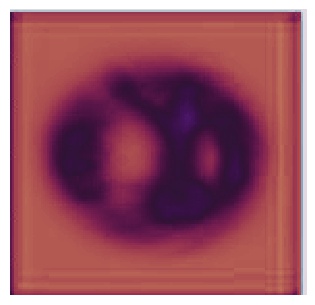}}
    \subfigure{\includegraphics[width=1.5cm,height=1.5cm]{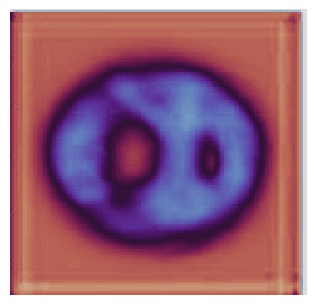}}
    \subfigure{\includegraphics[width=1.5cm,height=1.5cm]{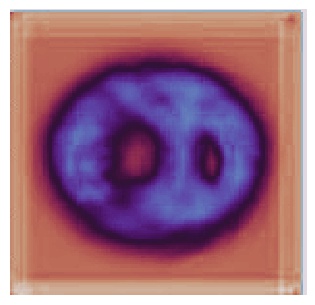}}
    \subfigure{\includegraphics[width=1.5cm,height=1.5cm]{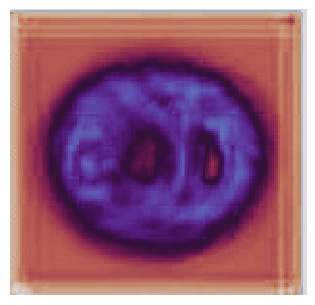}}
    \subfigure{\includegraphics[width=1.5cm,height=1.5cm]{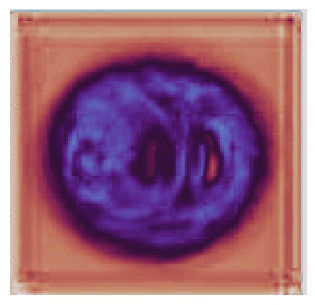}}
    \subfigure{\includegraphics[width=1.5cm,height=1.5cm]{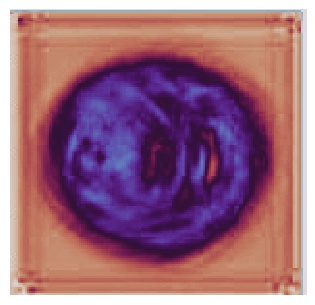}}
    \subfigure{\includegraphics[width=1.5cm,height=1.5cm]{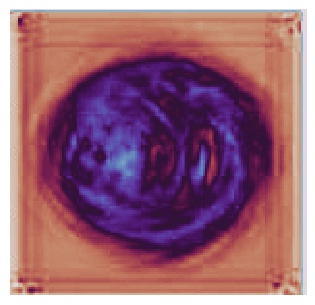}}
    \subfigure{\includegraphics[width=1.5cm,height=1.5cm]{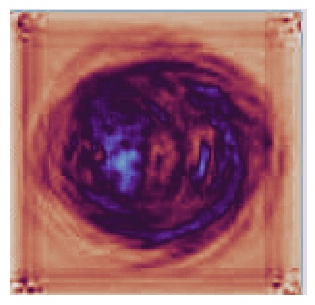}}
    \subfigure{\includegraphics[width=1.5cm,height=1.5cm]{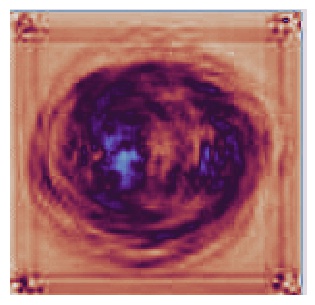}}
    \subfigure{\includegraphics[width=1.5cm,height=1.5cm]{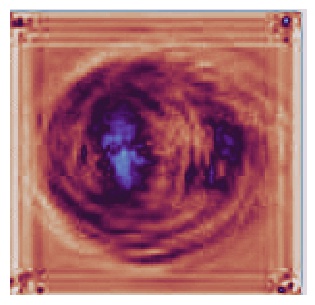}}

    \subfigure{\includegraphics[width=1.5cm,height=1.5cm]{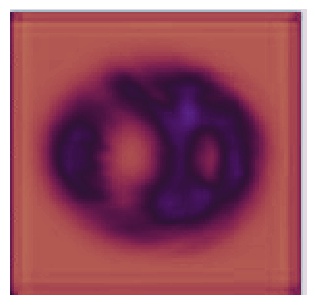}}
    \subfigure{\includegraphics[width=1.5cm,height=1.5cm]{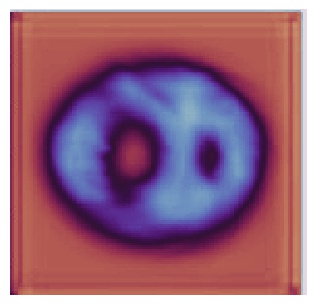}}
    \subfigure{\includegraphics[width=1.5cm,height=1.5cm]{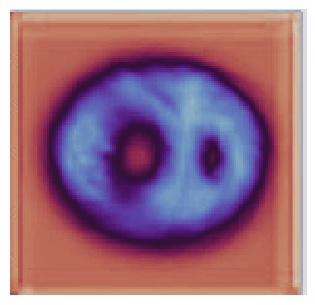}}
    \subfigure{\includegraphics[width=1.5cm,height=1.5cm]{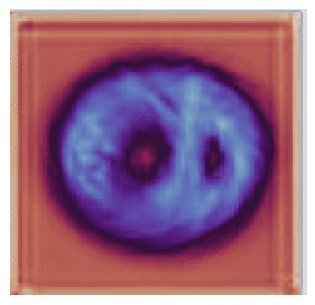}}
    \subfigure{\includegraphics[width=1.5cm,height=1.5cm]{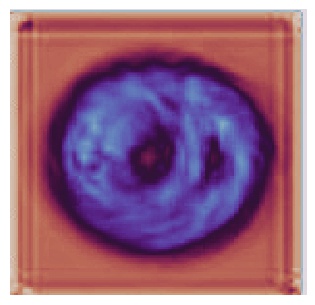}}
    \subfigure{\includegraphics[width=1.5cm,height=1.5cm]{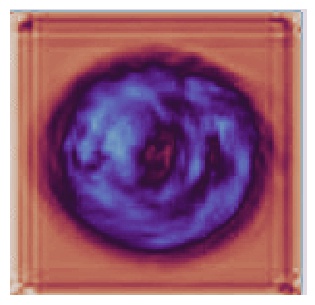}}
    \subfigure{\includegraphics[width=1.5cm,height=1.5cm]{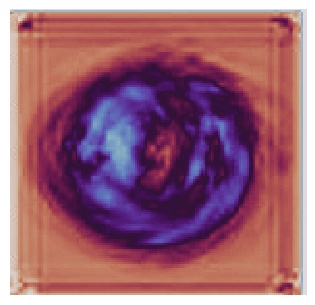}}
    \subfigure{\includegraphics[width=1.5cm,height=1.5cm]{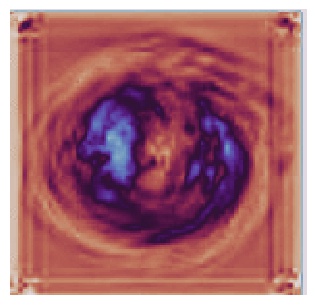}}
    \subfigure{\includegraphics[width=1.5cm,height=1.5cm]{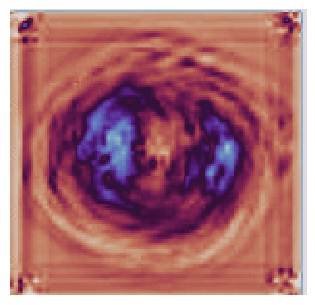}}
    \subfigure{\includegraphics[width=1.5cm,height=1.5cm]{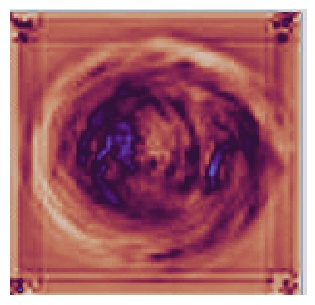}}

    \subfigure{\includegraphics[width=1.5cm,height=1.5cm]{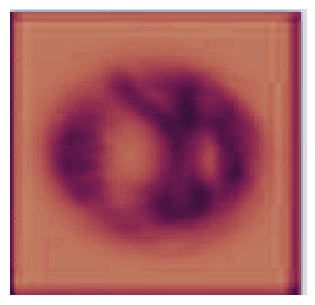}}
    \subfigure{\includegraphics[width=1.5cm,height=1.5cm]{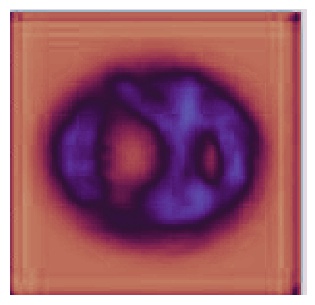}}
    \subfigure{\includegraphics[width=1.5cm,height=1.5cm]{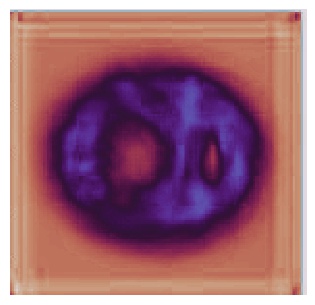}}
    \subfigure{\includegraphics[width=1.5cm,height=1.5cm]{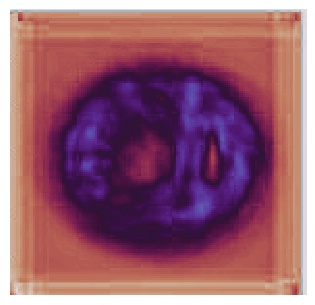}}
    \subfigure{\includegraphics[width=1.5cm,height=1.5cm]{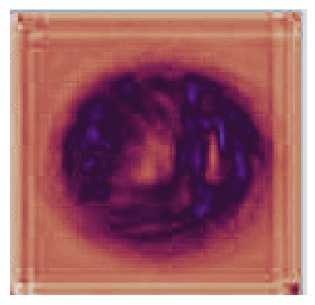}}
    \subfigure{\includegraphics[width=1.5cm,height=1.5cm]{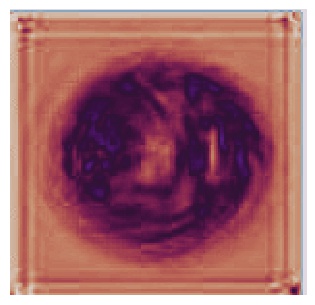}}
    \subfigure{\includegraphics[width=1.5cm,height=1.5cm]{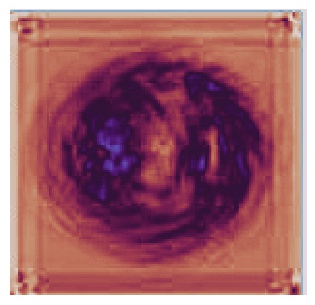}}
    \subfigure{\includegraphics[width=1.5cm,height=1.5cm]{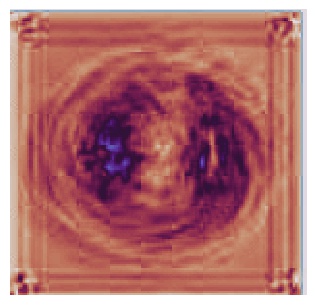}}
    \subfigure{\includegraphics[width=1.5cm,height=1.5cm]{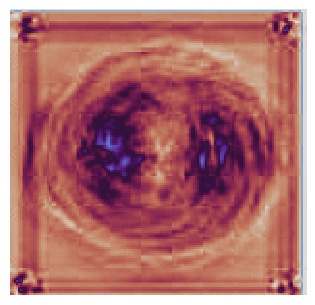}}
    \subfigure{\includegraphics[width=1.5cm,height=1.5cm]{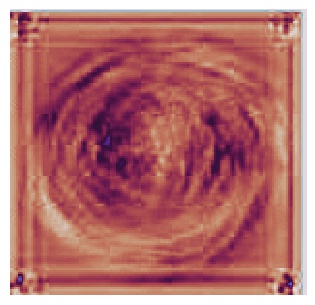}}   
    \label{phif}
    \caption{$10$ frames of each color channel - R, G, B of a TWILIGHT colormap highliting the idea of each colormap as a latent space. We evaluate CF from Eq-\ref{eqcf} combining all the latent space.}
\end{figure*}

\label{proof}
\begin{multline}
\phi(\alpha,\gamma) = \log \frac{1}{\sqrt{a\alpha^2+b\gamma^2}}  = -\frac{1}{2}\log({a\alpha^2+b\gamma^2})\\
d\phi(\alpha,\gamma)  = \frac{ \partial \phi}{\partial \alpha}d\alpha +  \frac{ \partial \phi}{\partial \gamma}d\gamma = 0 \\
\implies  {d\phi}(\alpha,\gamma) = \frac{-a\alpha d\alpha - b\gamma d\gamma}{a\alpha^2 + b\gamma^2} =0\\
\implies a\alpha d\alpha =  - b\gamma d\gamma
\implies d\alpha = \frac{f(x,y) df(x,y) -g(x,y) dg(x,y)}{\sqrt{f(x,y)^2-g(x,y)^2}} \\
\end{multline}
Now, let $\gamma = \eta_1 - \eta_2 \implies d\gamma = d\eta_1 - d\eta_2$ 
Thus on evaluating $d\eta_1$ and $d\eta_2$ and using the condition of minima $ a\alpha d\alpha =  - b\gamma d\gamma $ we get


\begin{multline}
	a [f(x,y)df(x,y) - g(x,y)dg(x,y)] = - b[\gamma(d\eta_1 - d\eta_2)]
\end{multline}

\end{appendix}
\end{document}